\documentclass[11pt]{article}
\pdfoutput = 1
\usepackage[utf8]{inputenc}
\renewcommand{\arraystretch}{2}
\usepackage[top=1in,bottom=1in,left=1in,right=1in]{geometry}
\makeatletter \g@addto@macro\@floatboxreset\centering \makeatother
\usepackage{color}
\definecolor{darkgreen}{rgb}{0,0.5,0}
\definecolor{darkblue}{rgb}{0,0,0.6}
\definecolor{purple}{rgb}{0.4,.2,0.7}
\usepackage[noEucal]{main}
\usepackage{aas_macros,empheq,fancybox,graphicx,multirow}
\usepackage{aas_macros}
\usepackage{dsfont}
\usepackage{esint}
\usepackage{float}
\usepackage{appendix}
\usepackage{dsfont}
\usepackage{stmaryrd}
\usepackage{changepage}
\usepackage{mathabx}
\usepackage{hhline}
\usepackage{booktabs}
\usepackage{tensor} 
\usepackage{footmisc}
\usepackage{tabularray}
\usepackage{float}
\usepackage{caption}
\usepackage{relsize}

\DeclareSymbolFont{cmlargesymbols}{OMX}{cmex}{m}{n}
\let\sumop\relax
\DeclareMathSymbol{\sumop}{\mathop}{cmlargesymbols}{"50}

\usepackage[framemethod=TikZ]{mdframed}

\newcommand{\overbar}[1]{\mkern 1.5mu\overline{\mkern-1.5mu#1\mkern-1.5mu}\mkern 1.5mu}

\makeatletter
\DeclareFontFamily{OMX}{MnSymbolE}{}
\DeclareSymbolFont{MnLargeSymbols}{OMX}{MnSymbolE}{m}{n}
\SetSymbolFont{MnLargeSymbols}{bold}{OMX}{MnSymbolE}{b}{n}
\DeclareFontShape{OMX}{MnSymbolE}{m}{n}{
    <-6>  MnSymbolE5
   <6-7>  MnSymbolE6
   <7-8>  MnSymbolE7
   <8-9>  MnSymbolE8
   <9-10> MnSymbolE9
  <10-12> MnSymbolE10
  <12->   MnSymbolE12
}{}
\DeclareFontShape{OMX}{MnSymbolE}{b}{n}{
    <-6>  MnSymbolE-Bold5
   <6-7>  MnSymbolE-Bold6
   <7-8>  MnSymbolE-Bold7
   <8-9>  MnSymbolE-Bold8
   <9-10> MnSymbolE-Bold9
  <10-12> MnSymbolE-Bold10
  <12->   MnSymbolE-Bold12
}{}

\let\llangle\@undefined
\let\rrangle\@undefined
\DeclareMathDelimiter{\llangle}{\mathopen}%
                     {MnLargeSymbols}{'164}{MnLargeSymbols}{'164}
\DeclareMathDelimiter{\rrangle}{\mathclose}%
                     {MnLargeSymbols}{'171}{MnLargeSymbols}{'171}
\makeatother

\newcommand{\bracenom}{\genfrac{\lbrace}{\rbrace}{0pt}{}}

\DeclareFontFamily{U}{jkpmia}{}
\DeclareFontShape{U}{jkpmia}{m}{it}{<->s*jkpmia}{}
\DeclareFontShape{U}{jkpmia}{bx}{it}{<->s*jkpbmia}{}
\DeclareMathAlphabet{\mathfrakalt}{U}{jkpmia}{m}{it}
\SetMathAlphabet{\mathfrakalt}{bold}{U}{jkpmia}{bx}{it}

\newcommand{\myw}{\mathfrakalt{w}}

\begin{document}

\thispagestyle{empty}

\begin{center}
    ~
    \vskip3cm

     {\LARGE  {\textbf{Symmetries of the Celestial Supersphere}}}
    
\vspace{2cm}
Adam Tropper \\
    \vskip1em
    {\it
        Center for the Fundamental Laws of Nature, \\
Harvard University, Cambridge, MA 02138\\ \vskip1mm
         \vskip1mm
    }
    \vskip5mm
    \tt{adam$\_$tropper@g.harvard.edu}
\end{center}
\vspace{10mm}

\begin{abstract}
\noindent

We study the celestial CFT dual to theories with bulk supersymmetry. The boundary theory realizes supersymmetry in the spirit of the Green-Schwarz superstring: there is manifest $4d$ super-Poincar\'e symmetry, but no $2d$ superconformal symmetry. Nevertheless, we can extend the celestial sphere itself to a supermanifold --- the celestial supersphere. This provides a unified framework for describing key features of celestial holography, including conformally soft theorems, OPEs, and chiral soft algebras. Using these tools, we demonstrate that the $\frak{bms}_{4}$ algebra extends to a novel $\frak{sbms}_{4|\mathcal{N}}$ algebra. We also relate the supersymmetric $\mathcal{L}(\myw_{1+\infty}^\wedge)$ algebra to Hamiltonian vector fields on $\mathbb{C}^{2|\mathcal{N}}$, consistent with the expectation from twistor theory, and deduce the deformation of this algebra by a cosmological constant, $\Lambda$. These results are all universal and independent of the specific details of the underlying theory.

\end{abstract}
\pagebreak

\setcounter{tocdepth}{2}
{\hypersetup{linkcolor=black}
\fontsize{11.4pt}{11pt}\selectfont
\tableofcontents
}

\section{Introduction}

AdS holography provides a robust framework for understanding quantum gravity; however, it is not obvious how to extend its successes to spacetimes that are asymptotically flat. One approach is afforded by the celestial holography program which aims to relate scattering amplitudes in $4d$ quantum gravity to correlation functions of a dual $2d$ \textit{celestial CFT} (CCFT) on the boundary of asymptotically flat spacetime (for reviews, see \cite{Pasterski:2021rjz, Raclariu:2021zjz, Donnay:2023mrd, Strominger:2017zoo}).

Supersymmetry imposes strong constraints on UV physics and, therefore, naturally plays an important part in top-down constructions of AdS holography. Moreover, the IR structure of physical theories commands a central role in celestial holography, and it is known that supersymmetry strongly restricts such features as well (see \cite{seiberg1994electric} for an example and \cite{Crawley:2024cak} for its interpretation in CCFT). Despite this, a systematic understanding of supersymmetry in celestial holography has not been developed. This article aims to address that gap in the literature.

What little is known suggests that the CCFT duals to supersymmetric theories are simplified in their own right. For example, supersymmetric theories obey the so-called \textit{double-residue condition} \cite{Ball:2023qim} which implies that OPEs have no branch cuts and chiral soft algebras satisfy the Jacobi identity \cite{Ball:2023sdz}. In this article, we also argue that the CCFT duals to supersymmetric theories have several additional compelling features making them desirable toy models to work with.

Such examples may illuminate how celestial holography works in a broader context. Indeed, supersymmetric theories are just ordinary theories whose couplings are tuned to certain values. One may hope to extract general lessons --- for example about the complicated and understudied role that fermions and their symmetries play (see \cite{Dumitrescu:2015fej, Lysov:2015jrs, Avery:2015iix, Pasterski:2021fjn, Pano:2021ewd, Agriela:2023dnw, Narayanan:2020amh} for some partial results) --- from such examples.

The most important aspect of this work is that we exclusively present \textit{universal} results concerning supersymmetry in celestial holography. I.e. instead of studying a specific bulk Lagrangian with some particular dynamics, we solely report on results which hold for \textit{all} supersymmetric theories. This is in contrast with previous papers which only study specific examples. Our approach clarifies which results are guaranteed to hold more broadly and which ones are artifacts of having chosen an overly simple model. For example, there are certain results which hold in \textit{minimal} $\mathcal{N} = 1$ supersymmetric gauge theory that are no longer valid in \textit{non-minimal} $\mathcal{N} = 1$ gauge theory (e.g. the soft photino theorem of \cite{Dumitrescu:2015fej} is corrected in \cite{Tropper:2024kxy}) -- one should be extremely careful about such things, because the simple examples give the wrong impression. 

A related benefit is that if the reader becomes interested in studying a specific example, this framework will shed light on some features which are guaranteed to hold by universality. Indeed, it is no exaggeration to say that the main results of many previous papers are actually universal and are, therefore, immediate consequences of the formalism described here. 

Finally, the framework offers an expedient way to organize and simplify results so supersymmetry covariance is manifest on the celestial sphere. For example, a crucial part of \cite{Crawley:2024cak} was computing (by hand) $4096$ a priori unknown OPE coefficients for a specific model of $\mathcal{N} = 2$ gauge theory. In Section \ref{sec: OPEs}, we describe strong constraints on these OPE coefficients from supersymmetry which imply that there are actually only two independent parameters from which all others can be determined.

\subsection{Outline and Main Results}

The aim of this article is to provide a systematic treatment of the CCFT duals to supersymmetric theories. We exclusively describe \textit{universal results}, i.e. those which apply to all models.

In Section \ref{sec: preliminaries}, we review the structure of massless supermultiplets. These are labeled by an annihilation operator $a(p)$ for a so-called \textit{highest-weight state}. Superpartners of this particle are labeled by the annihilation operators $a^{I_1 \cdots I_A}(p)$ where $I_a = 1, ..., \mathcal{N}$ runs over the number of supercharges in the theory. In CCFT, each species of particle in the bulk corresponds to a set of conformal primary operators on the boundary. Thus, we have a collection of operators $\mathcal{O}^{I_1 \cdots I_A}_{a,\hspace{.5pt} \Delta}(z,\overbar z)$ which live in a \textit{supercelestial family} (not to be confused with a superconformal family --- we shall see that the CCFT has no $2d$ superconformal symmetry). This family is generated by a \textit{supercelestial primary} $\mathcal{O}_{a,\hspace{.5pt} \Delta}(z,\overbar z)$ which is the Mellin transform of the highest-weight state $a(p)$ in the bulk.  Using the language of \textit{on-shell superspace} for scattering amplitudes, we show that the entire supercelestial family may be combined into a single superoperator parameterized by $\mathcal{N}$ fermionic coordinates $\eta_{I_a}$
\begin{equation}
    \mathbb{O}_{a,\Delta}(z,\overbar z\hspace{.5pt}|\hspace{.5pt}\eta) = \sumop_{A=0}^\mathcal{N} \frac{1}{A!} \hspace{2pt} \eta_{I_1} \cdots \eta_{I_A} \mathcal{O}^{I_1 \cdots I_A}_{a,\hspace{1pt}\Delta - \frac{A}{2}}(z,\overbar z)\hspace{2pt}.\\[-.3em]
\end{equation}
$\mathbb{O}_{a,\Delta}(z,\overbar z|\eta)$ should be viewed as living on $\mathbb{C}^{1|\mathcal{N}}$ --- i.e. the celestial sphere itself gets extended to a supermanifold with $\mathcal{N}$ fermionic coordinates. We call this the \textit{celestial supersphere}. All correlation functions in the CCFT can be extracted from correlation functions of such superoperators on this supermanifold. Moreover, bulk supercharges $\overbar{Q}^I_{\overbar{\alpha}}$ and $Q_{I,\alpha}$ act in a natural, geometric way
\begin{equation}
    \big[\overbar{Q}^I_{\overbar{\alpha}},\mathbb{O}_{a,\hspace{.5pt}\Delta}(z,\overbar z\hspace{.5pt}|\hspace{.5pt}\eta)\big] = \overbar{z}^{\overbar{\alpha}} \hspace{2pt} \frac{\partial}{\partial \eta_I} \mathbb{O}_{a,\hspace{.5pt}\Delta + 1}(z,\overbar z\hspace{.5pt}|\hspace{.5pt}\eta) \hspace{23pt} \big[Q_{I,\alpha},\mathbb{O}_{a,\hspace{.5pt}\Delta}(z,\overbar z\hspace{.5pt}|\hspace{.5pt}\eta)\big] = - z^\alpha \hspace{2pt} \eta_I \hspace{2pt} \mathbb{O}_{a,\hspace{.5pt}\Delta}(z,\overbar z\hspace{.5pt}|\hspace{.5pt}\eta)\hspace{2pt}.\\[-.2em]
\end{equation}

In Section \ref{sec: SUSY Extension of Conformally Soft Theorems}, we study factorization properties of correlation functions when the superoperators have conformal dimension $k \in \mathbb{Z}$ (or $k \in \mathbb{Z} + \frac{1}{2}$ for fermions). We show
\begin{equation}
    \lim_{\varepsilon \hspace{2pt} \rightarrow \hspace{2pt} 0} \varepsilon \hspace{2pt} \big\langle \mathbb{O}_{a,\hspace{.5pt}k + \varepsilon}(z,\overbar z\hspace{.5pt}|\hspace{.5pt}\eta)  \cdots \big\rangle = \mathbb{S}_{a,\hspace{.5pt}k}(z,\overbar z\hspace{.5pt}|\hspace{.5pt}\eta) \bullet \big\langle \cdots \big\rangle \hspace{2pt},
\end{equation}
where $\mathbb{S}_{a,k}(z,\overbar z\hspace{.5pt}|\hspace{.5pt}\eta)$ are fixed by momentum-space soft theorems. Such relations are called \textit{conformally soft theorems}; supersymmetry implies that we can leverage the conformally soft theorem for one particle in a supermultiplet to determine the conformally soft theorems of all others. For example, we show that the leading conformally soft gravitino and graviphoton theorems are universal and fixed completely by the subleading conformally soft graviton theorem. Such relations imply that the $\frak{bms}_{4}$ symmetries of flat space quantum gravity get extended to a universal $\frak{sbms}_{4|\mathcal{N}}$ algebra. We describe how this should be viewed as a local extension of the super-Poincar\'e algebra. The $\frak{sbms}_{4|\mathcal{N}}$ algebra encodes \textit{large SUSY transformations} and a \textit{new} asymptotic symmetry where central charges, $Z^{IJ}$, of the supersymmetry algebra are conserved at every angle of the celestial sphere.

In Section \ref{sec: OPEs}, we consider tree-level \textit{holomorphic OPEs} between $\mathcal{O}^{I_1 \cdots I_A}_{a,\hspace{.5pt} \Delta}(z,\overbar z)$ and $\mathcal{O}^{J_1 \cdots J_B}_{b,\hspace{.5pt} \Delta}(0,0)$. These are captured by the following OPE between superoperators
\begin{equation}
    \mathbb{O}_{a,\hspace{.5pt}\Delta_a}(z,\overbar z\hspace{.5pt}|\hspace{.5pt}\eta_1)\hspace{2pt} \mathbb{O}_{b,\hspace{.5pt} \Delta_b}(0,0\hspace{.5pt}|\hspace{.5pt}\eta_2) \sim -\frac{1}{z} \sumop_{c}  \hspace{1pt} g_{ab}^c \hspace{2pt}C_p(\overbar{h}_a, \overbar{h}_b) \hspace{1pt} \mathbb{O}_{c,\hspace{.5pt} \Delta_a + \Delta_b + p}(0,0\hspace{.5pt}|\hspace{.5pt}\eta_1 + \eta_2)\hspace{2pt},
\end{equation}
where $c$ sums over supercelestial primaries, $p \in \mathbb{Z}$ is defined in Equation \eqref{eqn: def of p simplified}, and the OPE coefficients $g_{ab}^c$ are tremendously constrained by supersymmetry.

In Section \ref{sec: chiral soft algebras}, we study \textit{chiral soft algebras} which encode an infinite tower of symmetries associated to soft theorems. The generators of these algebras, $\mathcal{R}^{I_1 \cdots I_A}_{a,\hspace{.5pt} q, \hspace{.5pt} \overbar{m},\hspace{.5pt} n}$ also organize themselves into a single object $\mathds{R}_{a,\hspace{.5pt} q, \hspace{.5pt} \overbar{m},\hspace{.5pt} n}(\eta)$ on superspace. The commutation relations are
\begin{equation}
    \begin{split}
        \big[\mathds{R}_{\hspace{1pt} a,\hspace{.5pt} q_1,\hspace{.5pt}\overbar{m}_1,\hspace{.5pt}n_1}(\eta_1),\hspace{1pt}&\mathds{R}_{\hspace{1pt} b,\hspace{.5pt} q_2,\hspace{.5pt} \overbar{m}_2,\hspace{.5pt}n_2}(\eta_2)\big] \\[.3em]
        &= -\sumop_{c} \hspace{1pt} g_{ab}^c \hspace{2pt} N_p(q_1,q_2,\overbar{m}_1,\overbar{m}_2) \hspace{2pt} \mathds{R}_{\hspace{1pt}c,\hspace{.5pt} q_1 + q_2 - p - 1,\hspace{.5pt} \hspace{1pt} \overbar{m}_1 + \overbar{m}_2,\hspace{.5pt} n_1 + n_2}(\eta_1 + \eta_2)\hspace{2pt}.\\[-.5em]
    \end{split}
    \label{eqn: chiral algebra superspace}
\end{equation}
We show that these algebras satisfy certain appealing Jacobi identities.

In Section \ref{sec: super w 1 + inf}, we use this formalism to examine how the celebrated $\mathcal{L}(\myw_{1+\infty}^\wedge)$ algebra \cite{Strominger:2021mtt} and its deformation by a cosmological constant $\Lambda$ \cite{Taylor:2023ajd, Bittleston:2024rqe} get extended in supergravity
\begin{align}
        \big[&\mathds{W}^{q_1}_{\overbar{m}_1,\hspace{.5pt}n_1}(\eta_1),\mathds{W}^{q_2}_{\overbar{m}_2,\hspace{.5pt}n_2}(\eta_2)\big]_\Lambda \nonumber \\[.7em]
        &\hspace{5pt}= \big((q_2-1)m_1 - (q_1-1)m_2\big) \mathds{W}^{q_1 + q_2 -2}_{\overbar{m}_1 + \overbar{m}_2,n_1+n_2}(\eta_1 + \eta_2) + 2 \sqrt{\Lambda} \hspace{2pt} \eta_{2,I}
        \hspace{2pt} \eta_{1,I} \mathds{W}^{q_1 + q_2 -1}_{\overbar{m}_1 + \overbar{m}_2,n_1+n_2}(\eta_1 + \eta_2) \nonumber \\[.4em]
        &\hspace{25pt}- \Lambda\Big(\Big(q_2-2+ \frac{\eta_{1,I}}{2}\frac{\partial}{\partial \eta_{1,I}}\Big)n_1 - \Big(q_1-2+ \frac{\eta_{2,I}}{2}\frac{\partial}{\partial \eta_{2,I}}\Big)n_2\Big) \mathds{W}^{q_1 + q_2 -1}_{\overbar{m}_1 + \overbar{m}_2,n_1+n_2}(\eta_1 + \eta_2)\hspace{2pt}.
\end{align}
In the non-supersymmetric case, this algebra is related to Hamiltonian vector fields on $\mathbb{C}^2$. Including the loop counting parameter, the algebra is $\frak{ham}_\Lambda(\mathbb{C}^2 \times \mathbb{C}^*)$. In this spirit, we argue that the supersymmetric generalization is $\frak{ham}_\Lambda(\mathbb{C}^{2|\mathcal{N}} \times \mathbb{C}^*)$. When $\Lambda = 0$, this algebra is totally different from the so-called super $\myw_{1+\infty}^\wedge$ algebras appearing in the $2d$ SCFT literature. The latter manifest $2d$ worldsheet supersymmetry while the supersymmetric generalizations of $\myw_{1+\infty}^\wedge$ relevant to celestial holography manifest $4d$ spacetime supersymmetry but no $2d$ supersymmetry on the celestial sphere. 

We conclude the article by further clarifying this point of confusion concerning the difference between $4d$ bulk supersymmetry and $2d$ boundary supersymmetry on the celestial sphere. In AdS holography, bulk supersymmetry implies boundary supersymmetry and vice versa. In celestial holography, bulk supersymmetry is realized in the spirit of the Green-Schwarz superstring which manifests target space supersymmetry at the cost of worldsheet supersymmetry.

\section{Spacetime Supersymmetry and the Celestial Supersphere}
\label{sec: preliminaries}

Four-dimensional supersymmetric theories feature $\mathcal{N}$ fermionic supercharges that can be chosen to transform as two-component Weyl spinors. These supercharges are denoted $Q_{I,\alpha}$ and $\overbar{Q}^I_{\overbar{\alpha}} = (Q_{I,\alpha})^\dagger$ with $I = 1, \ldots, \hspace{1pt}\mathcal{N}$ and $\alpha,\overbar \alpha = 0,1$. They have the following (anti-)commutation relations which enhance the usual Poincaré algebra
\begin{equation}
    \begin{split}
        \big[Q_{I,\alpha},J^{\mu \nu}\big] &= (\sigma^{\mu \nu})_\alpha^{~ \beta}\hspace{2pt} Q_\beta^I \hspace{39pt} \big[\overbar{Q}^I_{\overbar{\alpha}},J^{\mu \nu}\big] =(\overbar{\sigma}^{\mu \nu})_{\overbar{\alpha}}^{~ \overbar{\beta}} \hspace{2pt} \overbar{Q}^{I}_{\overbar{\beta}} \hspace{39pt} \big[Q_{I,\alpha},P_\mu\big] = 0 = \big[\overbar{Q}^I_{\overbar{\alpha}},P_\mu\big] \\[.3em]
        \big[Q_{I,\alpha},Q_{J,\beta}\big] &= \varepsilon_{\alpha \beta}\hspace{1pt} Z_{IJ} \hspace{55pt} \big[\hspace{2pt}\overbar{Q}_{\overbar{\alpha}}^I,\overbar{Q}_{\overbar{\beta}}^J\big] = \varepsilon_{\overbar{\alpha}\overbar{\beta}} \hspace{1pt} \overbar{Z}^{IJ} \hspace{52pt} \big[Q_{I,\alpha},\overbar{Q}_{\overbar{\alpha}}^J\big] = -2 \hspace{1pt} \delta_I^{J}\hspace{1pt} \sigma^{\mu}_{\alpha \overbar{\alpha}} P_\mu\hspace{2pt},
        \label{eqn: SUSY algebra}
    \end{split}
\end{equation}
where $\overbar{Z}^{IJ} = (Z_{IJ})^\dagger$ is an antisymmetric matrix of central charges.

The supersymmetry algebra is one example of a \textit{Lie superalgebra}, i.e. a Lie algebra with $\mathbb{Z}_2$ grading between bosonic and fermionic elements. For an element $X$ with definite parity,\footnote{In this article, we always safely assume that the elements we are working with have definite parity.} we write
\begin{equation}
    |X| = \begin{cases}
        ~ ~ 0 \hspace{30pt} X ~ ~ \text{is bosonic}\\
        ~ ~ 1 \hspace{30pt} X ~ ~ \text{is fermionic.}\\
    \end{cases}
\end{equation}
The algebra is equipped with a \textit{Lie superbracket} $[\hspace{1pt}\cdot\hspace{1pt},\hspace{1pt}\cdot\hspace{1pt}]$  satisfying 
\begin{equation}
    \big[X,Y\big] = -(-1)^{|X||Y|}\hspace{2pt}\big[Y,X\big] \hspace{50pt} \big[X,\big[Y,Z\big]\big] = \big[\big[X,Y\big],Z\big] + (-1)^{|X||Y|} \hspace{2pt} \big[Y,\big[X,Z\big]\big]\hspace{2pt}.
\end{equation}
These equations respectively imply that the superbracket degenerates to the usual (anti-)commutator according to the parity of its entries and that the superbracket enjoys a super Jacobi identity. When we write $[\hspace{1pt}\cdot\hspace{1pt},\hspace{1pt}\cdot\hspace{1pt}]$, it should \textit{always} be understood as a supercommutator satisfying the above identities, not to be confused with the ordinary commutator.

\subsection{Supermultiplets in Momentum Space}

In this article, we will focus exclusively on massless particles in the bulk for the pragmatic reason that only these states enjoy soft theorems (which encode symmetries on the celestial sphere). Such particles may be organized into supermultiplets labeled by a \textit{highest-weight state} of the supersymmetry algebra, i.e. a state which is annihilated by all the left-handed supercharges $Q_{I,\alpha}$. 

It is convenient to write this highest-weight state as an annihilation operator $a(p)$ acting on the out-state vacuum. The highest-weight condition translates to $\big[Q_{I,\alpha},a(p)\big] = 0$. The remaining particles in a supermultiplet are obtained by repeatedly acting on $a(p)$ with all possible combinations of the right-handed supersymmetry generators $\overbar{Q}^{I}_{\overbar \alpha}$. This procedure defines annihilation operators $a^{I_1 \cdots I_A}(p)$ for $0 \leq A \leq \mathcal{N}$ which are related to the highest-weight state according to
\begin{equation}
    \big[\overbar{Q}^{J}_{\overbar \alpha},a^{I_1 \cdots I_A}(p)\big] = [p|_{\overbar \alpha} \hspace{2pt} a^{J I_1 \cdots I_A}(p) \hspace{55pt} \big[Q_{J, \alpha}, a^{I_1 \cdots I_A}(p)\big] = - |p\rangle_{\alpha} \hspace{2pt} \delta{}^{[I_1}_J a^{I_2 \cdots I_A]}(p)\hspace{2pt}.
    \label{eqn: SUSY multiplet}
\end{equation}
Here, $[I_1 \cdots I_A]$ denotes anti-symmetrization of the indices, and $[p|_{\overbar \alpha}$ and $|p\rangle_{\alpha}$ are spinor brackets (see Appendix \ref{appendix: conventions} for conventions). Ultimately, the action of $\overbar{Q}^I_{\overbar \alpha}$ and $Q_{I \alpha}$ can be viewed as adding or subtracting an index followed with multiplying by a momentum spinor.

One can verify that the operator $a^{I_1 \cdots I_A}(p)$ is totally antisymmetric in its indices with each successive index lowering the helicity of the corresponding particle by $\frac{1}{2}$ and exchanging its statistics between fermionic and bosonic. The operators also satisfy canonical (anti-)commutation relations.

\subsection{CPT Conjugate Supermultiplets}

A supermultiplet is called \textit{CPT self-conjugate} if for each particle in the supermultiplet, its opposite helicity anti-particle is also represented in the supermultiplet. $\mathcal{N} = 4$ super Yang-Mills and $\mathcal{N} = 8$ supergravity are two such examples; however, generally supermultiplets are not CPT self-conjugate. To mollify this, one must add the CPT conjugates by hand.

We define $\widetilde{a}_{I_1 \cdots I_A}(p)$ as the anti-particle of $a^{I_1 \cdots I_A}(p)$
\begin{equation}
    a^{I_1 \cdots I_A}(p) ~ ~ ~ \xleftrightarrow{~ \text{antiparticles}~} ~ ~ ~ \widetilde{a}_{I_1 \cdots I_A}(p)\hspace{2pt}.
\end{equation}
The collection of these anti-particles form a supermultiplet of their own. We label the corresponding highest-weight state $\overbar{a}(p) ~ \propto ~ \widetilde{a}_{1\cdots \mathcal{N}}(p).$ The details are described in Appendix \ref{appendix: conventions}. Crucially, we may do this operation twice forming a highest-weight supermultiplet generated by $\overbar{\overbar{a}}(p)$. It is clear that this should be related to the original supermultiplet generated by $a(p).$ The normalization convention is
\begin{equation}
    \overbar{\overbar{a}}(p) = (-1)^{(\mathcal{N}+1)|a|} \hspace{2pt} a(p)\hspace{2pt}.
    \label{eqn: double CPT conjugate}
\end{equation}

\subsection{Supermultiplets on the Celestial Sphere}

In celestial CFT, it is typical to use crossing symmetry to write all momentum-space scattering amplitudes in an \textit{all-out formalism}. In particular, we use the crossing relations of Equation \eqref{eqn: crossing} to interpret an incoming particle as an outgoing one with $\epsilon = \text{sgn}(p^0) = -1.$ We parameterize null momenta of such particles as 
\begin{equation}
    p^\mu = \frac{\epsilon \hspace{2pt} \omega}{2}\hspace{2pt} \big(1+z \overbar z,z + \overbar z, - i(z - \overbar z), 1-z \overbar z\big)\hspace{2pt},
\end{equation}
where $\omega \in \mathbb{R}_+$ is an energy scale and $(z,\overbar z)$ labels a point on the celestial sphere.

The fundamental degrees of freedom in CCFT are related to momentum eigenstates via the Mellin transform
\begin{equation}
    \mathcal{O}^{I_1 \cdots I_A}_{\epsilon,\hspace{1pt} a,\hspace{1pt} \Delta}(z,\overbar z) = \int_0^\infty d\omega \hspace{2pt} \omega^{\Delta - 1} \hspace{2pt} a^{I_1 \cdots I_A}(\epsilon \hspace{2pt} \omega,z,\overbar z)\hspace{2pt}.
    \label{eqn: mellin transform}
\end{equation}
One can verify that $\mathcal{O}^{I_1 \cdots I_A}_{\epsilon,\hspace{1pt} a,\hspace{1pt} \Delta_a}(z,\overbar z)$ is a conformal primary operator located at the point $(z,\overbar z)$ on the celestial sphere. Its conformal dimension is $\Delta$, and its spin is equal to the bulk helicity of the momentum-space operator $a^{I_1 \cdots I_A}(p).$ This is summarized in Table \ref{tab: operator dimension}. Correlation functions in the CCFT are defined by Mellin transforming the external legs in a scattering amplitude individually
\begin{equation}
    \begin{split}
        \big\langle \mathcal{O}_{\epsilon_1,\hspace{1pt} a,\hspace{1pt} \Delta_1}^{I_1 \cdots I_{A}}(z_1,\overbar z_1) &\cdots \hspace{2pt} \mathcal{O}^{J_1 \cdots J_{B}}_{\epsilon_n,\hspace{1pt}b, \hspace{1pt} \Delta_n}(z_n,\overbar z_n)\big\rangle \\[.3em]
        &= \int \prod_{i=1}^n d\omega_i \hspace{2pt} \omega_i^{\Delta_i - 1} \hspace{2pt} \langle 0| a^{I_1 \cdots I_{A}}(\epsilon_1 \hspace{1pt} \omega_1,z_1,\overbar z_1)\cdots \hspace{2pt} b^{J_1 \cdots J_{B}}(\epsilon_n \hspace{1pt} \omega_n,z_n,\overbar z_n) \hspace{2pt} S \hspace{2pt}|0\rangle\hspace{2pt}.
    \end{split}
    \label{eqn: CCFT correlation functions definition}
\end{equation}
Henceforth, we drop the $\epsilon = \pm 1$ subscript focusing on outgoing particles without loss of generality.

Such operators obey commutation relations which follow from Equation \eqref{eqn: SUSY multiplet}
\begin{equation}
    \begin{split}
        \big[\overbar{Q}^{J}_{\overbar \alpha},\mathcal{O}_{a,\hspace{1pt}\Delta}^{I_1 \cdots I_A}(z,\bar z)\big] &= \overbar{z}^{\overbar{\alpha}} \hspace{2pt} \mathcal{O}_{a,\hspace{1pt}\Delta + \frac{1}{2}}^{J I_1 \cdots I_A}(z,\bar z) \hspace{30pt}
        \big[Q_{J, \alpha}, \mathcal{O}_{a,\hspace{1pt}\Delta}^{I_1 \cdots I_A}(z,\bar z)\big] = -\hspace{1pt} z^\alpha \hspace{1pt} \delta{}^{[I_1}_J \mathcal{O}_{a,\hspace{1pt}\Delta + \frac{1}{2}}^{I_2 \cdots I_A]}(z,\bar z)\hspace{2pt}.
        \label{eqn: action of SUSY on conformal primary operators}
    \end{split}
\end{equation}
Thus, the operators $\mathcal{O}_{a, \Delta}^{I_1 \cdots I_A}(z,\overbar{z})$ are organized into a supermultiplet of their own. This can be made precise with an appropriate notion of supercelestial primary operators in CCFT. Such supercelestial primaries are remniscent of superconformal primaries in a generic $2d$ SCFT; however, the representation theory is slightly different, so we choose a different name.

\begin{table}[H]
\centering
\begin{tabular}{||c|c|c|c||} 
\hhline{|t:====:t|}
\textbf{O\footnotesize PERATOR} & \textbf{D\footnotesize IMENSION} & \textbf{S\footnotesize PIN} & \textbf{C\footnotesize ONFORMAL \normalsize W\footnotesize EIGHTS} \\ 
\hhline{|:====:|}
   $\mathcal{O}^{I_1 \cdots I_A}_{a,\Delta}(z,\overbar{z})$      & $\Delta$          & $s_a - \frac{A}{2}$     & $(h,\overbar{h})=\big(\frac{1}{2}(\Delta+s),\frac{1}{2}(\Delta-s)\big)$         \\
\hhline{|b:====:b|}
\end{tabular}
\caption{Dimension, spin, and conformal weights of the conformal primary operator $\mathcal{O}^{I_1 \cdots I_A}_{a,\Delta}(z,\overbar{z})$. Here, $s_a$ is the bulk helicity of the highest-weight state, $a(p).$}
\label{tab: operator dimension}
\end{table}

\subsection{Supercelestial Primary Operators}

In celestial holography, generators of the conformal group on the boundary are related to generators of angular momentum in the bulk (see \cite{Pasterski:2021rjz} for a review). From the supersymmetry algebra, we deduce the following non-vanishing commutation relations
\begin{equation}
    \begin{split}
        \big[L_{-1},Q_{I,\alpha}\big] &= \binom{\hspace{3pt}0 ~ ~ 0}{\text{--}1 \hspace{5pt} 0}_{\alpha}^{~ \beta} \hspace{2pt} Q_{I,\beta} \hspace{50pt} \big[\overbar{L}_{-1},\overbar{Q}^I_{\overbar \alpha}\big] =\binom{\hspace{3pt}0 ~ ~ 0}{\text{--}1 \hspace{5pt} 0}_{\overbar \alpha}^{~ \overbar \beta} \hspace{2pt} Q^I_{\overbar \beta} \\
        \big[L_{0},Q_{I,\alpha}\big] &= \binom{\hspace{-2pt}\frac{1}{2} \hspace{8pt} 0}{\hspace{2pt}0 \hspace{6pt} \text{--}\frac{1}{2}}_{\alpha}^{~ \beta} \hspace{2pt} Q_{I,\beta} \hspace{52pt}\big[\overbar{L}_{0},\overbar{Q}^{I}_{\overbar \alpha}\big] = \binom{\hspace{-2pt}\frac{1}{2} \hspace{8pt} 0}{\hspace{2pt}0 \hspace{6pt} \text{--}\frac{1}{2}}_{\overbar \alpha}^{~ \overbar \beta} \hspace{2pt} \overbar{Q}^{I}_{\overbar \beta} \\ \big[L_{1},Q_{I,\alpha}\big] &= \binom{0 ~ ~ 1}{0 ~ ~ 0}_{\alpha}^{~ \beta} \hspace{2pt} Q_{I,\beta} \hspace{58pt} \big[\overbar{L}_{1},\overbar{Q}^{I}_{\overbar \alpha}\big] = \binom{0 ~ ~ 1}{0 ~ ~ 0}_{\overbar \alpha}^{~ \overbar \beta} \hspace{2pt} Q^{I}_{\overbar \beta}\hspace{2pt}.
        \label{eqn: Supercharge and L_n commutator}
    \end{split}
\end{equation}
In a non-supersymmetric CFT, a conformal primary operator is one which commutes with $L_1$ and $\overbar{L}_1$ at the origin. When we enlarge the conformal algebra with $4 \mathcal{N}$ fermionic symmetry generators satisfying the above commutation relations, we wish to consider operators which are \textit{supercelestial primaries} in an appropriate sense. Such operators may be defined by the highest-weight condition:
\begin{equation}
    \begin{split}
        \textit{Supercelestial Primary:} \hspace{40pt} \begin{cases}\hspace{20pt} \big[L_1,\mathcal{O}(0,0)\big] = 0 \hspace{24pt},\hspace{15pt} \big[\overbar{L}_1,\mathcal{O}(0,0)\big] = 0 \\[.3em]
        \hspace{20pt} \big[Q_{I,\alpha},\mathcal{O}(0,0)\big] = 0 \hspace{15pt},\hspace{15pt} \big[\overbar{Q}^I_{1},\mathcal{O}(0,0)\big] = 0\hspace{2pt}.\end{cases}
    \end{split}
\end{equation}

It follows from Equation \eqref{eqn: action of SUSY on conformal primary operators}, that $\mathcal{O}_{a,\Delta}(z,\overbar{z})$ (which is the Mellin transform of a highest-weight state in the bulk) is a supercelestial primary in the CCFT. Thus, there is a correspondence
\begin{equation}
    \bracenom{~ ~ ~ \text{Highest-weight states,}~ ~ ~}{\text{$a(p)$, in the bulk}} ~ ~ \longleftrightarrow ~ ~ \bracenom{\text{Supercelestial primaries,}}{~ ~ ~ \text{$\mathcal{O}_{a,\Delta}(z,\bar z)$, on the boundary}~ ~ ~ }
\end{equation}
If we begin with a supercelestial primary operator at the origin, $\mathcal{O}_{a, \Delta}(0,0)$, we can build the rest of the supercelestial family by acting on it with the generators which do not annihilate it: $L_{-1},\overbar{L}_{-1},$ and $\overbar{Q}^J_{0}$. In this way, we construct the supercelestial family
\begin{equation}
    L_{-1}^{n} \hspace{2pt} \overbar{L}_{-1}^{\overbar{m}} \hspace{2pt} \overbar{Q}^{I_1}_{0} \cdots \overbar{Q}^{I_A}_0 \hspace{2pt} |a;\Delta \rangle ~ ~ ~ ~\longleftrightarrow ~ ~ ~ ~ \partial^n \hspace{2pt} \overbar{\partial}^{\overbar{m}} \hspace{2pt} \mathcal{O}^{I_1 \cdots I_A}_{a,\hspace{1pt}\Delta + \frac{A}{2}}(0,0)\hspace{2pt}.
\end{equation}
One verifies from Equations \eqref{eqn: SUSY algebra} and \eqref{eqn: Supercharge and L_n commutator} that although $\mathcal{O}^{I_1 \cdots I_A}_{a,\hspace{1pt}\Delta}(0,0)$ are not supercelestial primaries, they are still conformal primaries, as expected.

We emphasize that \textit{this definition differs from the usual notion of a superconformal primary operator} in a $2d$ SCFT. The confusion arises because while the bulk Lorentz algebra is isomorphic to the boundary conformal algebra, the bulk super-Poincar\'e algebra is not isomorphic to the boundary superconformal algebra of a standard $2d$ SCFT.
The celestial CFT inherits the symmetries of the bulk, so it is not a supersymmetric CFT in the ordinary sense. This is reminiscent of the Green-Schwarz superstring which has manifest target-space supersymmetry but no worldsheet supersymmetry. The difference between supersymmetric celestial CFTs and ordinary SCFTs is further discussed in Section \ref{sec: conclusion}. Ultimately, because the symmetry algebras are different, so too is the representation theory, which is why ``superconformal primaries'' in CCFT are defined as above.

\subsection{Examples from Gauge Theory and Gravity}

Much of this article uses the aforementioned notation for supermultiplets on the celestial sphere to remain general and applicable to any supersymmetric theory. However, Table \ref{tab: list of operators} provides concrete examples from gauge theory and gravity for reference.

\begin{table}[H]
\centering
\begin{tabular}{||l||l|c||l|c||} 
\hhline{|t:=====:t|}
\multicolumn{1}{||c||}{\textbf{G\footnotesize AUGE}} & \multicolumn{2}{c||}{\textbf{Supermultiplet}} & \multicolumn{2}{c||}{\textbf{CPT Conjugates}}     \\ 
\hhline{|:=====:|}
\multicolumn{1}{||c||}{\textit{Particle}} &\multicolumn{1}{c|}{\textit{Operator}}& \textit{Spin}  & \multicolumn{1}{c|}{\textit{Operator}} & \textit{Spin}    \\ 
\hhline{|:=====:|}
                    Photon     & $\mathcal{O}_{\Delta}(z,\overbar{z})$     &  $+1$ & $\widetilde{\mathcal{O}}_{\Delta}(z,\overbar{z})$   & $-1$   \\ 

                    Photino    & $\mathcal{O}^I_{\Delta}(z,\overbar{z})$  &  $+\frac{1}{2}$ & $\widetilde{\mathcal{O}}_{I,\hspace{1pt}\Delta}(z,\overbar{z})$  & $-\frac{1}{2}$      \\ 

                    Scalar    & $\mathcal{O}^{IJ}_{\Delta}(z,\overbar{z})$ &    $0$ & $\widetilde{\mathcal{O}}_{IJ,\hspace{1pt}\Delta}(z,\overbar{z})$ & $0$     \\ 
\hhline{|:=====:|}
\multicolumn{1}{||c||}{\textbf{G\footnotesize RAVITY}} & \multicolumn{2}{c||}{\textbf{Supermultiplet}} & \multicolumn{2}{c||}{\textbf{CPT conjugates}}   \\ 
\hhline{|:=====:|}
\multicolumn{1}{||c||}{\textit{Particle}} & \multicolumn{1}{c|}{\textit{Operator}} & \textit{Spin} & \multicolumn{1}{c|}{\textit{Operator}} & \textit{Spin}      \\ 
\hhline{|:=====:|}
                    Graviton    & $\mathcal{O}_{\Delta}(z,\overbar{z})$ & $+2$ & $\widetilde{\mathcal{O}}_{\Delta}(z,\overbar{z})$  & $-2$       \\ 
                    
                    Gravitino    & $\mathcal{O}^I_{\Delta}(z,\overbar{z})$  &  $+\frac{3}{2}$ & $\widetilde{\mathcal{O}}_{I,\hspace{1pt}\Delta}(z,\overbar{z})$  & $-\frac{3}{2}$  \\ 
                    
                    Graviphoton    & $\mathcal{O}^{IJ}_{\Delta}(z,\overbar{z})$   & $+1$ & $\widetilde{\mathcal{O}}_{IJ,\hspace{1pt}\Delta}(z,\overbar{z})$ & $-1$      \\ 
                    
                    Dilatino    &$\mathcal{O}^{IJK}_{\Delta}(z,\overbar{z})$  & $+\frac{1}{2}$ & $\widetilde{\mathcal{O}}_{IJK,\hspace{1pt}\Delta}(z,\overbar{z})$ & $-\frac{1}{2}$       \\ 
                    
                    Scalar    & $\mathcal{O}^{IJKL}_{\Delta}(z,\overbar{z})$    &  $0$  & $\widetilde{\mathcal{O}}_{IJKL,\hspace{1pt}\Delta}(z,\overbar{z})$  & $0$     \\
\hhline{|b:=====:b|}
\end{tabular}
\caption{Supercelestial families and the corresponding bulk particle species for gauge theory and gravity. The left column enumerates the supercelestial family generated by a supercelestial primary encoding a positive-helicity photon/graviton. The supercelestial family consisting of CPT conjugates is also listed.}
\label{tab: list of operators}
\end{table}

\subsection{The Celestial Sphere as a Supermanifold}

We conclude this section by arguing that bulk supersymmetry enriches the geometry of the boundary theory by giving the celestial sphere the structure of a supermanifold.

To see this, we momentarily return to momentum space. By introducing $\mathcal{N}$ real Grassmann variables, $\eta_1,...,\eta_\mathcal{N}$, we may package all states in the supermultiplet generated by $a(p)$ into a single on-shell superfield $\mathds{A}(p|\eta)$
\begin{equation}
    \mathds{A}(p|\eta) = \sumop_{A=0}^\mathcal{N} \frac{1}{A!} \hspace{2pt} \eta_{I_A} \cdots \eta_{I_1} a^{I_1 \cdots I_A}(p)\hspace{2pt}.
\end{equation}
This is called the \textit{on-shell superspace formalism} \cite{Nair:1988bq, Elvang:2011fx, Elvang:2013cua, mol2024ads3dualsupersymmetricmhv}. Individual component fields may be extracted from $\mathds{A}(p|\eta)$ with appropriate Grassmann derivatives
\begin{equation}
    a^{I_1 \cdots I_A}(p) = \frac{\partial}{\partial \eta_{I_1}} \cdots \frac{\partial}{\partial \eta_{I_A}} \hspace{2pt} \mathds{A}(p|\eta) \bigg|_{\eta_I = 0}
    \label{eqn: grassmann derivitives extract component operators}
\end{equation}

This is a convenient description, because it effectively geometrizes statements about supersymmetry. For example, just like how the momentum operator acts via partial derivative on functions in spacetime, supercharges act via Grassmann derivative on $\mathds{A}(p|\eta)$
\begin{equation}
    \big[\mathds{A}(p|\eta),\overbar{Q}^{I}_{\overbar \alpha}\big] = [p|_{\overbar \alpha} \hspace{2pt} \frac{\partial}{\partial \eta_I} \hspace{2pt} \mathds{A}(p|\eta) \hspace{30pt} \big[Q_{I,\alpha},\mathds{A}(p|\eta)\big] = -|p\rangle_{\alpha} \hspace{2pt} \eta_I \hspace{2pt} \mathds{A}(p|\eta)\hspace{2pt}.
\end{equation}
This geometric language also allows one to assemble a collection of scattering amplitudes into a single object called a \textit{superamplitude}. For example, all three-point functions involving particles in the $a(p)$, $b(p)$, and $c(p)$ supermultiplets are encapsulated in $\langle 0| \hspace{1pt}\mathds{A}(p_1|\eta_1) \hspace{1pt} \mathds{B}(p_2|\eta_2) \hspace{1pt} \mathds{C}(p_3|\eta_3) |0\rangle$. Individual component amplitudes can be obtained by taking appropriate Grassmann derivatives.

We may lift this description onto the celestial sphere, packaging an entire supercelestial family into a single operator $\mathbb{O}_{a,\Delta}(z,\overbar z|\eta)$ which depends on one's location $(z,\overbar z)$ on the celestial sphere as well as the fermionic coordinates $\eta_I$ (see \cite{Jiang:2021xzy, Brandhuber:2021nez, mol2024ads3dualsupersymmetricmhv} for similar constructions)
\begin{equation}
    \begin{split}
        \mathbb{O}_{a,\Delta}(z,\overbar z|\eta) &= \sumop_{A=0}^\mathcal{N} \frac{1}{A!} \hspace{2pt} \eta_{I_A} \cdots \eta_{I_1} \mathcal{O}^{I_1 \cdots I_A}_{a,\hspace{1pt}\Delta - \frac{A}{2}}(z,\overbar z)\hspace{2pt}.
        \label{eqn: Superoperator}
    \end{split}
\end{equation}
We may choose $\eta_I$ to have conformal weights $(h,\overbar{h}) = (\frac{1}{2},0)$. This ensures that $\mathbb{O}_{a,\Delta}(z,\overbar z|\eta)$ has homogeneous conformal dimension $\Delta$ and spin $s_a.$ Again, individual operators may be extracted from this superoperator via Grassmann derivatives in a procedure analogous to Equation \eqref{eqn: grassmann derivitives extract component operators}. 

This makes manifest that one ought to view the CCFT dual to a theory with bulk supersymmetry as a theory which lives on the \textit{celestial supersphere}, a supermanifold with $\mathcal{N}$ fermionic coordinates.
\begin{equation}
    \textit{Celestial supersphere:} \hspace{60pt} \mathbb{C}^{1|\mathcal{N}} = \mathbb{C} \otimes \Lambda^\bullet(\eta_1,...,\eta_{\mathcal{N}})\hspace{2pt},
\end{equation}
where $\Lambda^\bullet(\eta_1,...,\eta_{\mathcal{N}})$ is the Grassmann algebra on $\mathcal{N}$ real anti-commuting variables and the reduced manifold, $\mathbb{C}$, is the ordinary celestial sphere. Correlation functions should be viewed as living on this celestial supersphere $\langle \mathbb{O}_{a,\Delta_a}(z_a,\overbar z_a|\eta_a) \cdots \mathbb{O}_{b,\Delta_b}(z_b,\overbar z_b|\eta_b)\rangle$ and are dual to bulk superamplitudes. 

In addition to the usual conformal symmetries, the boundary theory on the celestial supersphere enjoys $4\mathcal{N}$ fermionic symmetries which act on superoperators geometrically:
\begin{equation}
    \big[\overbar{Q}^I_{\overbar{\alpha}},\mathbb{O}_{a,\hspace{.5pt}\Delta}(z,\overbar z\hspace{.5pt}|\hspace{.5pt}\eta)\big] = \overbar{z}^{\overbar{\alpha}} \hspace{2pt} \frac{\partial}{\partial \eta_I} \mathbb{O}_{a,\hspace{.5pt}\Delta + 1}(z,\overbar z\hspace{.5pt}|\hspace{.5pt}\eta) \hspace{23pt} \big[Q_{I,\alpha},\mathbb{O}_{a,\hspace{.5pt}\Delta}(z,\overbar z\hspace{.5pt}|\hspace{.5pt}\eta)\big] = - z^\alpha \hspace{2pt} \eta_I \hspace{2pt} \mathbb{O}_{a,\hspace{.5pt}\Delta}(z,\overbar z\hspace{.5pt}|\hspace{.5pt}\eta)\hspace{2pt}.
    \label{eqn: supercharges acting on superspace operators}
\end{equation}

\section{Supersymmetric Conformally Soft Theorems and Asymptotic Symmetries}
\label{sec: SUSY Extension of Conformally Soft Theorems}

Celestial amplitudes are parameterized by both the location of conformal primary operators on the celestial (super) sphere and also the conformal dimensions, $\Delta$, of such operators. When $\Delta \in \mathbb{Z}$ (or $\Delta \in \mathbb{Z} + \frac{1}{2}$ for fermions), these amplitudes will often have poles due to soft theorems in momentum space. If one regulates these poles, the celestial amplitudes will factorize. This factorization is called a \textit{conformally soft theorem}. Such expressions are reminiscent of Ward identities for conserved currents in a generic CFT and imply that celestial CFT enjoys a rich set of symmetries which act non-trivially in the bulk despite being very simple on the boundary. Indeed, these currents correspond to the so-called \textit{asymptotic symmetries} of a bulk theory (for reviews, see \cite{Strominger:2017zoo, Pasterski:2021rjz, Donnay:2023mrd, Raclariu:2021zjz}).

Historically, the first example of asymptotic symmetries involved so-called \textit{large} diffeomorphisms. These diffeomorphisms are generated by vector fields which do not vanish near null infinity of an asymptotically flat spacetime; they encode bona fide symmetries of the theory rather than coordinate redundancies. Bondi, van der Burg, Metzner, and Sachs found that in addition to the usual Lorentz transformations, there was an infinite family of \textit{supertranslations} (no relation to supersymmetry), which generalize ordinary translations and can be organized into an infinite-dimensional symmetry algebra known as the $\frak{bms}_4$ algebra \cite{Bondi:1962px, Sachs:1962wk}. Subsequently, Barnich and Troessaert argued that Lorentz transformations should be similarly generalized to an infinite-dimensional family of \textit{superrotations}. The resulting algebra is known as the \textit{extended} $\frak{bms}_4$ algebra \cite{Barnich:2009se}.

The aim of this section is to study conformally soft theorems in supersymmetric theories. We demonstrate that if one knows a conformally soft theorem for one particle in a supermultiplet, one may deduce corresponding conformally soft theorems for all superpartners! 

Afterwards, we further extended the $\frak{bms}_4$ algebra to a supersymmetric setting. The result is an $\frak{sbms}_{4|\mathcal{N}}$ algebra which is a symmetry of all supergravity theories. This is not the first time that the $\frak{sbms}_{4|1}$ algebra has appeared in the literature. It was originally derived using twistor theory in 1985 \cite{Awada:1985by}. In the 2020s, it was re-derived with conformally soft theorems \cite{Fotopoulos:2020bqj, Banerjee:2022lnz} and then re$^2$-derived from a spacetime perspective involving both large diffeomorphisms and large supersymmetry transformations with judiciously chosen asymptotic fall-off conditions \cite{Henneaux:2020ekh, Fuentealba:2020aax, Fuentealba:2021xhn, Boulanger:2023gpw, Prabhu:2021bod}. For other related papers, see also \cite{Banks:2014iha, Banerjee:2022abf, Bagchi:2022owq}. So, why do it again for a fourth time? Though our results agree with these earlier analyses, we offer two crucial generalizations in \textit{scope}. 

\begin{enumerate}
    \item We do not assume a specific model (so far, $\mathcal{N} = 1$ minimal supergravity has been the only thing considered). Our results are, therefore, \textit{universal}, meaning they are independent of any theory-specific details. They also hold at loop level.
    \item We permit an arbitrary number of supercharges. The super-Poincar\'e algebra is qualitatively different when $\mathcal{N} > 1$ due to the presence of central charges. We will see that these central charges have an infinite-dimensional enhancement in $\frak{sbms}_{4|\mathcal{N}}$ when $\mathcal{N} > 1$. This corresponds to a \textit{new asymptotic symmetry} associated to large $U(1)$ transformations of the central charges.
\end{enumerate}

\subsection{Supersymmetric Soft Theorems}

Soft theorems relate momentum-space scattering amplitudes involving massless particles with extremely low energies to scattering amplitudes without these particles. Indeed, in the soft limit ($\omega \rightarrow 0$), tree-level amplitudes factorize according to the \textit{soft theorem}
\begin{equation}
    \langle 0| a^{I_1 \cdots I_A}(\omega,z,\overbar{z}) \cdots |0\rangle \xrightarrow{\omega \hspace{2pt} \rightarrow \hspace{2pt} 0} \sumop_{k \hspace{2pt} \in \hspace{2pt} \mathbb{Z} + s(a)} \omega^{-k} \hspace{2pt} \mathcal{S}_{a,k}^{I_1 \cdots I_A}(z,\overbar{z}) \bullet \langle 0| \cdots |0\rangle\hspace{2pt},
\end{equation}
where $\mathcal{S}^{I_1 \cdots I_A}_{a,k}(z,\overbar{z})$ are a set of operators which act linearly on the space of scattering amplitudes and `$\cdots$' refers to a collection of annihilation operators for the various other particles participating in the scattering process. We call the most singular term in the soft expansion the \textit{leading soft theorem}, the second most singular term the \textit{subleading soft theorem}, and so on.

Soft theorems are highly constrained by symmetries of the theory. For example, the leading and subleading soft theorems for positive-helicity gravitons take the following \textit{universal} form
\begin{equation}
    \begin{split}
        \mathcal{S}_{\text{graviton},\hspace{1pt}1}(z,\overbar{z}) &= \frac{\kappa}{2} \sumop_{n} \frac{(\varepsilon_+ \cdot p_n)^2}{\widehat{p}_s \cdot p_n}\\
        \mathcal{S}_{\text{graviton},\hspace{1pt}0}(z,\overbar{z}) &= -\frac{i\kappa}{2} \sumop_{n} \frac{(\varepsilon_+ \cdot p_n)\hspace{2pt} (\widehat{p}_s \cdot J_n \cdot \varepsilon_+)}{\widehat{p}_s \cdot p_n}\hspace{2pt},\\[-.3em]
    \end{split}
\end{equation}
where $\widehat{p}_s = \frac{1}{2}(1+z\overbar z, z+\overbar{z},-i(z-\overbar{z}),1-z\overbar{z})$ is the momentum of the soft particle with its $\omega$ dependence stripped off, its polarization vector is $\varepsilon_+^\mu = \sqrt{2} \hspace{2pt} \partial_z \hspace{1pt} \widehat{p}^{\hspace{1pt}\mu}$, and $\kappa = \sqrt{32\pi G_N}$. The sum over $n$ includes the other particles participating in the scattering process which are acted on by the momentum and angular momentum operators $p_n$ and $J_{n}$.

Soft theorems in supersymmetric theories have been well-studied in \cite{Tropper:2024kxy} where it was shown that knowing the soft theorem for one particle in a supermultiplet uniquely fixes soft theorems for all of its superpartners. Concretely, the soft theorems themselves form a representation of the supersymmetry algebra (compare with Equations \eqref{eqn: SUSY multiplet} and \eqref{eqn: action of SUSY on conformal primary operators})
\begin{equation}
    \begin{split}
        \big[\overbar{Q}^{J}_{\overbar \alpha},\mathcal{S}_{a,k}^{I_1 \cdots I_A}(z,\overbar{z})\big] = \overbar{z}^{\overbar{\alpha}} \hspace{2pt} \mathcal{S}_{a,\hspace{1pt}k+\frac{1}{2}}^{J I_1 \cdots I_A}(z,\overbar{z}) \hspace{45pt} \big[Q_{J, \alpha}, \mathcal{S}^{I_1 \cdots I_A}_{a,k}(z,\overbar{z})\big] = - z^{\alpha} \hspace{2pt} \delta{}^{[I_1}_J \mathcal{S}_{a,\hspace{1pt}k + \frac{1}{2}}^{I_2 \cdots I_A]}(z,\overbar{z})\hspace{2pt}.
        \label{eqn: SUSY soft theorems}
    \end{split}
\end{equation}
Here, the commutator between $\overbar{Q}^J_{\overbar{\alpha}}$ and $\mathcal{S}^{I_1 \cdots I_A}_{a,k}(z,\overbar z)$ should be viewed as a commutator of linear operators acting on the space of scattering amplitudes, i.e.
\begin{equation}
     \begin{split}
         \big[\overbar{Q}^{J}_{\overbar \alpha},\mathcal{S}_{a,k}^{I_1 \cdots I_A}(z,\overbar{z})\big] \bullet \langle 0|\cdots |0\rangle &= \overbar{Q}^{J}_{\overbar \alpha}\bullet \mathcal{S}_{a,k}^{I_1 \cdots I_A}(z,\overbar{z}) \bullet \langle 0|\cdots |0\rangle \\
         &\hspace{70pt} - (-1)^{|a^{I_1 \cdots I_A}|} \hspace{2pt} \mathcal{S}_{a,k}^{I_1 \cdots I_A}(z,\overbar{z}) \bullet\overbar{Q}^{J}_{\overbar \alpha} \bullet \langle 0|\cdots |0\rangle\hspace{2pt}.\\
     \end{split}
\end{equation}
The triumph of \cite{Tropper:2024kxy} is that if one knows a certain soft theorem explicitly, the commutators defining the left hand side of Equation \eqref{eqn: SUSY soft theorems} may be evaluated expeditiously using the commutation relations in the supersymmetry algebra (Equation \eqref{eqn: SUSY algebra}) alone. 

For example, the leading soft graviton theorem is built from some fixed kinematics and the momentum operator acting on the $n^{th}$ particle. These ingredients all commute with the supercharges, so the entire leading soft graviton theorem must have vanishing commutator as well. At subleading order, something non-trivial happens because the soft theorem includes the angular momentum operator, $J_n$, which doesn't commute with the supercharges. Thus, the commutator between $\overbar{Q}^I_{\overbar{\alpha}}$ and the subleading soft graviton theorem will be non-vanishing; it may be evaluated explicitly from the supersymmetry algebra in a one-line derivation. According to Equation \eqref{eqn: SUSY soft theorems}, this commutator is the leading soft gravitino theorem
\begin{equation}
    \mathcal{S}_{\text{gravitino},\hspace{1pt}\frac{1}{2}}^{I}(z,\overbar{z}) = \frac{\kappa}{2\sqrt{2}} \sumop_{n} \frac{\varepsilon_+ \cdot p_n}{\widehat{p} \cdot p_n}\hspace{1pt} \big[\widehat{p}_s \hspace{1pt} \overbar{Q}^I_n \big]\hspace{2pt}.
\end{equation}
Because the subleading soft graviton theorem is universal, so too is this expression. This derivation is but one example of how soft theorems for superpartners are fixed by representation theory; leading-order soft theorems for the entire photon, gluon, and graviton supermultiplets are chronicled in \cite{Tropper:2024kxy}.

\subsection{Supersymmetric Conformally Soft Theorems}

In celestial CFT, correlation functions will generally be divergent when $\Delta = k$ for $k \in \mathbb{Z} + s(a^{I_1 \cdots I_A})$ due to the presence of a non-vanishing soft theorem at order $\omega^{-k}$. Fortunately, the divergence may be regulated by defining \textit{conformally soft operators}
\begin{equation}
    \mathcal{O}_{a,\hspace{1pt}k,\hspace{1pt}\text{CS}}^{I_1 \cdots I_A}(z,\overbar z) = \lim_{\varepsilon \hspace{1pt} \rightarrow \hspace{1pt} 0} \varepsilon \hspace{2pt} \mathcal{O}_{a,\hspace{1pt} k + \varepsilon}^{I_1 \cdots I_A}(z,\overbar z) \hspace{40pt} \text{where} \hspace{40pt} k \in \mathbb{Z} + s(a^{I_1 \cdots I_A})\hspace{2pt}.
    \label{eqn: def conformally soft}
\end{equation}
Just as scattering amplitudes factorize in the soft limit according to soft theorems, correlation functions involving these conformally soft operators also factorize in an extremely simple way. In fact, these conformally soft operators may be removed from a correlation function at the cost of acting on that correlation function with the $k^{th}$ order term in the momentum-space soft expansion  
\begin{equation}
    \big\langle \mathcal{O}_{a,\hspace{1pt}k,\hspace{1pt}\text{CS}}^{I_1 \cdots I_A}(z,\overbar z) \hspace{2pt} \cdots \big\rangle = \mathcal{S}^{I_1 \cdots I_A}_{a,\hspace{.5pt}k}(z,\overbar{z}) \bullet \big\langle \cdots \big\rangle\hspace{2pt}.
\end{equation}
Such relations are called \textit{conformally soft theorems}. They may be regarded as \textit{Ward identities} for asymptotic symmetries acting on the celestial sphere  (see \cite{Pasterski:2021rjz} for a review).

Of course these expressions have a natural realization on the celestial supersphere as well 
\begin{equation}
    \big\langle \mathbb{O}_{a,\hspace{.5pt}k,\hspace{.5pt}\text{CS}}(z,\overbar z|\eta) \hspace{2pt} \cdots \big\rangle = \mathbb{S}_{a,\hspace{.5pt}k}(z,\overbar{z}|\eta) \bullet \big\langle \cdots \big\rangle\hspace{2pt}, 
\end{equation}
where we have defined the following conformally soft theorem in superspace
\begin{equation}
    \mathbb{S}_{a,\hspace{.5pt}k}(z,\overbar{z}|\eta) = \sumop_{A = 0}^\mathcal{N} \frac{1}{A!} \hspace{2pt} \eta_{I_A} \cdots \eta_{I_1} \mathcal{S}^{I_1 \cdots I_A}_{a,\hspace{1pt}k- \frac{A}{2}}(z,\overbar z)\hspace{2pt}.
\end{equation}
These obey (compare with Equation \eqref{eqn: supercharges acting on superspace operators})
\begin{equation}
    \big[\overbar{Q}^I_{\overbar{\alpha}}\hspace{1pt},\mathbb{S}_{a,k}(z,\overbar z|\eta)\big] = \overbar{z}^{\overbar{\alpha}} \hspace{2pt} \frac{\partial}{\partial \eta_I} \hspace{2pt} \mathbb{S}_{a,\hspace{.5pt}k + 1}(z,\overbar z|\eta) \hspace{30pt} \big[Q_{I,\alpha}\hspace{1pt},\mathbb{S}_{a,k}(z,\overbar z|\eta)\big] = - z^\alpha \hspace{2pt} \eta_I \hspace{2pt} \mathbb{S}_{a,\hspace{.5pt}k}(z,\overbar z|\eta)\hspace{2pt}.
\end{equation}

\subsection{Examples from Gauge Theory and Gravity}
\label{sec: examples of conformally soft theorems}

Because the leading-order soft theorems in momentum space have been computed for the entire photon, gluon, and graviton supermultiplets, we may directly translate these equations to determine the leading-order (i.e. largest $k$) conformally soft theorems in celestial holography. These are given in Tables \ref{tab: conformally soft theorems (gauge)} and \ref{tab: conformally soft theorems (grav)}.\footnote{\textbf{Conventions:} In these tables, the momentum, $p_n^\mu$, and angular momentum, $J_n^{\mu \nu}$, operators are defined according to the following commutators: $\big[a(p_n),P^\mu\big] = p_n^\mu \hspace{2pt}a(p_n)$ and $\big[a(p_n),J^{\mu\nu}\big] = J_n^{\mu\nu} a(p_n)$. This convention ensures that $P^\mu |p_n\rangle = p_n^\mu|p_n\rangle$ and $J^{\mu\nu} |p_n\rangle = J_n^{\mu\nu}|p_n\rangle$ for the corresponding single-particle states. Notice that the annihilation operator always appears on the left-hand side of these commutators. 

In keeping with this convention, we define the central charge, $Z_n^{IJ}$, as $\big[a(p_n),Z^{IJ}\big] = Z^{IJ}_n \hspace{2pt}a(p_n)$. On the other hand, $\overbar{Q}^I_n$ is defined with the \textit{opposite} ordering convention: $\big[\overbar{Q}^{I}_{\overbar{\alpha}},a(p_n)\big] = \overbar{Q}^{I}_{\overbar{\alpha},n} \hspace{2pt}a(p_n)$, i.e. with the annihilation operator on the right-hand side. This ensures that $\overbar{Q}^{J}_{\overbar{\alpha},n} \hspace{2pt} a^{I_1 \cdots I_A}(p_n) = [p|_{\overbar{\alpha}} \hspace{2pt}a^{J I_1 \cdots I_A}(p_n)$ mirroring Equation \eqref{eqn: SUSY multiplet}}

\begin{table}
\centering
\begin{tabular}{||l||l|l||} 
\hhline{|t:===:t|}
\multicolumn{3}{||c||}{\textbf{C\footnotesize ONFORMALLY \normalsize S\footnotesize OFT \normalsize  T\footnotesize HEOREMS IN \normalsize  G\footnotesize AUGE \normalsize  T\footnotesize HEORY}} \\ 
\hhline{|:=:t:==:|}
\hspace{-2pt}\textit{Particle} \hspace{-2pt} & \hspace{-1pt}\textit{Operator} \hspace{-1pt} & \hspace{81pt}\textit{Expression}\hspace{81pt}\color{white}.\color{black}\\[-.1em]
\end{tabular}
\renewcommand{\arraystretch}{2.5}
\begin{tabular}{||l||l|l||} 
\hhline{|:=::==:|}
Photon   &  $\mathcal{S}_{\hspace{1pt}1}(z,\overbar{z})$                        &   \hspace{13pt} $\displaystyle e \hspace{2pt} \sumop_n q_n \hspace{2pt} \frac{\varepsilon_+ \cdot p_n}{\widehat{p}_s \cdot p_n}$         \\
         &  $\mathcal{S}_{\hspace{1pt}0}(z,\overbar{z})$                        &    $\displaystyle \hspace{4.5pt} -ie \hspace{2pt} \sumop_n q_n \hspace{2pt} \frac{\widehat{p}_s \cdot J_n \cdot \varepsilon_+}{\widehat{p}_s \cdot p_n} + g_m \sumop_n \hspace{2pt} \frac{[\hspace{.5pt}\widehat{p}_s \hspace{2pt} p_n]}{\langle\hspace{.5pt}\widehat{p}_s \hspace{2pt} p_n\rangle} \hspace{2pt} \mathcal{F}_{m,n}$       \\[.75em] 
\hline
Photino  &  $\mathcal{S}^I_{\hspace{1pt}1/2}(z,\overbar{z})$                        &  $\displaystyle \hspace{5pt} \frac{e}{\sqrt{2}} \hspace{2pt} \sumop_n q_n \hspace{2pt} \frac{[\hspace{.5pt} \widehat{p}_s \hspace{1pt} \overbar{Q}^I_n]}{\widehat{p}_s \cdot p_n} + g_m \sumop_n \hspace{2pt} \frac{[\hspace{.5pt}\widehat{p}_s \hspace{2pt} p_n]}{\langle\hspace{.5pt}\widehat{p}_s \hspace{2pt} p_n\rangle} \hspace{2pt} \mathcal{F}^I_{m,n}$           \\[.75em]
\hline 
Scalar   &  $\mathcal{S}^{IJ}_{\hspace{1pt}1}(z,\overbar{z})$                        &  $\hspace{-4pt}-\displaystyle \frac{e}{\sqrt{2}} \hspace{2pt} \sumop_n q_n \hspace{2pt} \frac{\overbar{Z}_n^{IJ}}{\widehat{p}_s \cdot p_n} + g_m \sumop_n \hspace{2pt} \frac{[\hspace{.5pt}\widehat{p}_s \hspace{2pt} p_n]}{\langle\hspace{.5pt}\widehat{p}_s \hspace{2pt} p_n\rangle} \hspace{2pt} \mathcal{F}^{IJ}_{m,n}$           \\[.75em]
\hhline{|b:=:b:==:b|}
\end{tabular}
\caption{Conformally soft theorems for the photon, photino, and scalar in supersymmetric QED. Here, $\widehat{p}_s = \frac{1}{2}(1+z\overbar z, z+\overbar{z},-i(z-\overbar{z}),1-z\overbar{z})$ labels the location of the conformally soft operator on the celestial sphere.}
\label{tab: conformally soft theorems (gauge)}
\end{table}

Included in these tables are certain unspecified operators $\mathcal{F}_{m,n}$ and coefficients $g_m$. These are so-called \textit{particle changing operators} which are \textit{theory-dependent} but totally classified. For example, in gauge theory, if any of the following dimension-5 operators appear in the effective Lagrangian \cite{Elvang:2013cua}:\footnote{Here $\phi$ is a scalar, $\chi$ is a spin-1/2 field,  $F_{\mu \nu}$ and $\widetilde{F}_{\mu \nu}$ are the field strength and dual field strength for the gauge field, $\psi_\mu$ is a spin-3/2 field, and $h$ is a spin-2 field.}
\begin{equation}
    g_1 \hspace{1pt} \overbar{\chi} \gamma^{\mu \nu} F_{\mu \nu} \chi, \hspace{30pt} g_2 \hspace{1pt}\phi F_{\mu \nu}F^{\mu \nu}, \hspace{30pt} g_3 \hspace{1pt}\phi F_{\mu \nu} \widetilde{F}^{\mu \nu},\hspace{30pt} g_4 \hspace{1pt}\overbar{\psi}_\mu F_{\nu \rho} \gamma^{\mu \nu \rho} \chi,\hspace{30pt} g_5\hspace{1pt} h F^2\hspace{2pt},
\end{equation}
then there will be a corresponding particle changing operator, $\mathcal{F}_{m,n}$, appearing in the subleading soft photon theorem. The $m$ index labels this collection of dimension-5 operators, and the coefficient $g_m$ appearing in the (conformally) soft theorems is the coupling constant in the effective Lagrangian. 

\begin{table}[H]
\centering
\begin{tabular}{||l||l|l||} 
\hhline{|t:===:t|}
\multicolumn{3}{||c||}{\textbf{C\footnotesize ONFORMALLY \normalsize S\footnotesize OFT \normalsize  T\footnotesize HEOREMS IN \normalsize  G\footnotesize RAVITY}}              \\ 
\hhline{|:=:t:==:|}
\hspace{7.5pt} \textit{Particle} \hspace{7pt} & \hspace{3pt} \textit{Operator} \hspace{2.5pt} & \hspace{99.3pt} \textit{Expression} \hspace{99pt}  \\[-.1em]
\end{tabular}
\renewcommand{\arraystretch}{2.5}
\begin{tabular}{||l||l|l||} 
\hhline{|:=::==:|}
Graviton   &  $\mathcal{S}_{\hspace{1pt}1}(z,\bar z)$                        & \hspace{18pt} $\displaystyle \frac{\kappa}{2} \hspace{2pt} \sumop_n \hspace{2pt} \frac{(\varepsilon_+ \cdot p_n)^2}{\widehat{p}_s \cdot p_n}$          \\
         & $\mathcal{S}_{\hspace{1pt}0}(z,\bar z)$                         & \hspace{6pt} $\displaystyle - \frac{i\kappa}{2} \hspace{2pt} \sumop_n \hspace{2pt} \frac{(\varepsilon_+ \cdot p_n)\hspace{2pt} (\widehat{p}_s \cdot J_n \cdot \varepsilon_+)}{\widehat{p}_s \cdot p_n}$         \\
         & $\mathcal{S}_{\hspace{1pt}-1}(z,\bar z)$                        & \hspace{10pt} $\displaystyle -\frac{\kappa}{4} \hspace{2pt} \sumop_n \hspace{2pt} \frac{(\widehat{p}_s \cdot J_n \cdot \varepsilon_+)^2}{\widehat{p}_s \cdot p_n} + g_m \sumop_n \hspace{2pt} \frac{[\hspace{.5pt}\widehat{p}_s \hspace{2pt} p_n]^3}{\langle\hspace{.5pt}\widehat{p}_s \hspace{2pt} p_n\rangle} \hspace{2pt} \mathcal{F}_{m,n}$          \\[.75em]
\hline
Gravitino  & $\mathcal{S}^I_{\hspace{1pt}1/2}(z,\bar z)$                         & \hspace{8pt}$\displaystyle \frac{\kappa}{2\sqrt{2}} \hspace{2pt} \sumop_n \hspace{2pt} \frac{\varepsilon_+ \cdot p_n}{\widehat{p}_s \cdot p_n} \hspace{2pt} [\hspace{.5pt} \widehat{p}_s \hspace{1pt} \overbar{Q}^I_n]$      \\
         &  $\mathcal{S}^I_{\hspace{1pt}-1/2}(z,\bar z)$  &  \hspace{-4pt} $\displaystyle -\frac{i\kappa}{4\sqrt{2}} \hspace{2pt} \sumop_n \hspace{2pt} \frac{[\hspace{.5pt} \widehat{p}_s \hspace{1pt} \overbar{Q}^I_n]\hspace{2pt} (\widehat{p}_s \cdot J_n \cdot \varepsilon_+) + (\widehat{p}_s \cdot J_n \cdot \varepsilon_+)\hspace{2pt} [\hspace{.5pt} \widehat{p}_s \hspace{1pt} \overbar{Q}^I_n]}{\widehat{p}_s \cdot p_n}$ \\[-.75em]
         & & \hspace{80pt} $\displaystyle + g_m \sumop_n \hspace{2pt} \frac{[\hspace{.5pt}\widehat{p}_s \hspace{2pt} p_n]^3}{\langle\hspace{.5pt}\widehat{p}_s \hspace{2pt} p_n\rangle} \hspace{2pt} \mathcal{F}^I_{m,n}$ \\[.75em]
\hline
Graviphoton   & $\mathcal{S}^{IJ}_{1}(z,\bar z)$                         & \hspace{-5pt} $\displaystyle -\frac{\kappa}{2\sqrt{2}} \hspace{2pt} \sumop_n \hspace{2pt} \frac{\varepsilon_+ \cdot p_n}{\widehat{p}_s \cdot p_n} \hspace{2pt} \overbar{Z}_{n}^{IJ}$             \\
         &  $\mathcal{S}^{IJ}_{\hspace{1pt}0}(z,\bar z)$                        &  \hspace{17pt} $\displaystyle\frac{\kappa}{4} \hspace{2pt} \sumop_n \hspace{2pt} \frac{[\hspace{.5pt} \widehat{p}_s \hspace{1pt} \overbar{Q}^{[I}_n] \hspace{1pt} [\hspace{.5pt} \widehat{p}_s \hspace{1pt} \overbar{Q}^{J]}_n] + \sqrt{2} i \hspace{1pt} (\widehat{p}_s \cdot J_n \cdot \varepsilon_+) \hspace{1pt} \overbar{Z}^{IJ}_n}{\widehat{p}_s \cdot p_n} $ \\[-.75em]
         & & \hspace{80pt} $\displaystyle + g_m \sumop_n \hspace{2pt} \frac{[\hspace{.5pt}\widehat{p}_s \hspace{2pt} p_n]^3}{\langle\hspace{.5pt}\widehat{p}_s \hspace{2pt} p_n\rangle} \hspace{2pt} \mathcal{F}^{IJ}_{m,n}$ \\[.75em]
\hline
Dilatino         & $\mathcal{S}^{IJK}_{\hspace{1pt}1/2}(z,\bar z)$                         & \hspace{5pt}$-\displaystyle\frac{3\kappa}{4} \hspace{2pt} \sumop_n \hspace{2pt} \frac{\overbar{Z}^{[IJ}_n [\hspace{.5pt} \widehat{p}_s \hspace{1pt} \overbar{Q}^{K]}_n]}{\widehat{p}_s \cdot p_n} + g_m \sumop_n \hspace{2pt} \frac{[\hspace{.5pt}\widehat{p}_s \hspace{2pt} p_n]^3}{\langle\hspace{.5pt}\widehat{p}_s \hspace{2pt} p_n\rangle} \hspace{2pt} \mathcal{F}^{IJK}_{m,n}$            \\[.75em]
\hline
Scalar        &  $\mathcal{S}^{IJKL}_{\hspace{1pt}1}(z,\bar z)$                        & \hspace{9.5pt} $\displaystyle\frac{3\kappa}{2} \hspace{2pt} \sumop_n \hspace{2pt} \frac{\overbar{Z}^{[IJ}_n \hspace{1pt} \overbar{Z}^{KL]}_n}{\widehat{p}_s \cdot p_n} + g_m \sumop_n \hspace{2pt} \frac{[\hspace{.5pt}\widehat{p}_s \hspace{2pt} p_n]^3}{\langle\hspace{.5pt}\widehat{p}_s \hspace{2pt} p_n\rangle} \hspace{2pt} \mathcal{F}^{IJKL}_{m,n}$            \\[.75em]
\hhline{|b:=:b:==:b|}
\end{tabular}
\caption{Conformally soft theorems in supergravity. $\widehat{p}_s = \frac{1}{2}(1+z\overbar z, z+\overbar{z},-i(z-\overbar{z}),1-z\overbar{z})$ labels the location of the conformally soft operator on the celestial sphere.}
\label{tab: conformally soft theorems (grav)}
\end{table}

While $\mathcal{F}_{m,n}$ has an explicit expression, that expression depends on the specific dimension-5 operator (see \cite{Laddha:2017vfh} for examples). Nevertheless, because the expressions are explicit, one can take repeated commutators with $\overbar{Q}^{I}_{\overbar{\alpha}}$ to determine the $\mathcal{F}^{I_1 \cdots I_A}_{m,n}$ operators on a case by case basis. For a detailed explanation, see \cite{Tropper:2024kxy}.

We only report on the conformally soft theorems associated to operators with spin $s \geq 0$. The conformally soft theorems for the opposite spin conformal primary operators are related by bulk CPT conjugation. Explicit relations are described in \cite{Tropper:2024kxy}.

\subsection{A Universal $\frak{sbms}_{4|\mathcal{N}}$ Algebra of Asymptotic Symmetries}
\label{sec: sbms algebra}

\begin{table}[t]
\centering
\begin{tabular}{||l||l|c||} 
\hhline{|t:===:t|}
\multicolumn{3}{||c||}{\textbf{G\footnotesize ENERATORS OF \normalsize E\footnotesize XTENDED} $\frak{sbms}_{4|\mathcal{N}}$ \textbf{\normalsize A\footnotesize LGEBRA}}                                                   \\ 
\hhline{|:=:t:==:|}
\multicolumn{1}{||c||}{\textit{Bulk Symmetry}} & \multicolumn{1}{c|}{\textit{Generators}} & \multicolumn{1}{c||}{\textit{Weights}}  \\ 
\hhline{|:=::==:|}
\multirow{2.5}{*}{Supertranslation} & & \\[-2.25em]
 & $\displaystyle \hspace{9pt}P(z) = -\frac{2}{\kappa} \hspace{2pt} \overbar{\partial} \mathcal{O}_{1,\text{CS}}(z,\overbar{z})$ &    $\big(\frac{3}{2},-\frac{1}{2}\big)$ \\
 & & \\[-2.25em]
& $\displaystyle \hspace{9pt} \overbar{P}(\overbar{z}) = -\frac{2}{\kappa} \hspace{2pt} \partial \widetilde{\mathcal{O}}_{1,\text{CS}}(z,\overbar{z})$ &    $\big(\hspace{-3pt}-\hspace{-2pt}\frac{1}{2},\frac{3}{2}\big)$ \\[.5em]
\hline
\multirow{2.6}{*}{Superrotation} & & \\[-2.25em]
    &  \hspace{6pt} $\displaystyle T(z) = \frac{3}{\pi \kappa} \int \frac{d^2 w}{(z-w)^4} \hspace{2pt} \widetilde{\mathcal{O}}_{0,\text{CS}}(w,\overbar{w})$ & $\big(2,0\big)$ \\
    & & \\[-2.25em]
    & \hspace{6pt}  $\displaystyle \overbar{T}(\overbar z) = \frac{3}{\pi \kappa} \int \frac{d^2 w}{(\overbar{z}-\overbar{w})^4} \hspace{2pt} \mathcal{O}_{0,\text{CS}} (w,\overbar{w})$ & $\big(0,2\big)$ \\[.75em]
\hline
\multirow{2.6}{*}{Large SUSY} & & \\[-2.25em]
& \hspace{2.5pt} 
 $\displaystyle S_I(z) = -\frac{2}{\pi \kappa} \int \frac{d^2 w}{(z-w)^3}\hspace{2pt} \widetilde{\mathcal{O}}_{I,1/2,\text{CS}}(w,\overbar{w})$ & $\big(\frac{3}{2},0\big)$ \\
 & & \\[-2.25em]
   & \hspace{2.5pt} 
 $\displaystyle 
 \overbar{S}^I(\overbar z) = -\frac{2}{\pi \kappa} \int  \frac{d^2 w}{(\overbar{z} - \overbar{w})^3} \hspace{2pt} \mathcal{O}^I_{1/2,\text{CS}}(w,\overbar w)$ &  $\big(0,\frac{3}{2}\big)$\\[.75em]
\hline
\multirow{2.6}{*}{$\begin{aligned} &\text{Large Central} \\ &\text{Charge Symmetry} \end{aligned}$} & & \\[-2.25em]
 &\hspace{1.5pt}$\displaystyle J_{IJ}(z) = \frac{1}{\pi \kappa} \int \frac{d^2 w}{(z-w)^2} \hspace{2pt} \widetilde{\mathcal{O}}_{IJ,\hspace{1pt} 1,\text{CS}}(w,\overbar{w})$ & $\big(1,0\big)$ \\
 & & \\[-2.25em]
 &\hspace{1pt}$\displaystyle \overbar{J}^{IJ}(\overbar z) = \frac{1}{\pi \kappa} \int \frac{d^2w}{(\overbar{z}-\overbar{w})^2} \hspace{2pt} \mathcal{O}^{IJ}_{1, \text{CS}}(w,\overbar{w})$ & $\big(0,1\big)$ \\[.75em]
\hhline{|b:=:b:==:b|}
\end{tabular}
\caption{Definition of generators of the $\frak{sbms}_{4|\mathcal{N}}$ algebra in terms of derivatives and shadow transforms of conformally soft operators in the gravity supermultiplet (see Table \ref{tab: list of operators} for relation between CCFT operators and different species of bulk particle in the supergravity multiplet).}
\label{tab: sbms algebra generators}
\end{table}

As mentioned, conformally soft theorems give a natural language for describing asymptotic symmetries. The aim of this section is to use the conformally soft theorems we just derived in supergravity to construct a novel $\frak{sbms}_{4|\mathcal{N}}$ algebra, which is the universal supserymmetric extension of the $\frak{bms}_{4}$ algebra. This analysis is similar in spirit to \cite{Fotopoulos:2019vac, Fotopoulos:2020bqj}, though there are several key differences in scope highlighted in the opening paragraphs of Section \ref{sec: SUSY Extension of Conformally Soft Theorems}.

One begins by constructing a pair of weight $(\frac{3}{2},-\frac{1}{2})$ and $(-\frac{1}{2},\frac{3}{2})$ operators $P(z)$ and $\overbar{P}(\overbar z)$ which are related to the leading conformally soft graviton according to Table \ref{tab: sbms algebra generators}. Their OPE with an arbitrary conformal primary is fixed by the leading conformally soft graviton theorem; however, it is often more convenient to combine them into a single weight $(\frac{3}{2},\frac{3}{2})$ operator $\mathcal{P}(z,\overbar{z})$ obeying \cite{Fotopoulos:2019vac}
\begin{equation}
    \mathcal{P}(z,\overbar{z}) \hspace{2pt} \mathcal{O}_{a,\hspace{.5pt}\Delta}^{I_1 \cdots I_A}(0,0) \sim \frac{1}{z \overbar{z}} \hspace{2pt} \mathcal{O}^{I_1 \cdots I_A}_{a,\hspace{.5pt}\Delta + 1}(0,0) \hspace{45pt} \mathcal{P}(z,\overbar{z}) = \hspace{-7pt} \sumop_{k,\hspace{.5pt}l \hspace{1.5pt} \in\hspace{1.5pt} \mathbb{Z} + \frac{1}{2}} \hspace{-5pt} z^{-k - 3/2}\hspace{1pt} \overbar{z}^{-l - 3/2} \hspace{2pt} P_{k,\hspace{.5pt}l}\hspace{2pt},
\end{equation}
where we have additionally expanded the $\mathcal{P}(z,\overbar z)$ operator into modes $P_{k,\hspace{.5pt}l}.$

Following this, we construct $(2,0)$, $(\frac{3}{2},0)$, and $(1,0)$ currents $T(z)$, $S_I(z)$ and $J_{IJ}(z)$ by taking the shadow transform of the negative-helicity, $\overbar{h} = 1$, conformally soft graviton, gravitino, and graviphoton respectively. Their OPEs with a generic conformal primary are fixed by conformally soft theorems to be
\begin{equation}
    \begin{split}
        T(z) \hspace{2pt} \mathcal{O}_{a,\hspace{.5pt}\Delta}^{I_1 \cdots I_A}(0,0) &\sim \bigg(\frac{h}{z^2}  + \frac{1}{z} \partial\bigg) \mathcal{O}_{a,\hspace{.5pt}\Delta}^{I_1 \cdots I_A}(0,0) \hspace{55pt} T(z) = \hspace{3.5pt}\sumop_{n \hspace{1pt} \in\hspace{1pt} \mathbb{Z}} \hspace{6.5pt} z^{-n - 2} \hspace{2pt} L_n \\[.35em]
        S_I(z) \hspace{2pt} \mathcal{O}_{a,\hspace{.5pt}\Delta}^{I_1 \cdots I_A}(0,0) &\sim \frac{1}{z} \hspace{2pt} \delta_I^{[I_1}\mathcal{O}_{a,\hspace{.5pt}\Delta +\frac{1}{2}}^{I_2 \cdots I_A}(0,0) \hspace{78pt} S_I(z) = \hspace{-1.5pt} \sumop_{k \hspace{1pt} \in\hspace{1pt} \mathbb{Z} + \frac{1}{2}} z^{-k - 3/2} \hspace{2pt} G_{I,\hspace{.5pt}k} \\
        J_{IJ}(z) \hspace{2pt} \mathcal{O}_{a,\hspace{.5pt}\Delta}^{I_1 \cdots I_A}(0,0) &\sim \frac{1}{z}\hspace{2pt} Z_{IJ} \hspace{1pt} \mathcal{O}_{a,\hspace{.5pt}\Delta + 1}^{I_1 \cdots I_A}(0,0) \hspace{73pt} J_{IJ}(z) = \hspace{3.5 pt}\sumop_{n \hspace{1pt} \in\hspace{1pt} \mathbb{Z}} \hspace{6.5pt} z^{-n - 1} \hspace{2pt} \mathcal{Z}_{IJ,\hspace{.5pt}n}\hspace{2pt}. \\
    \end{split}
\end{equation}
There are similar expressions for $(0,2)$, $(0,\frac{3}{2})$, and $(0,1)$ currents $\overbar{T}(\overbar z)$, $\overbar{S}^I(\overbar{z})$, and $\overbar{J}^{IJ}(\overbar{z})$ given by shadow transforms of the positive-helicity, $h = 1$, conformally soft operators.

In Appendix \ref{appendix: derivation of sbms}, we use the aforementioned OPEs to deduce the commutators between these various modes. We report the following $\frak{sbms}_{4|\mathcal{N}}$ algebra
\begin{equation}
    \begin{split}
        \big[L_{n},P_{k,\hspace{.5pt}l}\big] &= \big(\tfrac{1}{2}n -k\big)\hspace{1pt} P_{n+k,\hspace{.5pt}l} \hspace{74pt} \big[\overbar{L}_{n},P_{k,\hspace{.5pt}l}\big] = \big(\tfrac{1}{2}n - l\big)\hspace{1pt} P_{k,\hspace{.5pt}n+l}\\[.7em]
        \big[L_{n},L_m\big] &= \big(n-m\big)\hspace{1pt} L_{n+m}\hspace{81pt} \big[\overbar{L}_{n},\overbar{L}_m\big] = \big(n-m\big) \hspace{1pt}\overbar{L}_{n+m} \\[.7em]
        \big[L_{n},G_{I,\hspace{.5pt}k}\big] &= \big(\tfrac{1}{2}n - k\big)\hspace{1pt} G_{I,\hspace{.5pt}n+k} \hspace{75pt} \big[\overbar{L}_{n},\overbar{G}^I_k\big] = (\tfrac{1}{2}n - k)\hspace{1pt} \overbar{G}^I_{n+k}\\[.7em]
        \big[L_{n},\mathcal{Z}_{IJ,\hspace{.5pt}m}\big] &= -m \hspace{1pt} \mathcal{Z}_{IJ,\hspace{.5pt}n+m}\hspace{88pt} \big[\overbar{L}_{n},\overbar{\mathcal{Z}}^{IJ}_m\big] = -m \hspace{1pt} \overbar{\mathcal{Z}}^{IJ}_m \\[.7em]
        \big[G_{I,\hspace{.5pt}k},G_{J,\hspace{.5pt}l}\big] = -\big(k-l\big)\hspace{1pt} &\mathcal{Z}_{IJ,\hspace{.5pt}k+l} \hspace{20pt}\hspace{12pt} \big[\overbar{G}^I_{k},\overbar{G}^J_l\big] = -\big(k-l\big)\hspace{1pt} \overbar{\mathcal{Z}}^{IJ}_{k+l} \hspace{33pt} \big[G_{I,\hspace{.5pt}k},\overbar{G}^J_{l}\big] = \delta^J_I \hspace{1pt} P_{k,\hspace{.5pt}l}\hspace{2pt}.\\
    \end{split}
    \label{eqn: extended super BMS algebra}
\end{equation}
We conclude this section with a few relevant comments and highlight several open questions.
\begin{itemize}
    \item \textbf{A Local Extension of the Super-Poincar\'e Algebra:} Various generators of this algebra may be identified with the usual super-Poincar\'e generators. For example
    \begin{equation}
        \begin{split}
            Q_{I,\hspace{.5pt}0} \hspace{3pt} &\leftrightsquigarrow \hspace{3pt}G_{I,\hspace{.5pt}-1/2} \hspace{39pt} Q_{I,\hspace{.5pt}1}\hspace{3pt} \leftrightsquigarrow\hspace{3pt} G_{I,\hspace{.5pt}+1/2} \hspace{39pt} Z_{IJ} \hspace{3pt}\leftrightsquigarrow\hspace{3pt} \mathcal{Z}_{IJ,\hspace{.5pt}0} \\[.3em]
            \overbar{Q}_{0}^I \hspace{3pt} &\leftrightsquigarrow \hspace{3pt}\overbar{G}_{-1/2}^I \hspace{54pt} \overbar{Q}_{1}^I \hspace{3pt} \leftrightsquigarrow\hspace{3pt} \overbar{G}^I_{+1/2} \hspace{45pt} \overbar{Z}^{IJ} \hspace{3pt}\leftrightsquigarrow\hspace{3pt} \overbar{\mathcal{Z}}^{IJ}_{0}\hspace{2pt}.
        \end{split}
        \label{eqn: conformal generators in sbms}
    \end{equation}
    These generators act on conformal primaries in the same way that the super-Poincar\'e generators do and form a closed subalgebra amongst themselves. The $\frak{sbms}_{4|\mathcal{N}}$ algebra should, therefore, be viewed as a local enhancement of the super-Poincar\'e algebra. This is consistent with asymptotic symmetries generalizing ordinary symmetries by having a non-trivial, local dependence on the celestial sphere. For example, the current $S_{I}(z)$ should be viewed as generating \textit{large supersymmetry transformations} in the bulk (see \cite{Awada:1985by,Ferber:1977qx, Avery:2015iix, Lysov:2015jrs, Henneaux:2020ekh, Fuentealba:2020aax, Fuentealba:2021xhn, Pasterski:2021fjn, Boulanger:2023gpw}).
    \item \textbf{Large Gauge Symmetry Involving Central Charges:} The $J_{IJ}(z)$ current generates a large gauge symmetry involving the central charges $Z^{IJ}$. To our knowledge such a symmetry has not been previously studied at all, though its existence seems completely natural from the perspective of conformally soft theorems. Indeed, the graviphoton is a $U(1)$ gauge field whose $U(1)$ charges are precisely the central charges, $Z_{IJ}$. Its asymptotic symmetries should mirror those of any other $U(1)$ gauge field \cite{Strominger:2017zoo, He:2014cra}, in this case generating a large gauge symmetry where there is a \textit{conservation of central charge at every angle} of the celestial sphere. We, therefore, call this a \textit{large central charge} symmetry.
    
    \hspace{13pt} Due to the BPS bound, central charges play a privileged role in supersymmetric theories. Moreover, $4d$ supergravity theories may often be viewed as compactifications of higher dimensional theories with central charges of the $4d$ theory encoding the geometry of the internal manifold \cite{Polchinski:1998rr}. We, thus, expect large central charge symmetry to have a higher-dimensional interpretation as a large diffeomorphism symmetry of this internal manifold (where the vector field generating diffeomorphisms has non-trivial dependence on both the celestial sphere and the internal manifold). For further discussions of asymptotic symmetries in compactifications, see \cite{Marotta:2019cip, Marotta:2021oiw, Miller:2022fvc, Tropper:2023fjr}.
    \item \textbf{Virasoro but no Super-Virasoro Subalgebra:} Another interesting subalgebra is the non-supersymmetric extended $\frak{bms}_4$ algebra. It is not surprising to find such a subalgebra because all theories of flat space quantum gravity --- supersymmetric or not --- should enjoy these symmetries. Included in the extended $\frak{bms}_4$ algebra are two copies of the Virasoro algebra generated by $L_n$ and $\overbar{L}_n.$ Curiously, the $\frak{sbms}_{4|\mathcal{N}}$ algebra does not have a super-Virasoro subalgebra. While the $G_{I}(z)$ currents have the appropriate weight for supercurrents in a $2d$ CFT, their modes commute to give supertranslations or large central charge symmetry generators rather than Virasoro generators. The difference between $4d$ spacetime supersymmetry and $2d$ supersymmetry on the celestial sphere is further discussed in Section \ref{sec: conclusion}.
    \item \textbf{Universality:} This algebra is universal, holding (even at loop level) for all theories of quantum gravity including those with an arbitrary number of supersymmetries. One might be worried that there should be more generators in this algebra when $\mathcal{N} > 2$. For example, one might wonder about the shadow transform of the dilatino. This ends up not being a concern. For one, generators in this algebra correspond to $\overbar{h} = 3/2$ or $\overbar{h} = 1$ negative helicity gravitons, gravitinos, and graviphotons as well as their $h = 3/2$ or $h = 1$ positive-helicity anti-particles. As shown in Table \ref{tab: conformally soft theorems (grav)}, the positive-helicity dilatino has trivial conformally soft theorems for both $\overbar{h} = 3/2$ and $\overbar{h} = 1$. Mathematically, this means that the algebra generated by the modes of the aforementioned currents closes on itself without any contributions from the dilatino. 
    
    \hspace{13pt} This is not surprising because the leading conformally soft dilatino theorem (appearing at $\overbar{h} = 1/2$) is not universal. The $\frak{sbms}_{4|\mathcal{N}}$ algebra only seems to encode the universal extension to the super-Poincar\'e algebra for which the dilatino plays no role. Of course, if we include the modes of all $\overbar{h} = 1/2$ conformally soft operators, we expect an even larger non-universal algebra. To our knowledge, this hasn't been understood either even in the simpler case of non-supersymmetric theories (see \cite{Banerjee:2021cly, Himwich:2023njb} for partial results).
    \item \textbf{Central Extensions:} Finally, the analysis presented in Appendix \ref{appendix: derivation of sbms} cannot fix any possible central terms appearing in this algebra, and we leave them unspecified.
\end{itemize}

\section{Supersymmetric OPEs}
\label{sec: OPEs}

In this section, we describe how another important structure in celestial holography generalizes to the supersymmetric setting. Namely, singular terms in the tree-level $\mathcal{O}^{I_1 \cdots I_A}_{a,\hspace{1pt}\Delta_a}(z,\overbar z) \hspace{2pt} \mathcal{O}^{J_1 \cdots J_B}_{b,\hspace{1pt}\Delta_b}(0,0)$ OPE in a holomorphic limit where one takes $z \rightarrow 0$ while holding $\overbar z =$ fixed.\footnote{In this article, we report on the simplest case of $2$-collinear OPEs. There is a broader framework of $n$-collinear OPEs which generally include branch cuts owing to contributions from multi-particle operators \cite{Ball:2023sdz, Ball:2024oqa, Guevara:2024ixn, Ball:2022bgg}. In supersymmetric theories, the double residue condition is satisfied \cite{Ball:2023qim}, so there are no branch cuts in the 3-collinear OPE. It is believed that this feature will persist for all $n$, meaning the singular part of the $2$-collinear OPEs genuinely captures all information about the singular terms in the full OPE of the theory.}

\subsection{Supersymmetry Relations Among OPE Coefficients}

At tree-level, holomorphic OPEs may be obtained using the formalism developed in \cite{Himwich:2021dau}. The \textit{OPE coefficient} for $\mathcal{O}^{K_1 \cdots K_C}_{c,\hspace{1pt}\Delta_c}$ is proportional to $g\big(a^{I_1 \cdots I_A},b^{J_1 \cdots J_B}\hspace{1pt};\hspace{1pt} c^{K_1 \cdots K_C}\big)$, a theory-dependent constant controlling the holomorphic collinear splitting for the scattering process $a^{I_1 \cdots I_A} + b^{J_1 \cdots J_B} \rightarrow c^{K_1 \cdots K_C}$ in the bulk. If there are $r$ highest-weight states, then there are $r \hspace{1pt} 2^\mathcal{N}$ different particle species in the theory, so OPE data is fixed in terms of $r^3 \hspace{1pt} 2^{3\mathcal{N}}$ a priori unknown OPE coefficients.

Fortunately, supersymmetry places extremely powerful constraints on these couplings. As shown in Appendix \ref{appendix: supersymmetry relations and covariance}, \textit{all} non-vanishing OPE coefficients are completely fixed by $r^3$ three-point coefficients $g_{ab}^c$ mediating the collinear splitting among \textit{highest-weight states} $a + b \rightarrow c.$ Specifically, the non-vanishing OPE coefficients take the form
\begin{equation}
    g\big(a^{I_1 \cdots I_A}, b^{J_1 \cdots J_B}\hspace{1pt};\hspace{1pt} c^{I_1 \cdots I_A J_1 \cdots J_B}\big) = (-1)^{B|a|} \hspace{2pt} g_{ab}^c\hspace{2pt}.
    \label{eqn: three-point proposition}
\end{equation}
Notice that the indices on the third particle exactly match the indices on the first two particles. When they do not match like this, the OPE coefficient is identically zero. Moreover, there can be no repetitions among the $I_1,...,I_A,J_1,...,J_B$ indices as they are totally anti-symmetric on $c^{I_1 \cdots I_A J_1 \cdots J_B}$. Furthermore, there are additional constraints coming from the symmetry conditions
\begin{equation}
    g^{c}_{ab} = (-1)^{|a||b| + p + 1} \hspace{2pt} g_{ba}^{c} \hspace{30pt}\text{and} \hspace{30pt}
        g_{ab}^c = g_{b\overbar c}^{\overbar a}\hspace{2pt}.
        \label{eqn: symmetry relations three point}
\end{equation}
Altogether, the number of \textit{independent} OPE coefficients is bounded by 
\begin{equation}
    \# \hspace{2pt}\big(\text{Independent} ~ g_{ab}^c\big)  \leq \hspace{2pt} \frac{1}{6} (r^3 + 3r^2 + 2r)\hspace{2pt}.
    \label{eqn: bound on number of independent three-point coefficients}
\end{equation}

Moreover, the three-point coefficients $g_{ab}^c$ only contribute to the \textit{holomorphic} OPE when the helicities of the highest-weight states satisfy \cite{Himwich:2021dau}
\begin{equation}
    p \equiv s_a + s_b - s_c - 1 \geq 0\hspace{2pt}.
    \label{eqn: def of p simplified}
\end{equation}
This further restricts the number of independent $g_{ab}^c$ which need to be specified. For example, as we shall see in Section \ref{sec: super w 1 + inf}, in a supergravity theory whose particle content involves a single gravity multiplet and its CPT conjugate, there is exactly one independent OPE coefficient from which all others can be determined (unless $\mathcal{N} = 4$, in which case there are two).

\subsection{Universal Structure of Supersymmetric OPEs}
\label{sec: SUSY OPE expressions}

OPEs may be read off from \cite{Himwich:2021dau} after a little massaging using Equations \eqref{eqn: three-point proposition} and \eqref{eqn: def of p simplified}.\footnote{Similar expressions in specific supersymmetric theories are reported in \cite{jiang2022holographic,Bu:2021avc}.}
\begin{equation}
    \begin{split}
        \mathcal{O}^{I_1 \cdots I_A}_{a,\hspace{1pt}\Delta_a}(z,\overbar z)\hspace{2pt} \mathcal{O}^{J_1 \cdots J_B}_{b,\hspace{1pt} \Delta_b}(0,0) &\sim -\frac{(-1)^{B|a|}}{z} \sumop_{c} \hspace{1pt} g_{ab}^c \hspace{2pt}C_{p}(\hspace{1pt}\overbar{h}_a, \overbar{h}_b) \hspace{1pt} \mathcal{O}_{c,\hspace{1pt} \Delta_a +\Delta_b + p-1}^{I_1 \cdots I_A J_1 \cdots J_B}(0,0) \hspace{2pt}.
    \end{split}
    \label{eqn: holomorphic OPE}
\end{equation}
Here, $c$ sums over highest-weight states with $s_c \leq s_a + s_b - 1$ (to ensure that $p \geq 0$) and
\begin{equation}
    C_{p}(\overbar{h}_a, \overbar{h}_a) \equiv \sumop_{m=0}^\infty \frac{1}{m!} \hspace{2pt} B( 2\overbar{h}_a + p + m, 2\overbar{h}_b + p) \hspace{2pt} \overbar z^{p+m} \hspace{2pt} \overbar{\partial}^m \hspace{2pt}.
\end{equation}

This expression is much simpler than it had any right to be thanks to Equations \eqref{eqn: three-point proposition} and \eqref{eqn: def of p simplified} which allow us to sum over just the supercelestial primaries rather than all conformal primaries. Moreover, the indices on the left-hand side get immediately carried to the right-hand side with their relative order preserved. An immediate corollary is that if $\mathcal{O}^{I_1 \cdots I_A}_{a,\hspace{.5pt}\Delta_a}$ and $\mathcal{O}^{J_1 \cdots J_B}_{b,\hspace{.5pt}\Delta_b}$ share an index, their OPE will be trivial due to the anti-symmetry of indices on the right-hand side. 

\subsection{OPEs on the Celestial Super Sphere}

There is a nice description of Equation \eqref{eqn: holomorphic OPE} in terms of OPEs for superoperators (defined in Equation \eqref{eqn: Superoperator}) living on the celestial supersphere
\begin{equation}
     \mathbb{O}_{a,\hspace{.5pt}\Delta_a}(z,\overbar z\hspace{.5pt}|\hspace{.5pt}\eta_1)\hspace{2pt} \mathbb{O}_{b,\hspace{.5pt} \Delta_b}(0,0\hspace{.5pt}|\hspace{.5pt}\eta_2) \sim -\frac{1}{z} \sumop_{c}  \hspace{1pt} g_{ab}^c \hspace{2pt}C_p(\overbar{h}_a, \overbar{h}_b) \hspace{1pt} \mathbb{O}_{c,\hspace{.5pt} \Delta_a + \Delta_b + p}(0,0\hspace{.5pt}|\hspace{.5pt}\eta_1 + \eta_2) \hspace{2pt}.
\end{equation}
It is worth mentioning that this equation is covariant under the action of supersymmetry in the sense that after acting on both sides with a bulk supercharge, the results are self-consistent.
\begin{equation}
    \begin{split}
        &\big[Q_{I,\alpha}, \mathbb{O}_{a,\hspace{.5pt}\Delta_a}(z,\overbar z\hspace{.5pt}|\hspace{.5pt}\eta_1) \hspace{1pt} \mathbb{O}_{b,\hspace{.5pt}\Delta_b}(0,0\hspace{.5pt}|\hspace{.5pt}\eta_2)\big] \\[.5em]
        &\hspace{30pt}\hspace{3pt}= \big[Q_{I,\alpha}, \mathbb{O}_{a,\hspace{.5pt}\Delta_a}(z,\overbar z\hspace{.5pt}|\hspace{.5pt}\eta_1)\big]\hspace{1pt} \mathbb{O}_{b,\hspace{.5pt}\Delta_b}(0,0\hspace{.5pt}|\hspace{.5pt}\eta_2) + (-1)^{|a|} \mathbb{O}_{a,\hspace{.5pt}\Delta_a}(z,\overbar z\hspace{.5pt}|\hspace{.5pt}\eta_1) \hspace{1pt} \big[Q_{I,\alpha}, \mathbb{O}_{b,\hspace{.5pt}\Delta_b}(0,0\hspace{.5pt}|\hspace{.5pt}\eta_2)\big]  \\[.2em]
        &\hspace{30pt}\hspace{3pt}\sim -\frac{1}{z} \sumop_{c}  \hspace{1pt} g_{ab}^c \hspace{1pt}C_p(\overbar{h}_a, \overbar{h}_b) \hspace{2pt} \big[Q_{I,\alpha},\mathbb{O}_{c,\hspace{.5pt}\Delta_a + \Delta_b + p-1}(0,0\hspace{.5pt}|\hspace{.5pt}\eta_1 + \eta_2)\big] \\[.75em]
        &\big[\overbar{Q}^I_{\overbar{\alpha}}, \mathbb{O}_{a,\hspace{.5pt}\Delta_a}(z,\overbar z\hspace{.5pt}|\hspace{.5pt}\eta_1) \hspace{1pt} \mathbb{O}_{b,\hspace{.5pt}\Delta_b}(0,0\hspace{.5pt}|\hspace{.5pt}\eta_2)\big] \\[.5em]
        &\hspace{30pt}\hspace{3pt}= \big[\overbar{Q}^I_{\overbar{\alpha}}, \mathbb{O}_{a,\hspace{.5pt}\Delta_a}(z,\overbar z\hspace{.5pt}|\hspace{.5pt}\eta_1)\big]\hspace{1pt} \mathbb{O}_{b,\hspace{.5pt}\Delta_b}(0,0\hspace{.5pt}|\hspace{.5pt}\eta_2) + (-1)^{|a|} \mathbb{O}_{a,\hspace{.5pt}\Delta_a}(z,\overbar z\hspace{.5pt}|\hspace{.5pt}\eta_1) \hspace{1pt} \big[\overbar{Q}^I_{\overbar{\alpha}}, \mathbb{O}_{b,\hspace{.5pt}\Delta_b}(0,0\hspace{.5pt}|\hspace{.5pt}\eta_2)\big]  \\[.2em]
        &\hspace{30pt}\hspace{3pt}\sim -\frac{1}{z} \sumop_{c}  \hspace{1pt} g_{ab}^c \hspace{2pt}C_p(\overbar{h}_a, \overbar{h}_b) \hspace{1pt} \big[\overbar{Q}^I_{\overbar{\alpha}},\mathbb{O}_{c,\hspace{.5pt}\Delta_a + \Delta_b + p-1}(0,0\hspace{.5pt}|\hspace{.5pt}\eta_1 + \eta_2)\big] \hspace{2pt}.
    \end{split}
    \label{eqn: OPE covariance under SUSY}
\end{equation}
This is ultimately related to the fact that the three-point coefficients are subject to such powerful constraints from supersymmetry; the verification of this identity isn't especially illuminating, so we defer it to Appendix \ref{appendix: supersymmetry relations and covariance}.

\section{Supersymmetric Chiral Soft Algebras}
\label{sec: chiral soft algebras}

In this section, we study how chiral soft algebras in CCFT generalize to the supersymmetric setting. \textit{Chiral soft algebras} concisely elucidate an infinite tower of symmetries associated to soft theorems. They can be systematically derived by studying holomorphic OPEs in the theory through a formalism developed in \cite{Guevara:2021abz, Strominger:2021mtt,Himwich:2021dau,Mago:2021wje}, which we briefly review.

\subsection{From Holomorphic OPEs to Chiral Soft Algebras}

The generators of chiral soft algebras are related to the conformally soft operators $\mathcal{O}_{a,\hspace{.5pt} k,\hspace{.5pt} \text{CS}}^{I_1 \cdots I_A}(z,\overbar z)$ defined in Equation \eqref{eqn: def conformally soft}. Such operators have $\overbar{h} = \frac{1}{2}(k - s(a^{I_1 \cdots I_A})) \in \frac{1}{2}\mathbb{Z}.$ For $\overbar{h} \leq 0$, they can be consistently expanded into a negative weight $(-2\overbar{h} + 1)-$dimensional representation of $\overbar{SL}_2(\mathbb{R})$:
\begin{equation}
    \mathcal{O}_{a,\hspace{1pt} \overbar{h},\hspace{1pt} \text{CS}}^{I_1 \cdots I_A}(z,\overbar z) = \sumop_{\overbar{m} \hspace{1pt} = \hspace{1pt} \overbar{h}}^{-\overbar h}  \hspace{3pt} \frac{\overbar z^{-\overbar{m} - \overbar{h}}}{\Gamma(1-\overbar{h}+\overbar{m})\Gamma(1-\overbar{h}-\overbar{m})} \hspace{2pt} \mathcal{R}^{I_1 \cdots I_A}_{a,\hspace{1pt}1-\overbar{h},\hspace{1pt} \overbar{m}}(z)\hspace{2pt}.
\end{equation}
Including the gamma functions gives a convenient mode redefinition which ultimately yields nicer expression for the chiral soft algebra  as originally observed in \cite{Strominger:2021mtt}. It is conventional to define the index $q = 1-\overbar{h}$ writing the operators in this mode expansion as $\mathcal{R}^{I_1 \cdots I_A}_{a,\hspace{1pt} q, \overbar{m}}(z)$. These operators are chiral currents with holomorphic weight $h = 1 - q + s(a^{I_1 \cdots I_A}).$ They may be further mode expanded 
\begin{equation}
    \mathcal{R}^{I_1 \cdots I_A}_{a,\hspace{1pt} q,\hspace{1pt} \overbar{m}}(z) = \sumop_{n \hspace{1pt} \in \hspace{1pt} \mathbb{Z}+h} z^{-n - h} \hspace{2pt} \mathcal{R}^{I_1 \cdots I_A}_{a,\hspace{1pt} q,\hspace{1pt} \overbar{m},\hspace{1pt} n}\hspace{2pt}.
\end{equation}
To summarize notation: $\mathcal{R}^{I_1 \cdots I_A}_{a,\hspace{1pt} q,\hspace{1pt} \overbar{m},\hspace{1pt} n}$ describes physics associated to the operator $\mathcal{O}^{I_1 \cdots I_A}_{a,\hspace{1pt} \Delta}(z,\overbar z)$ in the supercelestial primary multiplet of the highest-weight state $a$. The $q$ index refers to the anti-holomorphic weight of this operator, while the $\overbar{m}$ and $n$ indices respectively parameterize a mode expansion in $\overbar z$ and $z$. The various indices run over the range:
\begin{equation}
    q = 1, \hspace{2pt}\frac{3}{2},\hspace{2pt} 2,\hspace{2pt} \frac{5}{2},\hspace{2pt} ... \hspace{30pt},\hspace{30pt} 1-q \leq \overbar{m} \leq q-1 \hspace{30pt},\hspace{30pt} n \in \mathbb{Z} + h\hspace{2pt}.
\end{equation}
Finally, $\mathcal{R}^{I_1 \cdots I_A}_{a,\hspace{1pt} q,\hspace{1pt} \overbar{m},\hspace{1pt} n}$ is fermionic precisely when the corresponding bulk particle, $a^{I_1 \cdots I_A}(p),$ is fermionic.

One can verify that $\mathcal{R}^{I_1 \cdots I_A}_{a,\hspace{1pt} q,\hspace{1pt} \overbar{m},\hspace{1pt} n}$ satisfy the following commutation relations with bulk supercharges
\begin{equation}
    \begin{split}
        \big[\overbar{Q}^{J}_{\overbar \alpha},\hspace{1pt}\mathcal{R}^{I_1 \cdots I_A}_{a,\hspace{1pt} q,\hspace{1pt} \overbar{m},\hspace{1pt} n}\hspace{1pt}\big] &= \big((1-2\overbar \alpha)\hspace{.5pt} \overbar{m} + q - 1\big) R^{JI_1 \cdots I_A}_{a,\hspace{1pt} q-\frac{1}{2},\hspace{1pt} \overbar{m} + \overbar \alpha - \frac{1}{2},\hspace{1pt} n}\\[.3em]
        \big[Q_{J \alpha}\hspace{1pt},\hspace{1pt}\mathcal{R}^{I_1 \cdots I_A}_{a,\hspace{1pt} q,\hspace{1pt} \overbar{m},\hspace{1pt} n}\hspace{1pt}\big]  &= - \delta_J^{[I_1} \mathcal{R}^{I_2 \cdots I_A}_{a,\hspace{1pt}q,\hspace{1pt}\overbar{m},\hspace{1pt}n+ \alpha-\frac{1}{2}}\hspace{2pt}.
        \label{eqn: supercharge and chiral soft generators}
    \end{split}
\end{equation}
The reason that $\overbar{Q}^J_{\overbar{\alpha}}$ and $Q_{J,\alpha}$ act in such a different fashion is because we are working in a chiral limit. One might also be worried that the $\overbar{Q}^J_{\overbar{\alpha}}$ operators can take the $\mathcal{R}$ generators out of the wedge $1-q \leq \overbar{m} \leq q-1.$ However, when this would happen, the prefactor of $\big((1-2\overbar \alpha)\hspace{.5pt} \overbar{m} + q - 1\big)$ vanishes; thus, the action of supersymmetry constrains one to live inside the wedge.

These operators generate the chiral soft algebra for the theory. Their mode algebra is inherited from the standard commutator between holomorphic currents $\mathcal{R}^{I_1 \cdots I_A}_{a,\hspace{1pt} q_1,\hspace{1pt} \overbar{m}_1}(z)$ and $\mathcal{R}^{J_1 \cdots J_B}_{b,\hspace{1pt} q_2,\hspace{1pt} \overbar{m}_2}(z)$ defined by the contour integral
\begin{equation}
    \begin{split}
        \Big[\mathcal{R}^{I_1 \cdots I_A}_{a,\hspace{1pt} q_1,\hspace{1pt} \overbar{m}_1,n_1}, \mathcal{R}&^{J_1 \cdots I_B}_{b,\hspace{1pt} q_2,\hspace{1pt} \overbar{m}_2,n_2}\Big] = \oint_0 \hspace{2pt} \frac{dw}{2\pi i} \hspace{2pt} w^{n_2 + h_2 - 1} \oint_{w} \frac{dz}{2\pi i} \hspace{2pt} z^{n_1 + h_1 - 1} \hspace{2pt} \mathcal{R}^{I_1 \cdots I_A}_{a,\hspace{1pt}q_1,\hspace{1pt}\overbar{m}_1}(z) \hspace{2pt} \mathcal{R}^{J_1 \cdots I_B}_{b,\hspace{1pt}q_2,\hspace{1pt}\overbar{m}_2}(w)\hspace{2pt}.
    \end{split}
    \label{eqn: contour integral algebra}
\end{equation}
We should emphasize that these are \textit{boundary (anti-)commutators} among generators of the chiral soft algebra. They should not be confused with the commutators given in Equation \eqref{eqn: supercharge and chiral soft generators} involving the bulk supercharge, though we shall see that the two structures are consistent with one another.

\subsection{Universal Structure of Supersymmetric Chiral Soft Algebras}
\label{sec: SUSY Soft Algebra expressions}

We may compute this quantity by plugging the tree-level holomorphic OPE \eqref{eqn: holomorphic OPE} into the contour integral and simplifying the results using some mathematical machinery developed in \cite{Mago:2021wje}
\begin{equation}
    \begin{split}
        \Big[\mathcal{R}^{I_1 \cdots I_A}_{a,\hspace{1pt} q_1,\hspace{1pt} \overbar{m}_1,n_1}&, \mathcal{R}^{J_1 \cdots J_B}_{b,\hspace{1pt} q_2,\hspace{1pt} \overbar{m}_2,n_2}\Big] 
        = -(-1)^{B|a|} \sumop_{c} \hspace{2pt} g_{ab}^c \hspace{2pt} N_p(q_1,q_2,\overbar{m}_1,\overbar{m}_2) \hspace{2pt} \mathcal{R}^{I_1 \cdots I_A J_1 \cdots J_B}_{c,\hspace{1pt} q_1 + q_2 - p - 1, \hspace{1pt} \overbar{m}_1 + \overbar{m}_2,\hspace{1pt} n_1 + n_2} \hspace{2pt},
        \label{eqn: chiral algebra}
    \end{split}
\end{equation}
where $c$ sums over highest-weight states (i.e. supercelestial primaries). Again, the indices on the left-hand side get immediately carried over to the right-hand side with their order preserved. We have also defined the functions $N_p(q_1,q_2,\overbar{m}_1,\overbar{m}_2)$ as 

\begin{equation}
    \begin{split}
        N_p(q_1,q_2,\overbar{m}_1,\overbar{m}_2) &= \sumop_{x=0}^p (-1)^{p-x} \binom{p}{x} \frac{\Gamma(q_1 + \overbar{m}_1)}{\Gamma(q_1 + \overbar{m}_1 - p + x)} \frac{\Gamma(q_1 - \overbar{m}_1)}{\Gamma(q_1 -\overbar{m}_1 -x)}\\
        &\hspace{140pt} \frac{\Gamma(q_2 + \overbar{m}_2)}{\Gamma(q_2 + \overbar{m}_2 -x)} \frac{\Gamma(q_2 -\overbar{m}_2)}{\Gamma(q_2 - \overbar{m}_2 - p +x)} \hspace{2pt}.
    \end{split}
\end{equation}
These coefficients are only non-vanishing when $p \geq 0$, consistent with the constraint \eqref{eqn: def of p simplified}. The explicit expression for $p = 1$ is
\begin{equation}
    \begin{split}
        N_1(q_1,q_2,\overbar{m}_1,\overbar{m}_2) = -2 \big((q_2 - 1)\overbar{m}_1 - (q_1 - 1)\overbar{m}_2\big)\hspace{2pt}.
    \end{split}
\end{equation}
These are precisely the structure constants appearing in the $\myw^\wedge_{1+\infty}$ algebra, which have been shown to describe the chiral soft algebra appearing in Einstein gravity \cite{Strominger:2021mtt}. We will have more to say about this in Section \ref{sec: super w 1 + inf}.

\subsection{Chiral Algebras on the Celestial Supersphere}

The chiral algebra may also be recast into a superspace notation by defining generators
\begin{equation}
    \mathds{R}_{\hspace{1pt} a,\hspace{.5pt} q,\hspace{.5pt}\overbar{m},\hspace{.5pt}n}(\eta) = \sumop_{A\hspace{.5pt}=\hspace{.5pt}0}^\mathcal{N} \frac{1}{A!} \hspace{2pt} \eta_{I_A} \cdots \eta_{I_1} \hspace{1pt} \mathcal{R}^{I_1 \cdots I_A}_{a,\hspace{.5pt}q,\hspace{.5pt}\overbar{m},\hspace{.5pt} n}\hspace{2pt},
\end{equation}
which satisfy the commutation relations
\begin{equation}
    \begin{split}
        \big[\overbar{Q}^{J}_{\overbar \alpha} ,\hspace{1pt}\mathds{R}_{\hspace{1pt}a,\hspace{.5pt}q,\hspace{.5pt} \overbar{m},\hspace{.5pt} n}(\eta)\hspace{1pt}\big] &= \big((1-2\overbar \alpha)\hspace{.5pt} \overbar{m} + q - 1\big) \hspace{1pt} \frac{\partial}{\partial \eta_J} \hspace{1pt} \mathds{R}_{\hspace{1pt}a,\hspace{.5pt}q-\frac{1}{2},\hspace{.5pt} \overbar{m} + \overbar{\alpha} - \frac{1}{2},\hspace{.5pt} n}(\eta) \\
        \big[Q_{J,\alpha}\hspace{1pt},\hspace{1pt}\mathds{R}_{\hspace{1pt}a,\hspace{.5pt}q,\hspace{.5pt} \overbar{m},\hspace{.5pt}n}(\eta)\hspace{1pt}\big] &= - \eta_J \hspace{1pt} \mathds{R}_{\hspace{1pt}a,\hspace{.5pt}q,\hspace{.5pt}\overbar{m},n+\alpha - \frac{1}{2}}(\eta)\hspace{2pt},
    \end{split}
\end{equation}
and enjoy the following algebra which is a superspace generalization of Equation \eqref{eqn: chiral algebra}
\begin{equation}
    \begin{split}
        \big[\mathds{R}_{\hspace{1pt} a,\hspace{.5pt} q_1,\hspace{.5pt}\overbar{m}_1,\hspace{.5pt}n_1}(\eta_1),\hspace{1pt}&\mathds{R}_{\hspace{1pt} b,\hspace{.5pt} q_2,\hspace{.5pt} \overbar{m}_2,\hspace{.5pt}n_2}(\eta_2)\big] \\[.3em]
        &= -\sumop_{c} \hspace{1pt} g_{ab}^c \hspace{2pt} N_p(q_1,q_2,\overbar{m}_1,\overbar{m}_2) \hspace{2pt} \mathds{R}_{\hspace{1pt}c,\hspace{.5pt} q_1 + q_2 - p - 1,\hspace{.5pt} \hspace{1pt} \overbar{m}_1 + \overbar{m}_2,\hspace{.5pt} n_1 + n_2}(\eta_1 + \eta_2)\hspace{2pt}.
    \end{split}
    \label{eqn: chiral algebra superspace}
\end{equation}

\subsection{Jacobi Identities}

We conclude this section by noting the following two associativity properties of the chiral soft algebra. First, it was shown in \cite{Ball:2023qim} that all supersymmetric theories satisfy the double residue condition. It follows that chiral algebras in supersymmetric theories necessarily satisfy the following Jacobi identity \cite{Ball:2023sdz}
\begin{equation}
    \begin{split}
        \big[\mathds{R}_{\hspace{.5pt}a}, \big[\mathds{R}_{\hspace{.5pt}b},\mathds{R}_{\hspace{.5pt}c}\big]\big] &= \big[\big[\mathds{R}_{\hspace{.5pt}a},\mathds{R}_{\hspace{.5pt}b}\big],\mathds{R}_{\hspace{.5pt}c}\big]\big]  + (-1)^{|a||b|} \hspace{1pt} \big[\mathds{R}_{\hspace{.5pt}b},\big[\mathds{R}_{\hspace{.5pt}a},\mathds{R}_{\hspace{.5pt}c}\big]\big]\hspace{2pt},
    \end{split}
\end{equation}
where we have dropped non-essential subscripts for readability. There is a second set of interesting Jacobi identities originally pointed out in \cite{Crawley:2024cak} in the context of $\mathcal{N} = 2$ supersymmetric gauge theory
\begin{equation}
    \begin{split}
        \big[Q_{J,\alpha}, \big[\mathds{R}_{\hspace{.5pt}a},\mathds{R}_{\hspace{.5pt}b}\big]\big] &= \big[\big[Q_{J,\alpha},\mathds{R}_{\hspace{.5pt}a}\big],\mathds{R}_{\hspace{.5pt}b}\big] + (-1)^{|a|} \hspace{1pt}\big[\mathds{R}_{\hspace{.5pt}a},\big[Q_{J, \alpha},\mathds{R}_{\hspace{.5pt}b}\big]\big] \\[.5em] \vspace{3pt}
        \big[\hspace{3pt} \overbar{Q}^J_{\hspace{1pt} \overbar \alpha}\hspace{2.75pt}, \big[\mathds{R}_{\hspace{.5pt}a},\mathds{R}_{\hspace{.5pt}b}\big]\big] &= \big[\big[\hspace{3pt} \overbar{Q}^J_{\hspace{1pt} \overbar \alpha}\hspace{2.75pt},\mathds{R}_{\hspace{.5pt}a}\big],\mathds{R}_{\hspace{.5pt}b}\big] + (-1)^{|a|} \hspace{1pt}\big[\mathds{R}_{\hspace{.5pt}a},\big[\hspace{3pt} \overbar{Q}^J_{\hspace{1pt} \overbar \alpha}\hspace{2.75pt},\mathds{R}_{\hspace{.5pt}b}\big]\big]\hspace{2pt}.
        \label{eqn: fermionic derivation}
    \end{split}
\end{equation}
This set of Jacobi identities is checked using the relations among $g_{ab}^c$ in Appendix \ref{appendix: supersymmetry relations and covariance}. In fact, such careful checks are redundant, as these identities are actually consequences of the OPE relation \eqref{eqn: OPE covariance under SUSY}. 

Equation \eqref{eqn: fermionic derivation} implies that the fermionic generators $Q_{J,\alpha}$ and $\overbar{Q}^{J}_{\hspace{1pt}\overbar{\alpha}}$ constitute a set of $4 \mathcal{N}$ \textit{fermionic derivations} of the chiral soft algebra. For a generic CCFT, they are outer derivations of the Lie algebra. For a CCFT dual to a theory of quantum gravity, these become \textit{inner derivations} because the supercharges are related to the $\overbar{h} = 1$ negative-helicity conformally soft gravitino and the $h = 1$ positive-helicity conformally soft gravitino according to Equation \eqref{eqn: conformal generators in sbms}!

\section{Super $\mathcal{L}(\myw^\wedge_{1+\infty})$, its $\Lambda$-deformation, and a Twistorial Origin}
\label{sec: super w 1 + inf}

In this section, we study chiral soft algebras in supergravity as an explicit example of the above framework.  For simplicity, we assume that the only particles in the theory either live in the gravity multiplet or its CPT conjugate. In $\mathcal{N} = 8$ supergravity, the gravity multiplet is CPT self-conjugate, so there is only one highest-weight state in the theory --- call it $a(p)$. All chiral soft algebras in the theory are completely specified by the OPE coefficient $g_{aa}^a.$ The same is true for $\mathcal{N} = 7$ supergravity whose two supermultiplets organize themselves into a single $\mathcal{N} = 8$ supermultiplet. 

In a less constrained setting (i.e. $\mathcal{N} \leq 6$) where one needs to discuss both the gravity multiplet $a(p)$ and its CPT conjugate $\overbar{a}(p)$, there are $2^3 = 8$ OPE coefficients among the highest-weight states which completely specify the algebra
\begin{equation}
    \underbrace{~ g_{aa}^a \hspace{15pt},\hspace{15pt} g_{a\overbar a}^{\overbar a} \hspace{15pt},\hspace{15pt} g_{\overbar a a}^{\overbar{a}}~ }_{\text{identical}} \hspace{15pt},\hspace{15pt} g_{aa}^{\overbar{a}} \hspace{15pt},\hspace{15pt} \underbrace{~ g_{\overbar a \overbar a}^{\overbar a} \hspace{15pt},\hspace{15pt} g_{a\overbar a}^a \hspace{15pt},\hspace{15pt} g_{\overbar aa}^a~ }_{\text{identical}} \hspace{15pt},\hspace{15pt} g_{\overbar a\overbar a}^a\hspace{2pt}.
\end{equation}
The braces imply that certain collections of coefficients are actually identical, being fixed by the symmetry relations of Equation \eqref{eqn: symmetry relations three point} . This equation, furthermore, implies that the final two sets of coefficients all vanish automatically or fail to satisfy $p \geq 0$ --- in either case, they can be ignored for the study of \textit{chiral} soft algebras. Finally, $g_{aa}^{\overbar a}$ vanishes due to the symmetry conditions unless $\mathcal{N} \equiv 0$ mod $4.$ When $\mathcal{N} = 0$ it vanishes because it is proportional to the three positive-helicity graviton scattering amplitude, which is zero in flat space. For $\mathcal{N} = 8$, $g_{aa}^{\overbar{a}}$ is related to $g_{aa}^a$ because the supermultiplets are CPT self-conjugate, so we may ignore this case. Thus, we only have a possibly non-trivial structure when $\mathcal{N} = 4.$ We shall also ignore it for simplicity: $g_{aa}^{\overbar{a}} = 0$ henceforth. We normalize the generators of the chiral soft algebra so that $g_{aa}^a = 1/2$ without loss of generality. 

\subsection{Supersymmetric Extension of $\mathcal{L}(\myw^\wedge_{1+\infty})$}
Now, that we have classified the possibly non-trivial OPEs in the theory, we may write down the chiral soft algebra for supergravity theories using Equation \eqref{eqn: chiral algebra}
\begin{equation}
    \begin{split}
        \big[(w^{I_1 \cdots I_A})^{q_1}_{\overbar{m}_1,n_1},(w^{J_1 \cdots J_B})^{q_2}_{\overbar{m}_2,n_2}\big] &= \hspace{-1pt}\big((q_2 - 1)\overbar{m}_1 - (q_1 - 1)\overbar{m}_2\big)\hspace{1pt} (w^{I_1 \cdots I_A J_1 \cdots J_B})^{q_1+q_2-2}_{\overbar{m}_1+\overbar{m}_2,n_1+n_2} \\
        \big[(w^{I_1 \cdots I_A})^{q_1}_{\overbar{m}_1,n_1},(\overbar{w}^{J_1 \cdots J_B})^{q_2}_{\overbar{m}_2,n_2}\big] &= \hspace{-1pt}\big((q_2 - 1)\overbar{m}_1 - (q_1 - 1)\overbar{m}_2\big)\hspace{1pt} (\overbar{w}^{I_1 \cdots I_A J_1 \cdots J_B})^{q_1+q_2-2}_{\overbar{m}_1+\overbar{m}_2,n_1+n_2}\hspace{2pt},
    \end{split}
\end{equation}
where we have replaced the notation $\mathcal{R}^{I_1 \cdots I_A}_{q,\overbar{m},n}$ with $(w^{I_1 \cdots I_A})^q_{\overbar{m},n}$ to emphasize that the first line is a supersymmetric generalization of $\mathcal{L}(\myw_{1+\infty}^\wedge)$, i.e. the loop algebra of the wedge subalgebra of $\myw_{1+\infty}$ \cite{Bakas:1989xu, Hoppe:1988gk}.\footnote{Strictly speaking, the proper definition of $\mathcal{L}(\myw_{1+\infty}^\wedge)$ involves only $q = 1,2,3,...$ rather than $q = 1,\frac{3}{2},2,...$ The algebra presented here is actually $\mathcal{L}(\frak{ham}(\mathbb{C}^2))$, though it is commonplace to use the two names interchangeably.} Indeed, the generators $w^q_{\overbar{m},n}$ (i.e. no superscripts) correspond to positive helicity gravitons in the bulk; these are the usual generators of  $\mathcal{L}(\myw_{1+\infty}^\wedge)$, and they obey the standard $\mathcal{L}(\myw_{1+\infty}^\wedge)$ algebra \cite{Guevara:2021abz, Strominger:2021mtt, Himwich:2023njb}. Moreover, they form a closed subalgebra of this supersymmetric extension. Commutators of generators with superscripts obey the same general form --- the superscripts just get carried along for the ride over to the right-hand side. As such, the super $\mathcal{L}(\myw^\wedge_{1+\infty})$ algebra with $\mathcal{N}$ supersymmetries is a subalgebra of the one with $\mathcal{N} + 1$ supersymmetries as one would expect.

The second line of this algebra is also familiar. Recall that the barred generators are related to the opposite-helicity anti-particles of the states in the positive-helicity graviton multiplet. Again, there is a commutator involving a positive-helicity graviton generator and a negative-helicity one. The result is proportional to another negative-helicity graviton generator and exactly matches the familiar result from the non-supersymmetric setting.

The above algebra is almost completely universal. The one caveat is that it can be deformed when $\mathcal{N} = 4$ due to the presence of $g_{aa}^{\overbar{a}}.$ Ignoring this, we have completely determined how the $\mathcal{L}(\myw_{1+\infty}^\wedge)$ algebra gets extended in all supersymmetric theories.

\subsection{Hamiltonian Vector Fields on $\mathbb{C}^{2|\mathcal{N}}$}

Thus far, we have shown that the chiral symmetry algebras of supergravity are related to the $\myw_{1+\infty}^{\wedge}$ algebra and its supersymmetric extension which we reproduce below for convenience
\begin{align}
    \myw_{1+\infty}^{\wedge} &: \hspace{71pt} \big[w^{q_1}_{\overbar{m}_1},w^{q_2}_{\overbar{m}_2}\big] = \big((q_2 - 1)\overbar{m}_1 - (q_1 - 1)\overbar{m}_2\big)~ w^{q_1+q_2-2}_{\overbar{m}_1+\overbar{m}_2} \label{eqn: w 1+infinity}\\[.5em]
    \text{Super }\myw_{1+\infty}^{\wedge} &:  \hspace{1pt}\big[(w^{I_1 \cdots I_A})^{q_1}_{\overbar{m}_1},(w^{J_1 \cdots J_B})^{q_2}_{\overbar{m}_2}\big] = \hspace{-1pt}\big((q_2 - 1)\overbar{m}_1 - (q_1 - 1)\overbar{m}_2\big)\hspace{1pt} (w^{I_1 \cdots I_A J_1 \cdots J_B})^{q_1+q_2-2}_{\overbar{m}_1+\overbar{m}_2}\label{eqn: super w 1+infinity}
\end{align}
In both cases, the subscripts range from $q_i = 1, \frac{3}{2},2,...$ and $1-q_i\leq \overbar{m}_i \leq q_i-1$. We see that the $\myw_{1+\infty}^{\wedge}$ algebra is just the $\mathcal{N} = 0$ super $\myw_{1+\infty}^\wedge$ algebra. Note that the full chiral symmetry algebra in celestial holography involves taking the \textit{loop algebra} and adding additional commutators between $w$ and $\overbar{w}$ --- these algebras only capture the commutators between $w$ and $w$.

It is known that Equation \eqref{eqn: w 1+infinity} is just $\frak{ham}(\mathbb{C}^2)$ --- the algebra of Hamiltonian vector fields on $\mathbb{C}^2$. This algebra was originally appreciated in the context of a twistor theoretic description of self-dual gravity\cite{Penrose:1976js,Park:1989fz,Boyer:1985aj}. Its connection to celestial holography has been examined in \cite{Adamo:2021lrv}. We shall review this construction briefly. $\mathbb{C}^2$ is a symplectic manifold with holomorphic coordinates $(\mu^{\overbar 0},\mu^{\overbar 1})$, and its symplectic structure induces a Poison bivector
\begin{equation}
    \Pi_{\text{bosonic}} = \frac{\partial}{\partial \mu^{\overbar \alpha}} \wedge \frac{\partial}{\partial \mu_{\overbar{\alpha}}} \hspace{2pt},
    \label{eqn: Poisson bivector}
\end{equation}
and a corresponding Poisson bracket
\begin{equation}
    \big\{f,g\big\} = \sumop_{ij} \hspace{2pt} \Pi^{ij}_{\text{bosonic}} \hspace{2pt} \partial_i f \hspace{2pt} \partial_j g = \frac{\partial f}{\partial \mu^{\overbar 0}} \frac{\partial g}{\partial \mu^{\overbar 1}} - \frac{\partial f}{\partial \mu^{\overbar 1}} \frac{\partial g}{\partial  \mu^{\overbar 0}}\hspace{2pt}.
\end{equation}

For each smooth function $g$ on $\mathbb{C}^2$, one associates a Hamiltonian vector field $X_g$ which acts on the smooth function $f$ via $X_g(f) = \{f,g\}.$ These Hamiltonian vector fields form the Lie algebra $\frak{ham}(\mathbb{C}^2)$ defined by $[X_{f},X_g]= X_{\{f,g\}}$. To determine an explicit expression for the algebra, we must choose an appropriate basis of smooth functions on $\mathbb{C}^2.$ We select
\begin{equation}
    w^q_{\overbar{m}} = \frac{1}{2}(\mu^{\overbar 0})^{q-1+\overbar{m}}\hspace{2pt} (\mu^{\overbar 1})^{q-1-\overbar{m}}\hspace{2pt}.
\end{equation}
These functions must obey $q = 1, \frac{3}{2}, 2,...$ and $1-q \leq \overbar{m} \leq q-1$ to be non-singular at the origin. Because they have the following Poisson brackets among themselves
\begin{equation}
    \big\{w^{q_1}_{\overbar{m}_1},w^{q_2}_{\overbar{m}_2}\big\} = \big((q_2 - 1)\overbar{m}_1 - (q_1 - 1)\overbar{m}_2\big)\hspace{2pt} w^{q_1 + q_2 -2}_{\overbar{m}_1+\overbar{m}_2}\hspace{2pt}.
\end{equation}
The algebra of Hamiltonian vector fields on $\mathbb{C}^2$ with respect to the Poisson structure $\Pi_{\text{bosonic}}$ is precisely the $\myw_{1+\infty}^\wedge$ algebra with commutation relations given in Equation \eqref{eqn: w 1+infinity}.

In this spirit, we now wish to produce the super $\myw_{1+\infty}^\wedge$ algebra given in Equation \eqref{eqn: super w 1+infinity} with an appropriate generalization of this construction --- namely, as the Lie algebra of Hamiltonian vector fields on the supermanifold $\mathbb{C}^{2|\mathcal{N}}$. See \cite{Mason:2007ct, Wolf:2007tx} for a discussion of this algebra in the context of twistor theory. $\mathbb{C}^{2|\mathcal{N}}$ has local coordinates $(\mu^{\overbar{0}},\mu^{\overbar{1}},\theta^1,...,\theta^\mathcal{N})$. One should be careful not to confuse the fermionic coordinates $\theta^{I}$ with the fermionic coordiantes $\eta^I$ -- the former are coordinates on the supermanifold $\mathbb{C}^{2|\mathcal{N}}$ and have a geometric interpretation, while the latter are just a convenient tool to package a family generators into a single object.

A basis of functions on $\mathbb{C}^{2|\mathcal{N}}$ takes the form
\begin{equation}
    (w^{I_1 \cdots I_A})^q_{\overbar{m}} = \frac{1}{2^{1-A}}(\mu^{\overbar 0})^{q-1+\overbar{m}}\hspace{2pt} (\mu^{\overbar 1})^{q-1-\overbar{m}} \hspace{2pt} \theta^{I_1} \cdots \theta^{I_A}\hspace{2pt}.
\end{equation}
To compute $\frak{ham}(\mathbb{C}^{2|\mathcal{N}})$, we must first specify how the Poisson bivector generalizes to the supermanifold setting. If we demand that $\frak{ham}(\mathbb{C}^{2|0})$ reduces to $\frak{ham}(\mathbb{C}^2)$, we must have
\begin{equation}
    \Pi = \Pi_{\text{bosonic}} + \Pi_{\text{fermionic}}\hspace{2pt}.
\end{equation}

There is no a-priori obvious choice for $\Pi_{\text{fermionic}}$, but two natural candidates are
\begin{equation}
    \Pi_{\text{fermionic}} = 0 \hspace{40pt}\text{or}\hspace{40pt} \Pi'_{\text{fermionic}} = \delta^{IJ} \hspace{2pt} \frac{\partial}{\partial \theta^I} \wedge \frac{\partial}{\partial\theta^J}\hspace{2pt}.
    \label{eqn: poisson bivector fermionic options}
\end{equation}
It turns out that the \textit{first choice} (i.e. $\Pi_{\text{fermionic}} = 0$) is the Poisson structure relevant to celestial holography in spite of the fact that the Poisson bi-vector is degenerate, so it has no interpretation as a symplectic form.
The algebra of Hamiltonian vector fields with respect to this Poisson bracket is precisely Equation \eqref{eqn: super w 1+infinity}. 

Nevertheless, the physics associated to $\Pi_{\text{fermionic}}'$ is also interesting and worth briefly mentioning. For concreteness, let's compare the resulting algebras for the simple case of $\mathcal{N} = 1$ supersymmetry
\begin{equation}
    \begin{split}
        \Pi_{\text{fermionic}} : \hspace{20pt} \big[w^{q_1}_{m_1},w^{q_2}_{m_2}\big] &= \big((q_2 - 1)m_1 - (q_1 - 1)m_2\big) ~ w^{q_1+q_2 -2}_{m_1+m_2}\\[.2em]
        \big[w^{q_1}_{m_1},y^{q_2}_{m_2}\big] &= \big((q_2 - 1)m_1 - (q_1 - 1)m_2\big) ~ y^{q_1+q_2-2}_{m_1+m_2}\\[.2em]
        \big[y^{q_1}_{m_1},y^{q_2}_{m_2}\big] &= 0 \\[.5em]
        \Pi_{\text{fermionic}}' : \hspace{20pt} \big[w^{q_1}_{m_1},w^{q_2}_{m_2}\big] &= \big((q_2 - 1)m_1 - (q_1 - 1)m_2\big) ~ w^{q_1 + q_2 -2}_{m_1 + m_2}\\[.2em]
        \big[w^{q_1}_{m_1},y^{q_2}_{m_2}\big] &= \big((q_2 - 1)m_1 - (q_1 - 1)m_2\big) ~ y^{q_1 + q_2 -2}_{m_1 + m_2}\\[.2em]
        \big[y^{q_1}_{m_1},y^{q_2}_{m_2}\big] &= 2 \hspace{2pt} w^{q_1 + q_2 - 1}_{m_1 + m_2}\hspace{2pt}.
    \end{split}
\end{equation}
where we have defined $y^q_{m}$ as the Hamiltonian vector field generated by $(w^1)^q_{m}$. 

The only difference between these algebras is in the $[y^{q_1}_{m_1},y^{q_2}_{m_2}]$ commutator. Nevertheless, this difference is extremely meaningful. For example, the $\myw_{1+\infty}^\wedge$ algebra is known to have a $\text{SL}_2(\mathbb{R})$ subalgebra generated by $L_m = w^2_{m}$ for $m = -1,0,1$. These generators are related to subleading soft graviton operators, and their commutators form the algebra of global conformal symmetries on the celestial sphere. One might expect that the supersymmetric generalization of  $\myw_{1+\infty}^\wedge$ will, therefore, feature a subalgebra encoding global superconformal symmetries on the celestial sphere; however, the algebra induced by $\Pi_{\text{fermionic}}$ (i.e. the one relevant to celestial holography) has no such subalgebra. This is because the $y^q_{m}$ generators always commute. On the other hand, the Lie algebra induced by $\Pi_{\text{fermionic}}'$ does have a global superconformal subalgebra generated by $L_m = w^2_m$ for $m = -1,0,1$ and $G_m = y^{3/2}_{m}$ for for $m = \pm \frac{1}{2}$. For this reason, the second Lie algebra is the one which tends to appear with discussing supersymmetric extensions of $\myw_{1+\infty}^\wedge$ in ordinary $2d$ SCFT \cite{Inami:1988xy,Bergshoeff:1990yd,Buffon:1996dv}. 

One should be extremely careful about this distinction when trying to examine the relationship between super $\myw_{1+\infty}^\wedge$ algebras appearing in standard CFTs and the structure of spacetime symmetries in celestial holography (see \cite{Ahn:2021erj, Ahn:2022oor, Ahn:2024kpv}). While the super $\myw_{1+\infty}^\wedge$ algebra induced by $\Pi_{\text{fermionic}}'$ manifests $2d$ worldsheet symmetry in a conventional way, it does not manifest $4d$ bulk supersymmetry. We have already been alerted to the difference between $4d$ bulk spacetime supersymmetry and $2d$ boundary supersymmetry on the celestial sphere in the context of the $\frak{sbms}_{4|\mathcal{N}}$ algebra presented in Section \ref{sec: sbms algebra}. This example further illustrates this crucial distinction which we shall return to in Section \ref{sec: conclusion}.

\subsection{Supersymmetric Extension of $\mathcal{L}(\myw_{1+\infty}^\wedge)$ with a Cosmological Constant}

Recently, the $\mathcal{L}(\myw_{1+\infty}^\wedge)$ symmetries have been generalized to gravity in a curved background with cosmological constant $\Lambda.$ This was first noticed in \cite{Taylor:2023ajd} where the holomorphic OPE for tree-level graviton amplitudes was studied to leading order in $\Lambda$ corrections to the effective action. The authors deduced the following $\Lambda$-deformed algebra
\begin{equation}
    \begin{split}
        \big[w^{q_1}_{\overbar{m}_1,n_1},w^{q_2}_{\overbar{m}_2,n_2}\big]_\Lambda &= \big((q_2-1)m_1 - (q_1-1)m_2\big) w^{q_1 + q_2 -2}_{\overbar{m}_1 + \overbar{m}_2,n_1+n_2} \\[.3em]
        &\hspace{60pt}- \Lambda\big((q_2-2)n_1 - (q_1-2)n_2\big) w^{q_1 + q_2 -1}_{\overbar{m}_1 + \overbar{m}_2,n_1+n_2}\hspace{2pt},
        \label{eqn: Lambda deformed w}
    \end{split}
\end{equation}
The twistorial origins of this algebra were elucidated in \cite{Bittleston:2024rqe}, where it was shown that Equation \eqref{eqn: Lambda deformed w} was the algebra of Hamiltonian vector fields on $\mathbb{C}^2 \times \mathbb{C}^*$ with respect to the following $\Lambda$-deformed Poisson bivector
\begin{equation}
    \Pi_{\text{bosonic}}^{(\Lambda)} = \frac{\partial}{\partial \mu^{\overbar \alpha}} \wedge \frac{\partial}{\partial \mu_{\overbar{\alpha}}} + \Lambda \hspace{2pt} \frac{\partial}{\partial \lambda_\alpha} \wedge \frac{\partial}{\partial \lambda^\alpha}\hspace{2pt},
    \label{eqn: Lambda Poisson bivector}
\end{equation}
and we have chosen the following basis of holomorphic functions on $(\mu^{\overbar{\alpha}}/\lambda_1,\lambda_0/\lambda_1) \in \mathbb{C}^2 \times \mathbb{C}^*$
\begin{equation}
    w^q_{\overbar{m},n} = \frac{(\mu^{\overbar 0})^{q-1+\overbar{m}}\hspace{2pt} (\mu^{\overbar 1})^{q-1-\overbar{m}}}{2 \hspace{2pt} \lambda_0^{q-n-2} \lambda_1^{q+n-2}}\hspace{2pt}.
\end{equation}
This algebra is known as $\frak{ham}_{\Lambda}(\mathbb{C}^2 \times \mathbb{C}^*).$ When $\Lambda = 0$, one can verify that the Poisson bivector given in Equation \eqref{eqn: Lambda Poisson bivector} reduces to \eqref{eqn: Poisson bivector}, and $\frak{ham}_{\Lambda}(\mathbb{C}^2 \times \mathbb{C}^*)|_{\Lambda = 0}$ becomes the usual $\mathcal{L}(\frak{ham}(\mathbb{C}^2))$.

The supersymmetric generalization of this algebra has a similar twistorial description. Now, one considers $\frak{ham}_{\Lambda}(\mathbb{C}^{2|\mathcal{N}} \times \mathbb{C}^*)$. The Poisson bivector again becomes $\Pi_{\text{bosonic}}^{(\Lambda)} + \Pi_{\text{fermionic}}^{(\Lambda)}$ where
\begin{equation}
    \Pi_{\text{fermionic}}^{(\Lambda)} = \sqrt{\Lambda} \hspace{2pt} \delta^{IJ} \hspace{2pt} \frac{\partial}{\partial \theta^I} \wedge \frac{\partial}{\partial\theta^J}\hspace{2pt},
\end{equation}
as can be determined from the infinity twistor on the supersymmetric twistor space $\mathbb{CP}^{3|\mathcal{N}}$ \cite{Mason:2007ct, Wolf:2007tx,Adamo:2015ina}. We observe that the second Poisson bracket in Equation \eqref{eqn: poisson bivector fermionic options}, i.e. the one germane to $2d$ SCFT, becomes relevant in the presence of a cosmological constant. If we choose the following basis of holomorphic functions on $\mathbb{C}^{2|\mathcal{N}} \times \mathbb{C}^*$
\begin{equation}
    (w^{I_1 \cdots I_A})^q_{\overbar{m},n} = \frac{(\mu^{\overbar 0})^{q-1+\overbar{m}}\hspace{2pt} (\mu^{\overbar 1})^{q-1-\overbar{m}}}{2^{1-A} \hspace{2pt} \lambda_0^{q-n-2+A/2} \lambda_1^{q+n-2+A/2}} \hspace{2pt}\theta^{I_1} \cdots \theta^{I_A}\hspace{2pt},
\end{equation}
we observe that $\frak{ham}_{\Lambda}(\mathbb{C}^{2|\mathcal{N}} \times \mathbb{C}^*)$ is characterized by the commutation relations
\begin{equation}
    \begin{split}
        \big[(w^{I_1 \cdots I_A}&)^{q_1}_{\overbar{m}_1,n_1},(w^{J_1 \cdots J_B})^{q_2}_{\overbar{m}_2,n_2}\big]_\Lambda \\[.5em]
        &= \big((q_2-1)m_1 - (q_1-1)m_2\big) (w^{I_1 \cdots I_A J_1 \cdots J_B})^{q_1 + q_2 -2}_{\overbar{m}_1 + \overbar{m}_2,n_1+n_2} \\[.7em]
        &\hspace{54pt}- \Lambda\big((q_2-2+B/2)n_1 - (q_1-2+A/2)n_2\big) (w^{I_1 \cdots I_A J_1 \cdots J_B})^{q_1 + q_2 -1}_{\overbar{m}_1 + \overbar{m}_2,n_1+n_2} \\
        &\hspace{54pt} + 2 \sqrt{\Lambda} \hspace{2pt} \sumop_{a=1}^A \sumop_{b=1}^B (-1)^{A+a+b+1} \hspace{2pt} \delta^{I_a J_b} (w^{I_1 \cdots \widehat{I}_a \cdots I_A J_1 \cdots \widehat{J}_b \cdots J_B})^{q_1 + q_2 -1}_{\overbar{m}_1 + \overbar{m}_2,n_1+n_2} \hspace{2pt},
    \end{split}
\end{equation}
where $\widehat{I}_a,\widehat{J}_b$ appearing on the last line denote a superscript which has been removed.

Packaging the generators of $\frak{ham}_{\Lambda}(\mathbb{C}^{2|\mathcal{N}} \times \mathbb{C}^*)$ into generators $\mathds{W}^{q}_{\overbar{m},n}(\eta)$ on superspace via
\begin{equation}
    \mathds{W}^q_{\overbar{m},\hspace{.5pt}n}(\eta) = \sumop_{A\hspace{.5pt}=\hspace{.5pt}0}^\mathcal{N} \frac{1}{A!} \hspace{2pt} \eta_{I_A} \cdots \eta_{I_1} \hspace{1pt} (w^{I_1 \cdots I_A})^q_{\overbar{m},\hspace{.5pt} n} \hspace{2pt},
\end{equation}
we find the following supersymmetric extension of the $\Lambda$-deformed algebra
\begin{align}
        \big[&\mathds{W}^{q_1}_{\overbar{m}_1,\hspace{.5pt}n_1}(\eta_1),\mathds{W}^{q_2}_{\overbar{m}_2,\hspace{.5pt}n_2}(\eta_2)\big]_\Lambda \nonumber \\[1em]
        &\hspace{5pt}= \big((q_2-1)m_1 - (q_1-1)m_2\big) \mathds{W}^{q_1 + q_2 -2}_{\overbar{m}_1 + \overbar{m}_2,n_1+n_2}(\eta_1 + \eta_2) + 2 \sqrt{\Lambda} \hspace{2pt} \eta_{2,I}
        \hspace{2pt} \eta_{1,I} \mathds{W}^{q_1 + q_2 -1}_{\overbar{m}_1 + \overbar{m}_2,n_1+n_2}(\eta_1 + \eta_2) \nonumber \\[.4em]
        &\hspace{25pt}- \Lambda\bigg(\Big(q_2-2+ \frac{\eta_{1,I}}{2}\frac{\partial}{\partial \eta_{1,I}}\Big)n_1 - \Big(q_1-2+ \frac{\eta_{2,I}}{2}\frac{\partial}{\partial \eta_{2,I}}\Big)n_2\bigg) \mathds{W}^{q_1 + q_2 -1}_{\overbar{m}_1 + \overbar{m}_2,n_1+n_2}(\eta_1 + \eta_2)\hspace{2pt}.
\end{align}
Though we have derived this algebra from purely twistor theoretic arguments, at $\Lambda = 0$ it precisely reproduces the chiral soft algebra arising from holomorphic OPEs in flat space supergravity theories.

\section{Discussion: $4d$ Spacetime Supersymmetry and $2d$ Boundary Supersymmetry}
\label{sec: conclusion}

Supersymmetry imposes strong constraints on both UV and IR physics, making it a crucial tool for understanding the broader structure of flat-space holography. In this article, we have aimed to systematically explore how bulk supersymmetry is manifested in celestial holography. To this end, we report on \textit{universal results} which hold for any theory --- this is achieved by studying amplitudes in on-shell superspace rather than examining $4d$ spacetime Lagrangians which manifest off-shell supersymmetry. It is our view that $4d$ off-shell supersymmetry is a red herring which has little to do with celestial holography; such Lagrangians are often complicated, and off shell superfields are not directly related to supercelestial families on the boundary making this approach more challenging both conceptually and computationally.

A main accomplishment was demonstrating that the celestial sphere itself gets extended to a supermanifold, the \textit{celestial supersphere}, $\mathbb{C}^{1|\mathcal{N}}$. Various familiar structures in celestial holography have a natural description on this supermanifold. For example, the conformal primary operators associated to an entire bulk supermultiplet live in a single supercelestial family which may be assembled into the superoperator $\mathbb{O}_{a,\hspace{.5pt}\Delta}(z,\overbar{z}|\eta)$. Holomorphic OPEs between such superoperators take an elegant form and can be used to construct chiral soft algebras in terms of generators living on superspace, $\mathds{R}_{\hspace{.5pt}a,\hspace{.5pt}q,\hspace{.5pt}\overbar{m},\hspace{.5pt}n}(\eta)$.

Along the way, we have seen that conformally soft theorems for entire supermultiplets also organize themselves into a superoperator, $\mathbb{S}_{a,\hspace{.5pt}k}(z,\overbar{z}|\eta)$. This superoperator is tightly constrained by supersymmetry, so much so that knowing the conformally soft theorem for one particle in the supermultiplet allows one to read off conformally soft theorems for all others. We give examples of how this is done in gauge theory and gravity. This structure illuminates the fact that the $\frak{bms}_{4}$ algebra gets extended to a $\frak{sbms}_{4|\mathcal{N}}$ algebra in supersymmetric theories; miraculously, this extension is completely universal. Similarly, we have described the supergravity extension of the $\mathcal{L}(\myw_{1+\infty}^\wedge)$ algebra of Einstein gravity and interpreted it as an algebra of Hamiltonian vector fields on $\mathbb{C}^{2|\mathcal{N}}$, consistent with the expectation from twistor theory. We use additional tools from twistor theory to discuss the deformation of this algebra by a non-vanishing cosmological constant.

Time and again, these symmetries defy expectations from ordinary $2d$ SCFTs. For example, the extended $\frak{bms}_{4}$ algebra has a Virasoro subalgebra, but the $\frak{sbms}_{4|\mathcal{N}}$ algebra does not have a super-Virasoro subalgebra. Similarly, the super $\myw_{1+\infty}^\wedge$ algebras appearing in $2d$ SCFT constructions have a superconformal subalgebra, while those appearing in celestial holography do not. We illustrate the difference between these two supersymmetric extensions of $\myw_{1+\infty}^\wedge$ for the simple example of $\mathcal{N} = 1.$ 

In this way, we conclude that $4d$ bulk supersymmetry does not lead to a standard $2d$ boundary supersymmetry on the celestial sphere. In AdS holography, bulk supersymmetry implies boundary supersymmetry and vice versa. In celestial holography, bulk supersymmetry is realized in the spirit of the Green-Schwarz superstring which manifests target space supersymmetry at the cost of worldsheet supersymmetry.

At the heart of this discrepancy is the fact that supercharges square to translation generators in spacetime; however bulk translation symmetry has nothing to do with boundary conformal symmetry in celestial holography. Said another way, the supercharges act as weight-shifting operators on the celestial sphere rather than conformal generators
\begin{equation}
    \big[\overbar{Q}^I_{\overbar{\alpha}},Q_{J,\alpha}\big] = -\delta^I_J \hspace{1pt} z^\alpha \hspace{1pt}\overbar{z}^\alpha \hspace{1pt}e^{-\partial_\Delta}
\end{equation}
In AdS holography, there are no bulk translation generators, and supercharges square to AdS isometries. The collection of such isometries are isomorphic to the boundary conformal algebra, which is the reason AdS duals exhibit a manifest superconformal symmetry. This suggests that the crucial difference has to do with taking the flat limit wherein AdS isometry generators become flat space translations via an \.In\"on\"u-Wigner contraction \cite{VanProeyen:1999ni}. 

Nevertheless, just as the superstring has both Green-Schwarz and RNS descriptions, we might hope that explicit examples of celestial CFTs have a description in which $2d$ supersymmetry becomes manifest. For example, MHV amplitudes in celestial holography have recently been related to Liouville theory \cite{Taylor:2023bzj, Stieberger:2022zyk, Stieberger:2023fju, Melton:2024akx, mol2024ads3dualsupersymmetricmhv, Mol:2024etg, Mol:2024onu, Mol:2024qct}. An interesting extension of these papers links MHV superamplitudes to correlation functions in Liouville theory with $2d$ $(1,0)$ supersymmetry \cite{Taylor:2023bzj}! Understanding to what extent this phenomenon generalizes is a priority.

\section*{Acknowledgements}

I would like to thank Tim Adamo, Changhyun Ahn, Adam Ball, Wei Bu, Kevin Costello, Erin Crawley, Simon Heuveline, Matt Heydeman, Elizabeth Himwich, Walker Melton, Noah Miller, Julio Parra-Martinez, Andrew Strominger, Tomasz Taylor, Chiara Toldo, and Xi Yin for many useful discussions and helpful comments on the draft. I would like to especially thank Atul Sharma for patiently guiding me through the twistor theory literature and encouraging me to write Section 6.3. This work is supported by NSF GRFP grant DGE1745303.


\appendix

\section{Conventions}
\label{appendix: conventions}
\addtocontents{toc}{\protect\setcounter{tocdepth}{1}}  

\subsection{Spinor-Helicity}

In this article, we adopt a mostly-plus metric convention and choose our spinor conventions to ensure that the unbarred supersymmetry generator, $Q^I_\alpha$, acts holomorphically (vs anti-holomorphically) on the celestial sphere. We define Pauli-matrices $\sigma^\mu_{\alpha \overbar{\alpha}} = (\mathds{1}_{\alpha \overbar{\alpha}},-\sigma_{\alpha \overbar{\alpha}}^i)$ and $\overbar{\sigma}^{\mu, \overbar\alpha \alpha} = (\mathds{1}^{\overbar\alpha \alpha},\sigma^{i,\overbar\alpha \alpha})$ which are combined to give the following generators for the spinor representation of the Lorentz algebra $\sigma^{\mu \nu} = \frac{i}{2} \sigma^{[\mu} \overbar{\sigma}^{\nu]}$ and $\overbar{\sigma}^{\mu \nu} = -\frac{i}{2} \overbar{\sigma}^{[\mu} \sigma^{\nu]}$. 

While these spinor conventions are opposite those of \cite{Srednicki:2007qs, Elvang:2013cua} (which swap barred and unbarred objects), several other papers on celestial holography also break these conventions, and we agree that they are worth violating for the sake of making holomorphicity on the celestial sphere manifest.

We parameterize a null four-momentum $p^\mu$ as
\begin{equation}
    p^\mu = \hspace{2pt} \frac{\epsilon \hspace{2pt} \omega}{2}\hspace{2pt} \big(1+z \overbar z,z + \overbar z, - i(z - \overbar z), 1-z \overbar z\big)\hspace{2pt},
\end{equation}
where $\omega \in \mathbb{R}_+$ is an energy scale, $(z,\overbar z)$ labels a point on the celestial sphere that the momentum is pointing towards, and $\epsilon = +1,-1$ respectively label outgoing and incoming null particles. In this parametrization, we may write the spinor brackets as
\begin{equation}
    p_\mu \sigma^{\mu}_{\alpha \overbar{\alpha}} = -| p\rangle_{\alpha} [p|_{\overbar{\alpha}}~, \hspace{60pt}\text{where}\hspace{30pt} |p\rangle_\alpha = \epsilon \hspace{1pt} \sqrt{\omega} \hspace{2pt} z^\alpha \hspace{10pt},\hspace{10pt} [p|_{\overbar \alpha} = \sqrt{\omega} \hspace{2pt} \overbar{z}^{\overbar\alpha}\hspace{2pt}.
\end{equation}
For real momenta, these spinor brackets satisfy the Hermiticity condition $|p\rangle_\alpha^\dagger = \epsilon \hspace{2pt} [p|_{\overbar \alpha}$. Spinor indices are always raised and lowered as $\langle p|^{\alpha} = \varepsilon^{\alpha \beta} |p\rangle_\beta$ and $|p]^{\overbar{\alpha}} = \varepsilon{}^{\overbar{\alpha} \overbar{\beta}} [p|_{\overbar \beta}$ where $\varepsilon^{01} = +1 = \varepsilon_{10}$. 

\subsection{CPT Conjugates and Crossing Symmetry}

A supermultiplet is called \textit{CPT self-conjugate} if for each particle in the supermultiplet, its opposite helicity anti-particle is also in the supermultiplet. $\mathcal{N} = 4$ super Yang-Mills and $\mathcal{N} = 8$ supergravity are two such examples; however, generally supermultiplets are not CPT self-conjugate.

To mollify this, one must add the CPT conjugates by hand. We use the label $\widetilde{a}_{I_1 \cdots I_A \hspace{1pt}}(p)$ to denote the anti-particle of $a^{I_1 \cdots I_A}(p)$. These two operators are related by \textit{crossing symmetry}\footnote{Crossing symmetry is actually a very subtle thing involving careful analyticity assumptions and choices for contours of analytic continuation (see \cite{Mizera:2023tfe} for a nice overview). This equation is doing something less sophisticated. We have made the standard assumption that scattering amplitudes may be written in an all-out formalism with some momenta having $p^0 < 0$. From this assumption, \eqref{eqn: crossing} should be viewed a definition for the operator $\widetilde{a}^\dagger_{I_1 \cdots I_A}(-p)$. There is no physical content to this statement beyond introducing some helpful notation.}
\begin{equation}
    a^{I_1 \cdots I_A}(p) \hspace{3pt} \xleftrightarrow{~ \text{crossing}~ } \hspace{3pt} \widetilde{a}^\dagger_{I_1 \cdots I_A}(-p)\hspace{2pt}.
    \label{eqn: crossing}
\end{equation}
Often, when discussing crossing symmetry in QFT, crossing phases appear for fermions. These are due to the definition of single-particle states from fundamental quantum fields via the LSZ prescription. In the on-shell language that we adopt, such crossing phases are absorbed in the definition of $\widetilde{a}_{I_1 \cdots I_A}^\dagger(-p).$ Equation \eqref{eqn: crossing} implies the commutation relations
\begin{equation}
    \big[\widetilde{a}_{I_1 \cdots I_A}(p),Q_{J,\alpha}\big] = -\epsilon \hspace{2pt} |p\rangle_{\alpha} \hspace{2pt} \widetilde{a}_{J I_1 \cdots I_A}(p) \hspace{25pt},\hspace{25pt} \big[\widetilde{a}_{I_1 \cdots I_A}(p),\overbar{Q}^{J}_{\overbar \alpha}\big] = -\epsilon \hspace{2pt} [p|_{\overbar \alpha} \hspace{2pt} \delta^{J}_{[I_1} \widetilde{a}_{I_2 \cdots I_A]}(p)\hspace{2pt},
    \label{eqn: supermultiplet crossed particles}
\end{equation}
where $\epsilon = \text{sgn}(p^0)$, which is $+1$ for outgoing particles and $-1$ for incoming ones.

Contrasting Equation \eqref{eqn: supermultiplet crossed particles} with Equation \eqref{eqn: SUSY multiplet}, we conclude that the $\widetilde{a}_{I_1 \cdots I_A}(p)$ operators do not form a highest-weight representation of the supersymmetry algebra. In fact, because $\widetilde{a}(p)$ is annihilated by $\overbar{Q}^J_{\bar{\alpha}}$ but not $Q_{J,\alpha}$, they form a \textit{lowest-weight representation}. Fortunately, these operators may always be repackaged into a highest-weight representation generated by the highest-weight state $\overbar{a}(p) = \widetilde{a}_{1 \cdots \mathcal{N}}(p)$. 
\begin{equation}
    \overbar{a}^{I_1 \cdots I_A}(p) = \epsilon^{|a| + A} \hspace{2pt} \frac{(-1)^{A(|\overbar{a}|+1)}}{(\mathcal{N}-A)!} \hspace{2pt} \varepsilon^{I_1 \cdots I_A J_{A+1} \cdots J_{\mathcal{N}}} \hspace{2pt} \widetilde{a}_{J_{A+1} \cdots J_\mathcal{N}}(p)\hspace{2pt},
    \label{eqn: def CPT conjugate multiplet}
\end{equation}
where $\varepsilon^{I_1 \cdots I_\mathcal{N}}$ is the totally anti-symmetric tensor on $\mathcal{N}$ indices. For the most part, the multiplicative prefactors are fixed by demanding that the operators $\overbar{a}^{I_1 \cdots I_A}(p)$ satisfy Equation \eqref{eqn: SUSY multiplet} given the aforementioned commutation relations for $\widetilde{a}_{J_{A+1} \cdots J_{\mathcal{N}}}(p).$ However, there is some freedom to add extra multiplicative constants which we choose to ensure that Equation \eqref{eqn: double CPT conjugate} takes a simple form.


Finally, we note that we can take the CPT conjugate a second time forming a highest-weight supermultiplet generated by $\overbar{\overbar{a}}(p)$. It is clear that this should be related to the original supermultiplet generated by $a(p).$ The precise normalization can be obtained either with Equations \eqref{eqn: crossing} and \eqref{eqn: def CPT conjugate multiplet}
\begin{equation}
    \overbar{\overbar{a}}(p) = (-1)^{(\mathcal{N}+1)|a|} \hspace{2pt} a(p)\hspace{2pt}.
    \label{eqn: double CPT conjugate}
\end{equation}

\subsection{Antisymmetrization of Indices}

Finally, we discuss our notation for antisymmetrizing indices. We define
\begin{equation}
    \delta_J^{[I_1} a^{I_2 \cdots I_A]} = \begin{cases} (-1)^{n+1} \hspace{2pt} a^{I_1 \cdots I_{n-1} I_{n+1} \cdots I_A} \hspace{50pt} \text{if $J = I_n$} \\ \hspace{6pt} 0 \hspace{153.5pt} \text{if $J \neq I_n$ for all $n$}\hspace{2pt}.
    \end{cases}
\end{equation}
This may differ from other conventions by a multiplicative factor of $A$.

\newpage 

\section{Derivation of the $\frak{sbms}_{4|\mathcal{N}}$ Algebra From Conformally Soft Theorems}
\label{appendix: derivation of sbms}

In this appendix, we present the derivation of the $\frak{sbms}_{4|\mathcal{N}}$ algebra. We begin with the following OPEs which are derived immediately from the definition of the symmetry generators listed in Table \ref{tab: sbms algebra generators} and the conformally soft theorems listed in Table \ref{tab: conformally soft theorems (grav)}.
\begin{equation}
    \begin{split}
        \mathcal{P}(z,\overbar{z}) \hspace{2pt} \mathcal{O}_{a,\Delta}^{I_1 \cdots I_A}(0,0) &\sim \frac{1}{z \overbar{z}} \hspace{2pt} \mathcal{O}^{I_1 \cdots I_A}_{a,\Delta + 1}(0,0) \hspace{82pt} \mathcal{P}(z,\overbar{z}) = \hspace{-1.5pt} \sumop_{k,l \hspace{1pt} \in\hspace{1pt} \mathbb{Z} + \frac{1}{2}} \hspace{-5pt} z^{-k - 3/2}\hspace{1pt} \overbar{z}^{-l - 3/2} \hspace{2pt} P_{k,l} \\[-.2em]
        T(z) \hspace{2pt} \mathcal{O}_{a,\Delta}^{I_1 \cdots I_A}(0,0) &\sim \bigg(\frac{h}{z^2}  + \frac{1}{z} \partial\bigg) \mathcal{O}_{a,\Delta}^{I_1 \cdots I_A}(0,0) \hspace{52pt} T(z) = \hspace{3.5pt}\sumop_{n \hspace{1pt} \in\hspace{1pt} \mathbb{Z}} \hspace{6.5pt} z^{-n - 2} \hspace{2pt} L_n \\[.07em]
        \overbar{T}(\overbar z) \hspace{2pt} \mathcal{O}_{a,\Delta}^{I_1 \cdots I_A}(0,0) &\sim \bigg(\frac{\overbar h}{\overbar z^2}  + \frac{1}{\overbar z} \overbar{\partial}\bigg) \mathcal{O}_{a,\Delta}^{I_1 \cdots I_A}(0,0) \hspace{52pt} \overbar{T}(\overbar z) = \hspace{3.5pt}\sumop_{n \hspace{1pt} \in\hspace{1pt} \mathbb{Z}} \hspace{6.5pt} \overbar{z}^{-n - 2} \hspace{2pt} \overbar{L}_n \\[.35em]
        S_J(z) \hspace{2pt} \mathcal{O}_{a,\Delta}^{I_1 \cdots I_A}(0,0) &\sim \frac{1}{z} \hspace{2pt} \delta_J^{[I_1}\mathcal{O}_{a,\Delta +\frac{1}{2}}^{I_2 \cdots I_A]}(0,0) \hspace{73.5pt} S_J(z) = \hspace{-1.5pt} \sumop_{k \hspace{1pt} \in\hspace{1pt} \mathbb{Z} + \frac{1}{2}} z^{-k - 3/2} \hspace{2pt} G_{J,k} \\
        \overbar{S}^J(\overbar z) \hspace{2pt} \mathcal{O}_{a,\Delta}^{I_1 \cdots I_A}(0,0) &\sim \frac{1}{\overbar z} \hspace{2pt} \mathcal{O}_{a,\Delta+\frac{1}{2}}^{JI_1 \cdots I_A}(0,0) \hspace{87pt} \overbar{S}^J(\overbar z) = \hspace{-1.5pt} \sumop_{k \hspace{1pt} \in\hspace{1pt} \mathbb{Z} + \frac{1}{2}} \overbar{z}^{-k - 3/2} \hspace{2pt} \overbar{G}_{k}^J \\[-.2em]
        J_{JK}(z) \hspace{2pt} \mathcal{O}_{a,\Delta}^{I_1 \cdots I_A}(0,0) &\sim \frac{1}{z}\hspace{2pt} Z_{JK} \hspace{1pt} \mathcal{O}_{a,\Delta + 1}^{I_1 \cdots I_A}(0,0) \hspace{64pt} J_{JK}(z) = \hspace{3.5 pt}\sumop_{n \hspace{1pt} \in\hspace{1pt} \mathbb{Z}} \hspace{6.5pt} z^{-n - 1} \hspace{2pt} \mathcal{Z}_{IJ,n} \\[.4em]
        \overbar{J}^{JK}(\overbar z) \hspace{2pt} \mathcal{O}_{a,\Delta}^{I_1 \cdots I_A}(0,0) &\sim \frac{1}{\overbar z}\hspace{2pt} \overbar{Z}^{JK} \hspace{1pt} \mathcal{O}_{a,\Delta + 1}^{I_1 \cdots I_A}(0,0) \hspace{63pt} \overbar{J}^{JK}(\overbar z) = \hspace{3.5 pt}\sumop_{n \hspace{1pt} \in\hspace{1pt} \mathbb{Z}} \hspace{6.5pt} \overbar{z}^{-n - 1} \hspace{2pt} \overbar{\mathcal{Z}}^{IJ}_{n}\hspace{2pt}. \\
    \end{split}
\end{equation}
These OPEs hold when $\mathcal{O}^{I_1 \cdots I_A}_{a,\Delta}$ is a generic conformal primary operator in the theory. By taking contour integrals of the left hand side of this equation, one can extract how the modes individually act on such operators. We find
\begin{equation}
    \begin{split}
        P_{k,l} \bullet \mathcal{O}^{I_1 \cdots I_A}_{a,\Delta}(z,\overbar z) &= z^{k+1/2} \hspace{2pt} \overbar{z}^{l+1/2} \hspace{2pt} \mathcal{O}^{I_1 \cdots I_A}_{a,\Delta+1}(z,\overbar z) \\[.75em]
        L_n \bullet \mathcal{O}^{I_1 \cdots I_A}_{a,\Delta}(z,\overbar z) &= \big(h(n+1) z^n + z^{n+1} \partial\big) \mathcal{O}^{I_1 \cdots I_A}_{a,\Delta}(z,\overbar{z}) \\[.75em]
        \hspace{20pt} \overbar{L}_n \bullet \mathcal{O}^{I_1 \cdots I_A}_{a,\Delta +\frac{1}{2}}(z,\overbar z) &= \big(\overbar{h}(n+1) \overbar{z}^n + \overbar{z}^{n+1} \partial\big) \mathcal{O}^{I_1 \cdots I_A}_{a,\Delta}(z,\overbar{z})\\[.75em]
        G_{J,k} \bullet \mathcal{O}^{I_1 \cdots I_A}_{a,\Delta}(z,\overbar z) &= z^{k+1/2} \hspace{1pt} \delta_{J}^{[I_1} \mathcal{O}^{I_2 \cdots I_A]}_{a,\Delta +\frac{1}{2}}(z,\overbar{z}) \\[.75em] \overbar{G}^J_{k} \bullet \mathcal{O}^{I_1 \cdots I_A}_{a,\Delta}(z,\overbar z) &= \overbar{z}^{k+1/2} \hspace{1pt}\mathcal{O}^{J I_1 \cdots I_A}_{a,\Delta+1/2}(z,\overbar{z}) \\[.75em]
        J_{JK,n} \bullet \mathcal{O}^{I_1 \cdots I_A}_{a,\Delta}(z,\overbar z) &= z^n \hspace{1pt} Z_{JK} \mathcal{O}^{I_1 \cdots I_A}_{a,\Delta + 1}(z,\overbar{z}) \\[.75em] \overbar{J}_{n}^{JK} \bullet \mathcal{O}^{I_1 \cdots I_A}_{a,\Delta}(z,\overbar z) &= \overbar{z}^n \hspace{1pt} \overbar{Z}^{JK} \mathcal{O}^{I_1 \cdots I_A}_{a,\Delta + 1}(z,\overbar{z})\hspace{2pt}. 
    \end{split}
\end{equation}

To determine the commutation relations between the generators, we apply them successively to the conformal primary operators in an anti-symmetric way. For example:
\begin{equation}
    \begin{split}
        \big[L_n,G_{I,k}\big] \bullet \mathcal{O}^{I_1 \cdots I_A}_{a,\Delta}(z,\overbar z) &= \big(L_n\bullet G_{I,k} -G_{I,k}\bullet L_{n}\big) \bullet \mathcal{O}^{I_1 \cdots I_A}_{a,\Delta}(z,\overbar z) \\[.3em]
        &= z^{k+1/2} \big((h+1/2)(n+1) z^n + z^{n+1} \partial\big)\delta_{I}^{[I_1} \mathcal{O}^{I_2 \cdots I_A]}_{a,\Delta +\frac{1}{2}}(z,\overbar z) \\[.3em]
        &\hspace{50pt} - \big(h(n+1) z^n + z^{n+1} \partial\big) \Big(z^{k+1/2} \hspace{2pt} \delta_{I}^{[I_1} \mathcal{O}^{I_2 \cdots I_A]}_{a,\Delta +\frac{1}{2}}(z,\overbar z)\Big) \\[.3em]
        &= \big(\tfrac{n}{2} - k\big) z^{k+n + 1/2} \hspace{2pt} \delta_{I}^{[I_1} \mathcal{O}^{I_2 \cdots I_A]}_{a,\Delta +\frac{1}{2}}(z,\overbar z) \\[.3em]
        &= \big(\tfrac{n}{2} - k\big) \hspace{2pt} G_{I,k+n} \bullet \mathcal{O}^{I_1 \cdots I_A}_{a,\Delta}(z,\overbar z)\hspace{2pt}.
    \end{split}
    \label{eqn: L_n G_{I,k} commutator}
\end{equation}
Because the conformal primary $\mathcal{O}^{I_1 \cdots I_A}_{a,\Delta}(z,\overbar z)$ was arbitrary, we infer that
\begin{equation}
    \big[L_n,G_{I,k}\big] = \big(\tfrac{1}{2}n - k\big) \hspace{2pt} G_{I,k+n}\hspace{2pt}.
\end{equation}
This is a streamlined version of the technique used in \cite{Fotopoulos:2019vac, Fotopoulos:2020bqj}. Unfortunately, this calculation loses track of possible central terms (i.e. operators commuting with all conformal primaries) which may arise. A complete analysis would involve studying the $T(z) S_{I,k}(w)$ OPE directly, and extracting the mode algebra from this expression. However, this approach requires carefully regulating the shadow transform integrals defining these operators, and we will not discuss this issue here.

Performing calculations along the lines of Equation \eqref{eqn: L_n G_{I,k} commutator} yields most of the commutation relations in the $\frak{sbms}_{4|\mathcal{N}}$ algebra. There is one fly in the ointment: massless particles all have vanishing central charges due to the BPS bound, $Z_{IJ} = 0.$ One resolution is to carefully consider the action of the currents on massive conformal primaries which live in \textit{short multiplets} of the supersymmetry algebra. This is slightly tedious, so we shall provide a slicker argument which lands on the correct answer. The only difficult computation is showing that 
\begin{equation}
    \big[G_{I,k},G_{J,l}\big] = -(k-l) \hspace{2pt} \mathcal{Z}_{IJ,k+l}\hspace{2pt}. 
    \label{eqn: difficult sbms commutator}
\end{equation}

From the definition of the currents $S_I(z)$ in terms of the shadow transform, it follows that the commutator among $G_{I,k}$ modes can only give $\mathcal{Z}_{IJ,n}$ modes. This is consistent with the fact that the $\frak{sbms}_{4|\mathcal{N}}$ is a local extension of the super-Poincar\'e algebra as discussed in Section \ref{sec: sbms algebra} --- the SUSY generators must commute to give central charges. In particular, $[G_{I,-1/2},G_{J,1/2}] = Z_{IJ,\hspace{1pt} 0}$ according to Equation \eqref{eqn: conformal generators in sbms}.

The commutatore, therefore, takes the following form: $\big[G_{I,k},G_{J,l}\big] = f(k,l;n) \hspace{2pt} \mathcal{Z}_{IJ,n}$. Applying $L_0$ to both sides of this expression and using the Jacobi identity implies $f(k,l;n) = F(l-k)\delta_{k+l,n}$ where we have also used the fact that $\mathcal{Z}_{IJ,n} = - \mathcal{Z}_{JI,n}$ to determine that $f(k,l;n)$ only has dependence on $l-k$. The Equation $[G_{I,-1/2},G_{J,1/2}] = Z_{IJ,\hspace{1pt} 0}$ implies that $F(1) = 1.$ Repeatedly applying $L_1$ to both sides of the commutator $[G_{I,k},G_{J,1/2}] = F(\frac{1}{2}-k) \mathcal{Z}_{IJ,k+1/2}$ yields the recursion relation: $(\frac{1}{2}-k) F(-\frac{1}{2}-k) = -(\frac{1}{2}+k) F(\frac{1}{2}-k)$. This is solved uniquely by Equation \eqref{eqn: difficult sbms commutator}.

\section{SUSY Identities for Three-Point Coefficients, OPEs, and Soft Algebras}
\label{appendix: supersymmetry relations and covariance}

The aim of this appendix is to check that three-point functions, OPEs, and chirals soft algebras transform covariantly under the action of bulk supersymmetry, verifying some claims made in the body of the text. 

\subsection{SUSY Identities for Three-Point Coefficients}

We begin by studying the three-point functions $g\big(a^{I_1 \cdots I_A},b^{J_1 \cdots J_B}\hspace{1pt};\hspace{1pt}c^{K_1 \cdots K_C}\big)$ appearing in holomorphic OPEs and chiral algebras. They can be computed explicitly by evaluating the following three particle amplitude involving one incoming and two outgoing particles \cite{Himwich:2021dau}\footnote{On its face, this expression looks different from what is provided in \cite{Himwich:2021dau}. A first difference is that their expression for the three-point coefficient involves a three-particle amplitude where all particles are taken to be outgoing; the cost of this is that the three-point coefficient involves the first two particles and the CPT conjugate of the third. We use crossing symmetry to write $c^{K_1 \cdots K_C}(p_3)$ as an incoming particle, circumventing this annoyance. A second difference has to do with the fact that our massless momentum parameterizations and spinor-helicity conventions differ: $\omega_{\text{ours}} = \sqrt{2}\hspace{2pt} \omega_{\text{theirs}}$ and $[ij]_{\text{ours}} = 2 [ji]_{\text{theirs}}$. Finally, we rescale the three-point coefficients by $g_{\text{ours}} = g_{\text{theirs}}/2$ to avoid annoying factors of two which would otherwise crop up in expressions for the OPE and chiral soft algebras.}
\begin{equation}
    \begin{split}
        \langle 0| &a^{I_1 \cdots I_A}(p_1) \hspace{1pt} b^{J_1 \cdots J_B}(p_2) \hspace{2pt} S \hspace{2pt} c^{K_1 \cdots K_C  \dagger}(p_3) |0\rangle \\[.3em]
        &\hspace{28pt}= 2^{-p} \hspace{2pt} [21]^{s_1 + s_2 + s_3}[31]^{s_1 - s_2 - s_3}[23]^{s_2 - s_1 - s_3} g\big(a^{I_1 \cdots I_A}, b^{J_1 \cdots J_B}\hspace{1pt};\hspace{1pt} c^{K_1 \cdots K_C}\big)\hspace{2pt},
    \end{split}
    \label{eqn: definition of three-point function from amplitude amplitude}
\end{equation}
where the spins $s_1, s_2$ and $s_3$ are respectively defined as the spins of $a^{I_1 \cdots I_A},$ $b^{J_1 \cdots J_B}$, and $c^{K_1 \cdots K_C}$. Note that in this expression $c^{K_1 \cdots K_C}(p_3)$ is an incoming particle with a positive energy. One could use crossing symmetry to write it as an outgoing particle with negative energy, but this will only make the subsequent derivation messier. 

Our ultimate goal is to prove the proposition stated in Section \ref{sec: OPEs}, which culminates in Equations \eqref{eqn: three-point proposition}, \eqref{eqn: symmetry relations three point}, and \eqref{eqn: bound on number of independent three-point coefficients}. We first set up some notation. From Equation \eqref{eqn: SUSY multiplet}, we observe that the action of $\overbar{Q}^I_{\overbar \alpha}$ and $Q_{I \alpha}$ can be viewed as adding or subtracting an index followed with multiplying by a momentum spinor. It will be convenient to isolate these two operations by defining $\overbar{q}^I$ and $q_I$ which act solely by adding and subtracting indices
\begin{equation}
    \begin{split}
        \big[\overbar{q}^J,a^{I_1 \cdots I_A}(p)\big] &= a^{JI_1 \cdots I_A}(p) \hspace{30pt},\hspace{30pt} \big[q_J, a^{I_1 \cdots I_A}(p)\big] = \delta^{[I_1}_J a^{I_2 \cdots I_A]}(p)\hspace{2pt}. \\
    \end{split}
\end{equation}
These are fermionic operators satisfying $[\overbar{q}^I,q_J] = \delta^I_J$. While they are not symmetries of the theory, they are a helpful bookkeeping device which allows us to write $\overbar{Q}^I_{\overbar{\alpha}} = |p]_{\overbar{\alpha}} \hspace{2pt} \overbar{q}^I$ and $Q_{I,\alpha} = -\langle p|_{\alpha} \hspace{2pt} q_I$. We begin with the following lemma.

\vspace{10pt}

\noindent \textbf{Lemma:} The momentum space SUSY Ward identities imply that the couplings are subject to the following algebraic identities which relate three-point coefficients among superpartners
\begin{equation}
    \begin{split}
        g\big(a^{I_1 \cdots I_A},b^{J_1 \cdots J_B}\hspace{1pt};\hspace{1pt}\big[q_L,c^{K_1 \cdots K_C}\big]\big) &= g\big(\big[\overbar{q}^L,a^{I_1 \cdots I_A}\big], b^{J_1 \cdots J_B}\hspace{1pt};\hspace{1pt} c^{K_1 \cdots K_C}\big) \\[.3em]
        &= (-1)^{|a^{I_1 \cdots I_A}|} \hspace{2pt} g\big(a^{I_1 \cdots I_A}, \big[\overbar{q}^L,b^{J_1 \cdots J_B}\big]\hspace{1pt};\hspace{1pt} c^{K_1 \cdots K_C}\big) \hspace{24pt}
    \end{split}
    \label{eqn: susy three-point 2}
\end{equation} 
\begin{equation}
    \begin{split}
        \hspace{11pt} g\big(a^{I_1 \cdots I_A},b^{J_1 \cdots J_B}\hspace{1pt};\hspace{1pt}\big[\overbar{q}^L,c^{K_1 \cdots K_C}\big]\big) &= g\big(\big[q_L,a^{I_1 \cdots I_A}\big], b^{J_1 \cdots J_B}\hspace{1pt};\hspace{1pt} c^{K_1 \cdots K_C}\big) \\[.3em]
        &\hspace{35pt} + (-1)^{|a^{I_1 \cdots I_A}|} \hspace{2pt} g\big(a^{I_1 \cdots I_A}, \big[q_L,b^{J_1 \cdots J_B}\big]\hspace{1pt};\hspace{1pt} c^{K_1 \cdots K_C}\big)\hspace{2pt}.  \\
    \end{split}
    \label{eqn: susy three-point 1}
\end{equation}
We say that two-three point functions live in the same \textit{supersymmetry orbit} if they can be determined from one another by these algebraic relations.

\begin{adjustwidth}{40pt}{40pt}
\noindent------------------------------------------------------------------------------------------------------------
\textit{Proof:} This lemma follows from application of the supersymmetric Ward identities to the three-particle amplitudes defining the three-point coefficients. For example
\begin{equation}
    \begin{split}
        0 &= \big[\xi \overbar{Q}^L\big] \bullet \langle 0| a^{I_1 \cdots I_A} \hspace{1pt} b^{J_1 \cdots J_B} \hspace{2pt} S \hspace{2pt} c^{K_1 \cdots K_C  \dagger} |0\rangle \\[10pt]
        &= \langle 0| \big[\overbar{q}^L,a^{I_1 \cdots I_A}\big] \hspace{1pt} b^{J_1 \cdots J_B} \hspace{2pt} S \hspace{2pt} c^{K_1 \cdots K_C  \dagger} |0\rangle \hspace{2pt} [\xi 1] \\
        &\hspace{20pt} + (-1)^{|a^{I_1 \cdots I_A}|} \hspace{2pt} \langle 0| a^{I_1 \cdots I_A} \hspace{1pt} \big[\overbar{q}^L,b^{J_1 \cdots J_B}\big] \hspace{2pt} S \hspace{2pt} c^{K_1 \cdots K_C  \dagger} |0\rangle \hspace{2pt} [\xi 2] \\
        &\hspace{20pt} -  \langle 0| a^{I_1 \cdots I_A} \hspace{1pt} b^{J_1 \cdots J_B} \hspace{2pt} S \hspace{2pt} \big[q_L, c^{K_1 \cdots K_C}\big]^\dagger |0\rangle \hspace{2pt} [\xi 3] \\[10pt]
        &= g\big(\big[\overbar{q}^L,a^{I_1 \cdots I_A}\big], b^{J_1 \cdots J_B}\hspace{1pt};\hspace{1pt} c^{K_1 \cdots K_C}\big) \hspace{2pt} [\xi 1] \hspace{2pt} [23] \\
        &\hspace{20pt} + (-1)^{|a^{I_1 \cdots I_A}|} \hspace{2pt} g\big(a^{I_1 \cdots I_A},\big[\overbar{q}^L, b^{J_1 \cdots J_B}\big]\hspace{1pt};\hspace{1pt} c^{K_1 \cdots K_C}\big) \hspace{2pt} [\xi 2] \hspace{2pt} [31] \\
        &\hspace{20pt} - g\big(a^{I_1 \cdots I_A}, b^{J_1 \cdots J_B}\hspace{1pt};\hspace{1pt} \big[q_L,c^{K_1 \cdots K_C}\big]\big) [\xi 3] \hspace{2pt} [21]\hspace{2pt},
    \end{split}
\end{equation}
from which Equation \eqref{eqn: susy three-point 2} follows immediately. A similar derivation yields
\begin{equation}
    \begin{split}
        \hspace{12pt} 0 &= \big\langle\xi Q_L\big\rangle \bullet \langle 0| a^{I_1 \cdots I_A} \hspace{1pt} b^{J_1 \cdots J_B} \hspace{2pt} S \hspace{2pt} c^{K_1 \cdots K_C  \dagger} |0\rangle \\[10pt]
        &= -\langle 0| \big[q_L,a^{I_1 \cdots I_A}\big] \hspace{1pt} b^{J_1 \cdots J_B} \hspace{2pt} S \hspace{2pt} c^{K_1 \cdots K_C  \dagger} |0\rangle \hspace{2pt} \langle \xi 1\rangle \\
        &\hspace{20pt} - (-1)^{|a^{I_1 \cdots I_A}|} \hspace{2pt} \langle 0| a^{I_1 \cdots I_A} \hspace{1pt} \big[q_L,b^{J_1 \cdots J_B}\big] \hspace{2pt} S \hspace{2pt} c^{K_1 \cdots K_C  \dagger} |0\rangle \hspace{2pt} \langle \xi 2\rangle \\
        &\hspace{20pt} +  \langle 0| a^{I_1 \cdots I_A} \hspace{1pt} b^{J_1 \cdots J_B} \hspace{2pt} S \hspace{2pt} \big[\overbar{q}^L, c^{K_1 \cdots K_C}\big]^\dagger |0\rangle \hspace{2pt} \langle \xi 3\rangle \\[10pt]
        &= -g\big(\big[q_L,a^{I_1 \cdots I_A}\big], b^{J_1 \cdots J_B}\hspace{1pt};\hspace{1pt} c^{K_1 \cdots K_C}\big) \hspace{2pt} \langle \xi 1\rangle \hspace{2pt} [23]^{-1} \\
        &\hspace{20pt} - (-1)^{|a^{I_1 \cdots I_A}|} \hspace{2pt} g\big(a^{I_1 \cdots I_A},\big[q_L, b^{J_1 \cdots J_B}\big]\hspace{1pt};\hspace{1pt} c^{K_1 \cdots K_C}\big) \hspace{2pt} \langle \xi 2\rangle \hspace{2pt} [31]^{-1} \\
        &\hspace{20pt} + g\big(a^{I_1 \cdots I_A}, b^{J_1 \cdots J_B}\hspace{1pt};\hspace{1pt} \big[\overbar{q}^L,c^{K_1 \cdots K_C}\big]\big) \langle \xi 3\rangle \hspace{2pt} [21]^{-1}\hspace{2pt},
    \end{split}
\end{equation}
from which one can derive Equation \eqref{eqn: susy three-point 1}.

We caution the reader that in this derivation one needs to be very careful about one's convention for spinor-helicity variables as outlined in Appendix \ref{appendix: conventions}. Normally, the spinor-helicity formalism is developed in an all-out prescription where $p_1 + p_2 + p_3 = 0$. Here, we emphasize that $c^{K_1 \cdots K_C}(p_3)$ is an incoming particle. Momentum conservation now reads $p_1 + p_2 - p_3 = 0$. This will ruin certain spinor-helicity identities which one often uses intuitively. Now, $|3\rangle^\dagger = -[3|$ and $\langle \xi 1 \rangle \hspace{2pt} [21] = +\langle \xi 3 \rangle \hspace{2pt} [23].$ A careful accounting of these features yields the correct answer.

\noindent------------------------------------------------------------------------------------------------------------
\end{adjustwidth}

\vspace{10pt}

\noindent \textbf{Proposition:} The three-point coefficient $g\big(a^{I_1 \cdots I_A}, b^{J_1 \cdots J_B}\hspace{1pt};\hspace{1pt} c^{K_1 \cdots K_C}\big)$ vanishes if any of the indices $I_1,... I_A,J_1,...,J_B$ are repeated or if this collection of indices does not match the indices $K_1,...,K_C$. I.e. the only potentially non-vanishing three-point coefficients are $g\big(a^{I_1 \cdots I_A}, b^{J_1 \cdots J_B}\hspace{1pt};\hspace{1pt} c^{I_1 \cdots I_A J_1 \cdots J_B}\big)$ where none of the indices $I_1,... I_A,J_1,...,J_B$ are repeated. Moreover, such three-point coefficients are entirely fixed by the three-point coefficient $g(a,b\hspace{1pt};\hspace{1pt}  c)$ among highest-weight states (corresponding to supercelestial primaries in the CCFT) according to
\begin{equation}
    g_{ab}^c \equiv g(a, b\hspace{1pt};\hspace{1pt}  c) = (-1)^{B|a|} \hspace{2pt} g\big(a^{I_1 \cdots I_A}, b^{J_1 \cdots J_B}\hspace{1pt};\hspace{1pt} c^{I_1 \cdots I_A J_1 \cdots J_B}\big)\hspace{2pt}.
\end{equation}
\vspace{-14pt}
\begin{adjustwidth}{40pt}{40pt}
\noindent------------------------------------------------------------------------------------------------------------
\textit{Proof:} We first show by way of example why certain three-point coefficients vanish automatically when the above conditions are not met. Consider $g\big(a^I,b\hspace{1pt}\big|\hspace{1pt}c\big)$. We may use Equation \eqref{eqn: susy three-point 2} to remove the $I$ index from the left-hand side (LHS) at the cost of acting on the right-hand side (RHS) with $q_I$
\begin{equation}
    g\big(a^{I},b\hspace{1pt};\hspace{1pt}c\big) = g\big(\big[\overbar{q}^I,a\big],b\hspace{1pt};\hspace{1pt}c\big) = g\big(a,b\hspace{1pt};\hspace{1pt}\big[q_I,c\big]\big) = 0\hspace{2pt}.
\end{equation}
This three-point function is automatically equal to zero because $c$ is a highest-weight state, so $[q_I,c] = 0$. Generically, one can use Equation \eqref{eqn: susy three-point 2} to remove all superscripts on the LHS of $g\big(a^{I_1 \cdots I_A},b^{J_1 \cdots J_B}\hspace{1pt};\hspace{1pt}c^{K_1 \cdots K_C}\big)$ at the cost of acting with $q_{I_1}, ..., q_{I_A},q_{J_1},...,q_{J_B}$ on the RHS. If the indices $I_1,...,I_A,J_1,...,J_B$ are not all represented among $K_1,...,K_C$, then the expression vanishes. Thus,
\begin{equation}
    \{I_1,...,I_A,J_1,...,J_B\} \subset \{K_1,...,K_C\}\hspace{2pt}.
\end{equation}

We can also use Equation \eqref{eqn: susy three-point 1} to play this game in reverse, removing indices from the RHS at the cost of acting with $q_{K_1},...,q_{K_C}$ on the LHS. Thus, if the indices $K_1,...,K_C$ are not all represented among $I_1,...,I_A,J_1,...,J_B$ the coupling will also vanish. Therefore,
\begin{equation}
    \{I_1,...,I_A,J_1,...,J_B\} \supset \{K_1,...,K_C\}\hspace{2pt}.
\end{equation}
The only way these two constraints can be simultaneously met is if the collection of indices are equal. Because $c^{K_1 \cdots K_C}(p)$ is totally anti-symmetric in its indices, it follows that $K_1,...,K_C$ are distinct. Therefore, $I_1,...,I_A,J_1,...,J_B$ are distinct too.

Altogether, we have shown that the only potentially non-vanishing three-point coefficients take the form $g\big(a^{I_1 \cdots I_A},b^{J_1 \cdots J_B}\hspace{1pt};\hspace{1pt}c^{I_1 \cdots I_A J_1 \cdots J_B}\big)$ where none of the indices $I_1,... I_A,J_1,...,J_B$ are repeated. We compute them directly from Equation \eqref{eqn: susy three-point 2}
\begin{equation}
    \begin{split}
        g_{ab}^c &\equiv g\big(a,b\hspace{1pt};\hspace{1pt}c\big) \\[.5em]
        &= (-1)^{B|a|}\hspace{2pt} g\big(a,b^{J_1 \cdots J_B}\hspace{1pt};\hspace{1pt}c^{J_1 \cdots J_B}\big)\\[.5em]
        &= (-1)^{B|a|}\hspace{2pt} g\big(a^{I_1 \cdots I_A},b^{J_1 \cdots J_B}\hspace{1pt};\hspace{1pt}c^{I_1 \cdots I_A J_1 \cdots J_B}\big)\hspace{2pt}.\\[-.5em]
    \end{split}
\end{equation}
\noindent------------------------------------------------------------------------------------------------------------
\end{adjustwidth}
\noindent\textbf{Proposition:} The three-point coefficient obeys the following symmetry properties
\begin{equation}
    \begin{split}
        g^{c}_{ab} = (-1)^{|a||b| + p + 1} \hspace{2pt} g_{ba}^{c} \hspace{30pt}\text{and} \hspace{30pt}
        g_{ab}^c = g_{b\overbar c}^{\overbar a}\hspace{2pt}.
    \end{split}
    \label{eqn: appendix B second identity}
\end{equation}
\begin{adjustwidth}{40pt}{40pt}
\noindent------------------------------------------------------------------------------------------------------------
\textit{Proof:} These identities follow from Equation \eqref{eqn: definition of three-point function from amplitude amplitude} which defines the three-point coefficients in terms of the scattering amplitudes. For example
\begin{equation}
    \begin{split}
        \langle 0| a(p_1) &b(p_2) S \hspace{1pt} c(p_3) |0\rangle \\[.3em]
        &= 2^{-p} \hspace{2pt} g_{ab}^c \hspace{2pt} [21]^{s_a + s_b + s_c} [31]^{s_a - s_b - s_c} [23]^{s_b - s_a - s_c}\hspace{2pt}, \hspace{62pt}
    \end{split} 
\end{equation}
but also
\begin{equation}
    \begin{split}
         \langle 0| a(p_1) &b(p_2) S \hspace{1pt} c(p_3) |0\rangle \\[.3em]
         &= (-1)^{|a||b|} \hspace{2pt} \langle 0| b(p_2) a(p_1) S \hspace{1pt} c(p_3) |0\rangle\\[.3em]
         &= 2^{-p} \hspace{1pt} (-1)^{|a||b|} \hspace{2pt} g_{ba}^c \hspace{2pt} [12]^{s_a + s_b + s_c} [32]^{s_b - s_a - s_c} [13]^{s_a - s_b - s_c} \\[.3em]
         &= 2^{-p} \hspace{1pt} (-1)^{|a||b| + p + 1} \hspace{2pt} g_{ba}^c \hspace{2pt} [21]^{s_a + s_b + s_c} [31]^{s_a - s_b - s_c} [23]^{s_b - s_a - s_c}\hspace{2pt}.
    \end{split}
\end{equation}
Comparing the two expressions yields the first identity in Equation \eqref{eqn: appendix B second identity}. The second identity is a similar derivation following from the crossing symmetry conventions of Equation \eqref{eqn: crossing}.

\noindent------------------------------------------------------------------------------------------------------------
\end{adjustwidth}

\noindent\textbf{Corollary:} If there are $r$ highest-weight representations (counting a CPT self-conjugate representation only once) in the theory, there are $r^3$ three-point coefficients $g_{ab}^c$ which fully determine all OPEs and chiral algebras in the theory. However, \eqref{eqn: appendix B second identity} imposes symmetry relations among these three-point coefficients. The number of \textit{independent} coefficients is bounded from above by
\begin{equation}
    \# \hspace{2pt}\Big(\text{Independent} ~ g_{ab}^c\Big)  \leq \hspace{2pt} \frac{1}{6} (r^3 + 3r^2 + 2r)\hspace{2pt}.
    \label{eqn: appendix bound}
\end{equation}
\begin{adjustwidth}{40pt}{40pt}
------------------------------------------------------------------------------------------------------------
\textit{Proof:} We consider the most general theory with $2n$ highest-weight representations which are not CPT self-conjugate --- label them $a_1,...,a_n,\overbar{a}_1,...,\overbar{a}_n$ --- and $m$ highest-weight representations which are CPT self conjugate --- label them $b_1,...,b_m$. Denote the set of all highest-weight states 
\begin{equation}
    \Omega = \big\{a_1,...,a_n,\overbar{a}_1,...,\overbar{a}_n,b_1,...,b_m\big\}\hspace{2pt}.
\end{equation}
We define $\overbar{(a_i)} = \overbar{a}_i, \overbar{(\overbar{a}_i)} = a_i, \overbar{(b_i)} = b_i$ which encapsulates how the highest-weight states transform among themselves under CPT conjugation.  

We view the three-point coefficients as a map $g: \Omega^3 \rightarrow \mathbb{C}.$ Equation \eqref{eqn: appendix B second identity} induces a group action by $S_3 = \langle s,t | s^2 = 1, t^3 = 1, s t s = t^{-1} \rangle$ on $\Omega^3$
\begin{equation}
    t: (x,y,z) \longmapsto (y,\overbar{z},\overbar{x}) \hspace{40pt} s: (x,y,z) \longmapsto (y,x,z)\hspace{2pt},
    \label{eqn: group action}
\end{equation}
where $(x,y,z) \in \Omega^3$. Indeed, for all permutations $\sigma \in S_3$, the three-point coefficients $g((x,y,z))$ and $g(\sigma(x,y,z))$ are related by Equation \eqref{eqn: appendix B second identity}. In this way, one should view $S_3$ as the symmetry group generated by such identities.

It follows that the total number of independent three-point coefficients is bounded from above by the number of elements in $\Omega^3$ modulo the group action by $S^3$\hspace{1pt}\footnote{Here, the absolute value signs denote the cardinality of the set. Of course, it's possible that there are even fewer independent components because some of them may be fixed to zero identically: for example, if $g_{ab}^c = - g_{ba}^c$, then $g_{aa}^c = 0$ automatically. However, in general, such spurious cancellations depend non-trivially on which specific highest-weight states are present as well as how many supersymmetries the theory enjoys. For example, as discussed in Section \ref{sec: super w 1 + inf}, $g_{aa}^{\overbar{a}}$ necessarily vanishes in pure supergravity except perhaps when $\mathcal{N} = 4$.}
\begin{equation}
    \# \hspace{2pt}\Big(\text{Independent} ~ g_{ab}^c\Big)  \leq |\Omega^3/S_3|\hspace{2pt}.
\end{equation}
The number of orbits of $\Omega^3$ under $S_3$ may be computed with Burnside's lemma
\begin{equation}
    |\Omega^3/S_3| = \frac{1}{|S_3|} \sum_{\sigma \in S_3} |\text{Fix}(\sigma)|\hspace{2pt},
\end{equation}
where $\text{Fix}(\sigma) = \{(x,y,z) \in \Omega^3| (x,y,z) = \sigma(x,y,z)\}.$ From Equation \eqref{eqn: group action}, one can easily compute each term in the sum. The result is Equation \eqref{eqn: appendix bound}.

\noindent------------------------------------------------------------------------------------------------------------
\end{adjustwidth}
\vspace{-20pt}

\subsection{SUSY Covariance of OPEs}
\label{appendix: SUSY covariance subsection}

In this section, we verify that the OPE relations derived in Section \ref{sec: SUSY OPE expressions}
\begin{equation}
    \mathcal{O}^{I_1 \cdots I_A}_{a,\hspace{1pt}\overbar{h}_a}(z,\overbar z)\hspace{2pt} \mathcal{O}^{J_1 \cdots J_B}_{b,\hspace{1pt} \overbar{h}_b}(0,0) \sim -\frac{(-1)^{B|a|}}{z} \sumop_{c} \hspace{1pt} g_{ab}^c \hspace{2pt}C_{p}(\hspace{1pt}\overbar{h}_a, \overbar{h}_b) \hspace{1pt} \mathcal{O}_{c,\hspace{1pt} \overbar{h}_a +\overbar{h}_b + p}^{I_1 \cdots I_A J_1 \cdots J_B}(0,0) 
\end{equation}
are covariant under the action of bulk supersymmetry. For example, applying the supercharge $Q_{K,0}$ to the RHS of this expression versus applying it to the LHS and then simplifying using OPE relations yields matching answers
\begin{align}
        \big[Q_{K,0},\text{RHS}]
        &= \frac{1}{z} (-1)^{B|a|} \sumop_{c} \hspace{1pt} g_{ab}^c \hspace{2pt}C_{p}(\hspace{1pt}\overbar{h}_a, \overbar{h}_b) \hspace{1pt} \delta^{[I_1}_{K} \mathcal{O}_{c,\hspace{1pt} \overbar{h}_a +\overbar{h}_b + p}^{I_2 \cdots I_A J_1 \cdots J_B]}(0,0) \nonumber \\[.75em]
        [Q_{K,0},\text{LHS}\big] &= -\delta^{[I_1}_{K} \mathcal{O}^{I_2 \cdots I_A]}_{a,\hspace{1pt}\overbar{h}_a}(z,\overbar z) \mathcal{O}^{J_1 \cdots J_B}_{b,\hspace{1pt} \overbar{h}_b}(0,0) - (-1)^{|a^{I_1 \cdots I_A}|} \mathcal{O}^{I_1 \cdots I_A}_{a,\hspace{1pt}\overbar{h}_a}(z,\overbar z)\hspace{2pt} \delta^{[J_1}_K \mathcal{O}^{J_2 \cdots J_B]}_{b,\hspace{1pt} \overbar{h}_b}(0,0) \nonumber \\[.4em]
        &= \frac{1}{z} (-1)^{B|a|} \sumop_{c} \hspace{1pt} g_{ab}^c \hspace{2pt}C_{p}(\hspace{1pt}\overbar{h}_a, \overbar{h}_b) \hspace{1pt} \delta^{[I_1}_{K} \mathcal{O}_{c,\hspace{1pt} \overbar{h}_a +\overbar{h}_b + p}^{I_2 \cdots I_A] J_1 \cdots J_B}(0,0) \\[-.2em]
        &\hspace{40pt} + \frac{1}{z} (-1)^{B|a| + A}  \sumop_{c} \hspace{1pt} g_{ab}^c \hspace{2pt}C_{p}(\hspace{1pt}\overbar{h}_a, \overbar{h}_b) \hspace{1pt} \delta^{J_1]}_{K} \mathcal{O}_{c,\hspace{1pt} \overbar{h}_a +\overbar{h}_b + p}^{I_1 \cdots I_A [J_2 \cdots J_B}(0,0)\hspace{2pt}. \nonumber
\end{align}
A similar analysis for $\overbar{Q}^K_0$ gives
\begin{equation}
    \begin{split}
        \big[\overbar{Q}^K_{0},\text{RHS}]
        &= -\frac{1}{z} (-1)^{B|a|} \sumop_{c} \hspace{1pt} g_{ab}^c \hspace{2pt}C_{p}(\hspace{1pt}\overbar{h}_a, \overbar{h}_b) \hspace{1pt} \mathcal{O}_{c,\hspace{1pt} \overbar{h}_a +\overbar{h}_b + p + \frac{1}{2}}^{K I_1 \cdots I_A J_1 \cdots J_B]}(0,0) \\[.75em]
        [\overbar{Q}^K_{0},\text{LHS}\big] &=  \mathcal{O}^{K I_1 \cdots I_A}_{a,\hspace{1pt}\overbar{h}_a + \frac{1}{2}}(z,\overbar z) \mathcal{O}^{J_1 \cdots J_B}_{b,\hspace{1pt} \overbar{h}_b}(0,0) + (-1)^{|a^{I_1 \cdots I_A}|} \mathcal{O}^{I_1 \cdots I_A}_{a,\hspace{1pt}\overbar{h}_a}(z,\overbar z)\hspace{2pt} \mathcal{O}^{K J_1 \cdots J_B]}_{b,\hspace{1pt} \overbar{h}_b + \frac{1}{2}}(0,0) \\[.4em]
        &= -\frac{1}{z} (-1)^{B|a|} \sumop_{c} \hspace{1pt} g_{ab}^c \hspace{2pt}C_{p}(\hspace{1pt}\overbar{h}_a + \tfrac{1}{2}, \overbar{h}_b) \hspace{1pt} \mathcal{O}_{c,\hspace{1pt} \overbar{h}_a +\overbar{h}_b + p}^{K I_1 \cdots I_A J_1 \cdots J_B}(0,0) \\[-.2em]
        &\hspace{40pt} -\frac{1}{z} (-1)^{B|a| + A}  \sumop_{c} \hspace{1pt} g_{ab}^c \hspace{2pt}C_{p}(\hspace{1pt}\overbar{h}_a, \overbar{h}_b + \tfrac{1}{2}) \hspace{1pt} \mathcal{O}_{c,\hspace{1pt} \overbar{h}_a +\overbar{h}_b + p}^{I_1 \cdots I_A K J_1 \cdots J_B}(0,0)\hspace{2pt}.
    \end{split}
\end{equation}
The claimed equality follows from the identity $C_{p}(\hspace{1pt}\overbar{h}_a + \frac{1}{2}, \overbar{h}_b) + C_{p}(\hspace{1pt}\overbar{h}_a, \overbar{h}_b + \frac{1}{2}) = C_{p}(\hspace{1pt}\overbar{h}_a, \overbar{h}_b)$ \cite{Himwich:2021dau}.

Because $Q_{K,1} = [L_1,Q_{K,0}]$ and we know that both $L_1$ and $Q_{K,0}$ are symmetries of the OPE, it follows that the OPE will transform covariantly under $Q_{K,1}$. A similar analysis applies to $\overbar{Q}^K_{1}$.

\subsection{SUSY Jacobi Identity for Chiral Soft Algebras}
\label{appendix: Chiral soft algebra covariance subsection}
In this section, we verify that the chiral soft algebra derived in Section \ref{sec: SUSY Soft Algebra expressions}
\begin{equation}
    \begin{split}
        \Big[\mathcal{R}^{I_1 \cdots I_A}_{a,\hspace{1pt} q_1,\overbar{m}_1,n_1}, \mathcal{R}^{J_1 \cdots J_B}_{b,\hspace{1pt} q_2,\hspace{1pt} \overbar{m}_2,n_2}\Big] 
        = -(-1)^{B|a|} \sumop_{c} \hspace{2pt} g_{ab}^c \hspace{2pt} N_p(q_1,q_2,\overbar{m}_1,\overbar{m}_2) \hspace{2pt} \mathcal{R}^{I_1 \cdots I_A J_1 \cdots J_B}_{c,\hspace{1pt} q_1 + q_2 - p - 1, \hspace{1pt} \overbar{m}_1 + \overbar{m}_2,\hspace{1pt} n_1 + n_2}
    \end{split}
\end{equation}
is covariant under the action of bulk supersymmetry in the sense that the following Jacobi identity (and a similar one with $\overbar{Q}^K_{\overbar{\alpha}}$ replacing $Q_{K,\alpha}$) is respected
\begin{equation}
    \begin{split}
        \big[Q_{K,\alpha},\big[\mathcal{R}^{I_1 \cdots I_A}_{a,\hspace{1pt} q_1,\overbar{m}_1,n_1}, \mathcal{R}^{J_1 \cdots J_B}_{b,\hspace{1pt} q_2,\hspace{1pt} \overbar{m}_2,n_2}\big]\big] &= \big[\big[Q_{K,\alpha},\mathcal{R}^{I_1 \cdots I_A}_{a,\hspace{1pt} q_1,\overbar{m}_1,n_1}\big], \mathcal{R}^{J_1 \cdots J_B}_{b,\hspace{1pt} q_2,\hspace{1pt} \overbar{m}_2,n_2}\big]\big] \\[.3em]
        &\hspace{40pt} + (-1)^{|a^{I_1 \cdots I_A}|} \big[\mathcal{R}^{I_1 \cdots I_A}_{a,\hspace{1pt} q_1,\overbar{m}_1,n_1}, \big[Q_{K,\alpha}, \mathcal{R}^{J_1 \cdots J_B}_{b,\hspace{1pt} q_2,\hspace{1pt} \overbar{m}_2,n_2}\big]\big]\hspace{2pt}.
    \end{split}
\end{equation}
Again, the trick is to verify the relations for $Q_{K,0}$ and $\overbar{Q}^K_0$ and use the fact that $Q_{K,1} = [L_1,Q_{K,0}]$ and $\overbar{Q}_{K,1} = [\overbar{L}_1,\overbar{Q}_{K,0}]$ to infer that they also hold for $Q_{K,1}$ and $\overbar Q_{K,1}.$

The calculation for $Q_{K,0}$ proceeds as follows: 
\begin{equation}
    \begin{split}
        \text{LHS} &= -(-1)^{B|a|} \sumop_{c} \hspace{2pt} g_{ab}^c \hspace{2pt} N_p(q_1,q_2,\overbar{m}_1,\overbar{m}_2) \hspace{2pt} \big[Q_{K,0}, \mathcal{R}^{I_1 \cdots I_A J_1 \cdots J_B}_{c,\hspace{1pt} q_1 + q_2 - p - 1, \hspace{1pt} \overbar{m}_1 + \overbar{m}_2,\hspace{1pt} n_1 + n_2} \big] \\
        &= (-1)^{B|a|} \sumop_{c} \hspace{2pt} g_{ab}^c \hspace{2pt} N_p(q_1,q_2,\overbar{m}_1,\overbar{m}_2) \hspace{2pt} \delta_K^{[I_1} \mathcal{R}^{I_2 \cdots I_A J_1 \cdots J_B]}_{c,\hspace{1pt} q_1 + q_2 - p - 1, \hspace{1pt} \overbar{m}_1 + \overbar{m}_2,\hspace{1pt} n_1 + n_2 - \frac{1}{2}} \\[.75em]
        \text{RHS} &= - \big[\delta_K^{[I_1}\mathcal{R}^{I_2 \cdots I_A]}_{a,\hspace{1pt} q_1,\overbar{m}_1,n_1 - \frac{1}{2}}, \mathcal{R}^{J_1 \cdots J_B}_{b,\hspace{1pt} q_2,\hspace{1pt} \overbar{m}_2,n_2}\big] \\[.2em]
        &\hspace{24pt} - (-1)^{|a^{I_1 \cdots I_A}|} \big[\mathcal{R}^{I_1 \cdots I_A}_{a,\hspace{1pt} q_1,\overbar{m}_1,n_1}, \delta_{K}^{[J_1} \mathcal{R}^{J_2 \cdots J_B]}_{b,\hspace{1pt} q_2,\hspace{1pt} \overbar{m}_2,n_2 - \frac{1}{2}}\big] \\
        &= (-1)^{B|a|} \sumop_{c} \hspace{2pt} g_{ab}^c \hspace{2pt} N_p(q_1,q_2,\overbar{m}_1,\overbar{m}_2) \hspace{2pt} \delta^{[I_1}_{K} \mathcal{R}^{I_2 \cdots I_A] J_1 \cdots J_B}_{c,\hspace{1pt} q_1 + q_2 - p - 1, \hspace{1pt} \overbar{m}_1 + \overbar{m}_2,\hspace{1pt} n_1 + n_2 - \frac{1}{2}} \\
        &\hspace{24pt} +(-1)^{B|a| + A} \sumop_{c} \hspace{2pt} g_{ab}^c \hspace{2pt} N_p(q_1,q_2,\overbar{m}_1,\overbar{m}_2) \hspace{2pt} \delta^{J_1]}_{K} \mathcal{R}^{I_1 \cdots I_A [J_2 \cdots J_B}_{c,\hspace{1pt} q_1 + q_2 - p - 1, \hspace{1pt} \overbar{m}_1 + \overbar{m}_2,\hspace{1pt} n_1 + n_2 - \frac{1}{2}}\hspace{2pt},
    \end{split}
\end{equation}
while the calculation for $\overbar{Q}^K_0$ is
\begin{align}
    \text{LHS} &= -(-1)^{B|a|} \sumop_{c} \hspace{2pt} g_{ab}^c \hspace{2pt} N_p(q_1,q_2,\overbar{m}_1,\overbar{m}_2) \hspace{2pt} \big[\overbar{Q}^K_{0}, \mathcal{R}^{I_1 \cdots I_A J_1 \cdots J_B}_{c,\hspace{1pt} q_1 + q_2 - p - 1, \hspace{1pt} \overbar{m}_1 + \overbar{m}_2,\hspace{1pt} n_1 + n_2} \big] \nonumber\\
    &= -(-1)^{B|a|} \sumop_{c} \hspace{2pt} g_{ab}^c \hspace{2pt} N_p(q_1,q_2,\overbar{m}_1,\overbar{m}_2) \hspace{2pt} (\overbar{m}_1 + \overbar{m}_2 + q_1 + q_2 - p - 2) \mathcal{R}^{K I_1 \cdots I_A J_1 \cdots J_B}_{c,\hspace{1pt} q_1 + q_2 - p - \frac{3}{2}, \hspace{1pt} \overbar{m}_1 + \overbar{m}_2 - \frac{1}{2},\hspace{1pt} n_1 + n_2} \nonumber\\[.75em]
    \text{RHS} &= (\overbar{m}_1 + q_1 - 1) \hspace{2pt} \big[\mathcal{R}^{K I_1 \cdots I_A}_{a,\hspace{1pt} q_1-\frac{1}{2},\overbar{m}_1-\frac{1}{2},n_1}, \mathcal{R}^{J_1 \cdots J_B}_{b,\hspace{1pt} q_2,\hspace{1pt} \overbar{m}_2,n_2}\big] \nonumber \\[.2em]
    &\hspace{24pt} + (-1)^{|a^{I_1 \cdots I_A}|} (\overbar{m}_2 + q_2 - 1) \hspace{2pt} \big[\mathcal{R}^{I_1 \cdots I_A}_{a,\hspace{1pt} q_1,\overbar{m}_1,n_1}, \mathcal{R}^{K J_1 \cdots J_B}_{b,\hspace{1pt} q_2-\frac{1}{2},\hspace{1pt} \overbar{m}_2-\frac{1}{2},n_2}\big] \\
    &= -(-1)^{B|a|} \sumop_{c} \hspace{2pt} g_{ab}^c \hspace{2pt} N_p(q_1-\tfrac{1}{2},q_2,\overbar{m}_1-\tfrac{1}{2},\overbar{m}_2) \hspace{2pt} (\overbar{m}_1 + q_1 - 1)\hspace{2pt} \mathcal{R}^{K I_1 \cdots I_A J_1 \cdots J_B}_{c,\hspace{1pt} q_1 + q_2 - p - \frac{3}{2}, \hspace{1pt} \overbar{m}_1 + \overbar{m}_2 - \frac{1}{2},\hspace{1pt} n_1 + n_2} \nonumber \\
    &\hspace{24pt} -(-1)^{B|a| + A} \sumop_{c} \hspace{2pt} g_{ab}^c \hspace{2pt} N_p(q_1,q_2-\tfrac{1}{2},\overbar{m}_1,\overbar{m}_2-\tfrac{1}{2}) \hspace{2pt} (\overbar{m}_2 + q_2 - 1)\hspace{2pt} \mathcal{R}^{I_1 \cdots I_A K J_1 \cdots J_B}_{c,\hspace{1pt} q_1 + q_2 - p - \frac{3}{2}, \hspace{1pt} \overbar{m}_1 + \overbar{m}_2 - \frac{1}{2},\hspace{1pt} n_1 + n_2}\hspace{2pt}. \nonumber
\end{align}
The claimed equality follows from the identity
\begin{equation}
    \begin{split}
         (\overbar{m}_1 + q_1 - 1) N_p(q_1-\tfrac{1}{2},&q_2,\overbar{m}_1-\tfrac{1}{2},\overbar{m}_2) + (\overbar{m}_2 + q_2 - 1)\hspace{2pt} N_p(q_1,q_2-\tfrac{1}{2},\overbar{m}_1,\overbar{m}_2-\tfrac{1}{2}) \\[.3em]
         &= (\overbar{m}_1 + \overbar{m}_2 + q_1 + q_2 - p - 2 )N_p(q_1,q_2,\overbar{m}_1,\overbar{m}_2)\hspace{2pt}. 
    \end{split}
\end{equation}

\bibliography{bib.bib}

\begin{thebibliography}{10}%
\makeatletter
\providecommand \@ifxundefined [1]{%
 \ifx #1\undefined \expandafter \@firstoftwo
 \else \expandafter \@secondoftwo
\fi
}%
\providecommand \@ifnum [1]{%
 \ifnum #1\expandafter \@firstoftwo
 \else \expandafter \@secondoftwo
\fi
}%
\providecommand \enquote [1]{``#1''}%
\providecommand \bibnamefont  [1]{#1}%
\providecommand \bibfnamefont [1]{#1}%
\providecommand \citenamefont [1]{#1}%
\providecommand\href[0]{\@sanitize\@href}%
\providecommand\@href[1]{\endgroup\@@startlink{#1}\endgroup\@@href}%
\providecommand\@@href[1]{#1\@@endlink}%
\providecommand \@sanitize [0]{\begingroup\catcode`\&12\catcode`\#12\relax}%
\@ifxundefined \pdfoutput {\@firstoftwo}{%
 \@ifnum{\z@=\pdfoutput}{\@firstoftwo}{\@secondoftwo}%
}{%
 \providecommand\@@startlink[1]{\leavevmode\special{html:<a href="#1">}}%
 \providecommand\@@endlink[0]{\special{html:</a>}}%
}{%
 \providecommand\@@startlink[1]{%
  \leavevmode
  \pdfstartlink
   attr{/Border[0 0 1 ]/H/I/C[0 1 1]}%
   user{/Subtype/Link/A<</Type/Action/S/URI/URI(#1)>>}%
  \relax
 }%
 \providecommand\@@endlink[0]{\pdfendlink}%
}%
\providecommand \url  [0]{\begingroup\@sanitize \@url }%
\providecommand \@url [1]{\endgroup\@href {#1}{\urlprefix}}%
\providecommand \urlprefix [0]{URL }%
\providecommand \Eprint[0]{\href }%
\@ifxundefined \urlstyle {%
  \providecommand \doi [1]{doi:\discretionary{}{}{}#1}%
}{%
  \providecommand \doi [0]{doi:\discretionary{}{}{}\begingroup \urlstyle{rm}\Url }%
}%
\providecommand \doibase [0]{http://dx.doi.org/}%
\providecommand \Doi[1]{\href{\doibase#1}}%
\providecommand \bibAnnote [3]{%
  \BibitemShut{#1}%
  \begin{quotation}\noindent
    \textsc{Key:}\ #2\\\textsc{Annotation:}\ #3%
  \end{quotation}%
}%
\providecommand \bibAnnoteFile [2]{%
  \IfFileExists{#2}{\bibAnnote {#1} {#2} {\input{#2}}}{}%
}%
\providecommand \typeout [0]{\immediate \write \m@ne }%
\providecommand \selectlanguage [0]{\@gobble}%
\providecommand \bibinfo [0]{\@secondoftwo}%
\providecommand \bibfield [0]{\@secondoftwo}%
\providecommand \translation [1]{[#1]}%
\providecommand \BibitemOpen[0]{}%
\providecommand \bibitemStop [0]{}%
\providecommand \bibitemNoStop [0]{.\EOS\space}%
\providecommand \EOS [0]{\spacefactor3000\relax}%
\providecommand \BibitemShut [1]{\csname bibitem#1\endcsname}%
\bibitem{Pasterski:2021rjz}%
  \BibitemOpen
  \bibfield{author}{%
  \bibinfo {author} {\bibfnamefont{Sabrina}\ \bibnamefont{Pasterski}},\ }%
  \bibfield{title}{%
  \enquote{\bibinfo {title} {{Lectures on celestial amplitudes}},}\ }%
  \bibfield{journal}{%
  \Doi{10.1140/epjc/s10052-021-09846-7}{\bibinfo {journal} {Eur. Phys. J. C}}\ }%
  \textbf{\bibinfo {volume} {81}},\ \bibinfo {pages} {1062} (\bibinfo {year} {2021}),\ \Eprint{http://arxiv.org/abs/2108.04801}{arXiv:2108.04801 [hep-th]}%
  \bibAnnoteFile{NoStop}{Pasterski:2021rjz}%
\bibitem{Raclariu:2021zjz}%
  \BibitemOpen
  \bibfield{author}{%
  \bibinfo {author} {\bibfnamefont{Ana-Maria}\ \bibnamefont{Raclariu}},\ }%
  \bibfield{title}{%
  \enquote{\bibinfo {title} {{Lectures on Celestial Holography}},}\ }%
   (\bibinfo {month} {7}\ \bibinfo {year} {2021}),\ \Eprint{http://arxiv.org/abs/2107.02075}{arXiv:2107.02075 [hep-th]}%
  \bibAnnoteFile{NoStop}{Raclariu:2021zjz}%
\bibitem{Donnay:2023mrd}%
  \BibitemOpen
  \bibfield{author}{%
  \bibinfo {author} {\bibfnamefont{Laura}\ \bibnamefont{Donnay}},\ }%
  \bibfield{title}{%
  \enquote{\bibinfo {title} {{Celestial holography: An asymptotic symmetry perspective}},}\ }%
  \bibfield{journal}{%
  \Doi{10.1016/j.physrep.2024.04.003}{\bibinfo {journal} {Phys. Rept.}}\ }%
  \textbf{\bibinfo {volume} {1073}},\ \bibinfo {pages} {1--41} (\bibinfo {year} {2024}),\ \Eprint{http://arxiv.org/abs/2310.12922}{arXiv:2310.12922 [hep-th]}%
  \bibAnnoteFile{NoStop}{Donnay:2023mrd}%
\bibitem{Strominger:2017zoo}%
  \BibitemOpen
  \bibfield{author}{%
  \bibinfo {author} {\bibfnamefont{Andrew}\ \bibnamefont{Strominger}},\ }%
  \emph{\bibinfo {title} {{Lectures on the Infrared Structure of Gravity and Gauge Theory}}}\ (\bibinfo {year} {2017})\ ISBN \bibinfo {isbn} {978-0-691-17973-5},\ \Eprint{http://arxiv.org/abs/1703.05448}{arXiv:1703.05448 [hep-th]}%
  \bibAnnoteFile{NoStop}{Strominger:2017zoo}%
\bibitem{seiberg1994electric}%
  \BibitemOpen
  \bibfield{author}{%
  \bibinfo {author} {\bibfnamefont{Nathan}\ \bibnamefont{Seiberg}}\ and\ \bibinfo {author} {\bibfnamefont{Edward}\ \bibnamefont{Witten}},\ }%
  \bibfield{title}{%
  \enquote{\bibinfo {title} {{Electric-magnetic duality, monopole condensation, and confinement in N= 2 supersymmetric Yang-Mills theory}},}\ }%
  \bibfield{journal}{%
  \bibinfo {journal} {Nuclear Physics B}\ }%
  \textbf{\bibinfo {volume} {426}},\ \bibinfo {pages} {19--52} (\bibinfo {year} {1994})%
  \bibAnnoteFile{NoStop}{seiberg1994electric}%
\bibitem{Crawley:2024cak}%
  \BibitemOpen
  \bibfield{author}{%
  \bibinfo {author} {\bibfnamefont{Erin}\ \bibnamefont{Crawley}}, \bibinfo {author} {\bibfnamefont{Andrew}\ \bibnamefont{Strominger}},\ and\ \bibinfo {author} {\bibfnamefont{Adam}\ \bibnamefont{Tropper}},\ }%
  \bibfield{title}{%
  \enquote{\bibinfo {title} {{Chiral Soft Algebras for $\mathcal{N} = 2$ Gauge Theory}},}\ }%
   (\bibinfo {month} {7}\ \bibinfo {year} {2024}),\ \Eprint{http://arxiv.org/abs/2407.16752}{arXiv:2407.16752 [hep-th]}%
  \bibAnnoteFile{NoStop}{Crawley:2024cak}%
\bibitem{Ball:2023qim}%
  \BibitemOpen
  \bibfield{author}{%
  \bibinfo {author} {\bibfnamefont{Adam}\ \bibnamefont{Ball}}, \bibinfo {author} {\bibfnamefont{Marcus}\ \bibnamefont{Spradlin}}, \bibinfo {author} {\bibfnamefont{Akshay}\ \bibnamefont{Yelleshpur~Srikant}},\ and\ \bibinfo {author} {\bibfnamefont{Anastasia}\ \bibnamefont{Volovich}},\ }%
  \bibfield{title}{%
  \enquote{\bibinfo {title} {{Supersymmetry and the celestial Jacobi identity}},}\ }%
  \bibfield{journal}{%
  \Doi{10.1007/JHEP04(2024)099}{\bibinfo {journal} {JHEP}}\ }%
  \textbf{\bibinfo {volume} {04}},\ \bibinfo {pages} {099} (\bibinfo {year} {2024}),\ \Eprint{http://arxiv.org/abs/2311.01364}{arXiv:2311.01364 [hep-th]}%
  \bibAnnoteFile{NoStop}{Ball:2023qim}%
\bibitem{Ball:2023sdz}%
  \BibitemOpen
  \bibfield{author}{%
  \bibinfo {author} {\bibfnamefont{Adam}\ \bibnamefont{Ball}}, \bibinfo {author} {\bibfnamefont{Yangrui}\ \bibnamefont{Hu}},\ and\ \bibinfo {author} {\bibfnamefont{Sabrina}\ \bibnamefont{Pasterski}},\ }%
  \bibfield{title}{%
  \enquote{\bibinfo {title} {{Multicollinear singularities in celestial CFT}},}\ }%
  \bibfield{journal}{%
  \Doi{10.1007/JHEP02(2024)219}{\bibinfo {journal} {JHEP}}\ }%
  \textbf{\bibinfo {volume} {02}},\ \bibinfo {pages} {219} (\bibinfo {year} {2024}),\ \Eprint{http://arxiv.org/abs/2309.16602}{arXiv:2309.16602 [hep-th]}%
  \bibAnnoteFile{NoStop}{Ball:2023sdz}%
\bibitem{Dumitrescu:2015fej}%
  \BibitemOpen
  \bibfield{author}{%
  \bibinfo {author} {\bibfnamefont{Thomas~T.}\ \bibnamefont{Dumitrescu}}, \bibinfo {author} {\bibfnamefont{Temple}\ \bibnamefont{He}}, \bibinfo {author} {\bibfnamefont{Prahar}\ \bibnamefont{Mitra}},\ and\ \bibinfo {author} {\bibfnamefont{Andrew}\ \bibnamefont{Strominger}},\ }%
  \bibfield{title}{%
  \enquote{\bibinfo {title} {{Infinite-dimensional fermionic symmetry in supersymmetric gauge theories}},}\ }%
  \bibfield{journal}{%
  \Doi{10.1007/JHEP08(2021)051}{\bibinfo {journal} {JHEP}}\ }%
  \textbf{\bibinfo {volume} {08}},\ \bibinfo {pages} {051} (\bibinfo {year} {2021}),\ \Eprint{http://arxiv.org/abs/1511.07429}{arXiv:1511.07429 [hep-th]}%
  \bibAnnoteFile{NoStop}{Dumitrescu:2015fej}%
\bibitem{Lysov:2015jrs}%
  \BibitemOpen
  \bibfield{author}{%
  \bibinfo {author} {\bibfnamefont{Vyacheslav}\ \bibnamefont{Lysov}},\ }%
  \bibfield{title}{%
  \enquote{\bibinfo {title} {{Asymptotic Fermionic Symmetry From Soft Gravitino Theorem}},}\ }%
   (\bibinfo {month} {12}\ \bibinfo {year} {2015}),\ \Eprint{http://arxiv.org/abs/1512.03015}{arXiv:1512.03015 [hep-th]}%
  \bibAnnoteFile{NoStop}{Lysov:2015jrs}%
\bibitem{Avery:2015iix}%
  \BibitemOpen
  \bibfield{author}{%
  \bibinfo {author} {\bibfnamefont{Steven~G.}\ \bibnamefont{Avery}}\ and\ \bibinfo {author} {\bibfnamefont{Burkhard U.~W.}\ \bibnamefont{Schwab}},\ }%
  \bibfield{title}{%
  \enquote{\bibinfo {title} {{Residual Local Supersymmetry and the Soft Gravitino}},}\ }%
  \bibfield{journal}{%
  \Doi{10.1103/PhysRevLett.116.171601}{\bibinfo {journal} {Phys. Rev. Lett.}}\ }%
  \textbf{\bibinfo {volume} {116}},\ \bibinfo {pages} {171601} (\bibinfo {year} {2016}),\ \Eprint{http://arxiv.org/abs/1512.02657}{arXiv:1512.02657 [hep-th]}%
  \bibAnnoteFile{NoStop}{Avery:2015iix}%
\bibitem{Pasterski:2021fjn}%
  \BibitemOpen
  \bibfield{author}{%
  \bibinfo {author} {\bibfnamefont{Sabrina}\ \bibnamefont{Pasterski}}, \bibinfo {author} {\bibfnamefont{Andrea}\ \bibnamefont{Puhm}},\ and\ \bibinfo {author} {\bibfnamefont{Emilio}\ \bibnamefont{Trevisani}},\ }%
  \bibfield{title}{%
  \enquote{\bibinfo {title} {{Celestial diamonds: conformal multiplets in celestial CFT}},}\ }%
  \bibfield{journal}{%
  \Doi{10.1007/JHEP11(2021)072}{\bibinfo {journal} {JHEP}}\ }%
  \textbf{\bibinfo {volume} {11}},\ \bibinfo {pages} {072} (\bibinfo {year} {2021}),\ \Eprint{http://arxiv.org/abs/2105.03516}{arXiv:2105.03516 [hep-th]}%
  \bibAnnoteFile{NoStop}{Pasterski:2021fjn}%
\bibitem{Pano:2021ewd}%
  \BibitemOpen
  \bibfield{author}{%
  \bibinfo {author} {\bibfnamefont{Yorgo}\ \bibnamefont{Pano}}, \bibinfo {author} {\bibfnamefont{Sabrina}\ \bibnamefont{Pasterski}},\ and\ \bibinfo {author} {\bibfnamefont{Andrea}\ \bibnamefont{Puhm}},\ }%
  \bibfield{title}{%
  \enquote{\bibinfo {title} {{Conformally soft fermions}},}\ }%
  \bibfield{journal}{%
  \Doi{10.1007/JHEP12(2021)166}{\bibinfo {journal} {JHEP}}\ }%
  \textbf{\bibinfo {volume} {12}},\ \bibinfo {pages} {166} (\bibinfo {year} {2021}),\ \Eprint{http://arxiv.org/abs/2108.11422}{arXiv:2108.11422 [hep-th]}%
  \bibAnnoteFile{NoStop}{Pano:2021ewd}%
\bibitem{Agriela:2023dnw}%
  \BibitemOpen
  \bibfield{author}{%
  \bibinfo {author} {\bibfnamefont{Adri\'an}\ \bibnamefont{Agriela}}\ and\ \bibinfo {author} {\bibfnamefont{Miguel}\ \bibnamefont{Campiglia}},\ }%
  \bibfield{title}{%
  \enquote{\bibinfo {title} {{Fermionic asymptotic symmetries in massless QED}},}\ }%
  \bibfield{journal}{%
  \Doi{10.1103/PhysRevD.108.065011}{\bibinfo {journal} {Phys. Rev. D}}\ }%
  \textbf{\bibinfo {volume} {108}},\ \bibinfo {pages} {065011} (\bibinfo {year} {2023}),\ \Eprint{http://arxiv.org/abs/2307.11171}{arXiv:2307.11171 [hep-th]}%
  \bibAnnoteFile{NoStop}{Agriela:2023dnw}%
\bibitem{Narayanan:2020amh}%
  \BibitemOpen
  \bibfield{author}{%
  \bibinfo {author} {\bibfnamefont{Sruthi~A.}\ \bibnamefont{Narayanan}},\ }%
  \bibfield{title}{%
  \enquote{\bibinfo {title} {{Massive Celestial Fermions}},}\ }%
  \bibfield{journal}{%
  \Doi{10.1007/JHEP12(2020)074}{\bibinfo {journal} {JHEP}}\ }%
  \textbf{\bibinfo {volume} {12}},\ \bibinfo {pages} {074} (\bibinfo {year} {2020}),\ \Eprint{http://arxiv.org/abs/2009.03883}{arXiv:2009.03883 [hep-th]}%
  \bibAnnoteFile{NoStop}{Narayanan:2020amh}%
\bibitem{Tropper:2024kxy}%
  \BibitemOpen
  \bibfield{author}{%
  \bibinfo {author} {\bibfnamefont{Adam}\ \bibnamefont{Tropper}},\ }%
  \bibfield{title}{%
  \enquote{\bibinfo {title} {{Supersymmetric Soft Theorems}},}\ }%
   (\bibinfo {month} {4}\ \bibinfo {year} {2024}),\ \Eprint{http://arxiv.org/abs/2404.03717}{arXiv:2404.03717 [hep-th]}%
  \bibAnnoteFile{NoStop}{Tropper:2024kxy}%
\bibitem{Strominger:2021mtt}%
  \BibitemOpen
  \bibfield{author}{%
  \bibinfo {author} {\bibfnamefont{Andrew}\ \bibnamefont{Strominger}},\ }%
  \bibfield{title}{%
  \enquote{\bibinfo {title} {{$w_{1+\infty}$ Algebra and the Celestial Sphere: Infinite Towers of Soft Graviton, Photon, and Gluon Symmetries}},}\ }%
  \bibfield{journal}{%
  \Doi{10.1103/PhysRevLett.127.221601}{\bibinfo {journal} {Phys. Rev. Lett.}}\ }%
  \textbf{\bibinfo {volume} {127}},\ \bibinfo {pages} {221601} (\bibinfo {year} {2021}),\ \Eprint{http://arxiv.org/abs/2105.14346}{arXiv:2105.14346 [hep-th]}%
  \bibAnnoteFile{NoStop}{Strominger:2021mtt}%
\bibitem{Taylor:2023ajd}%
  \BibitemOpen
  \bibfield{author}{%
  \bibinfo {author} {\bibfnamefont{Tomasz~R.}\ \bibnamefont{Taylor}}\ and\ \bibinfo {author} {\bibfnamefont{Bin}\ \bibnamefont{Zhu}},\ }%
  \bibfield{title}{%
  \enquote{\bibinfo {title} {{w1+\ensuremath{\infty} Algebra with a Cosmological Constant and the Celestial Sphere}},}\ }%
  \bibfield{journal}{%
  \Doi{10.1103/PhysRevLett.132.221602}{\bibinfo {journal} {Phys. Rev. Lett.}}\ }%
  \textbf{\bibinfo {volume} {132}},\ \bibinfo {pages} {221602} (\bibinfo {year} {2024}),\ \Eprint{http://arxiv.org/abs/2312.00876}{arXiv:2312.00876 [hep-th]}%
  \bibAnnoteFile{NoStop}{Taylor:2023ajd}%
\bibitem{Bittleston:2024rqe}%
  \BibitemOpen
  \bibfield{author}{%
  \bibinfo {author} {\bibfnamefont{Roland}\ \bibnamefont{Bittleston}}, \bibinfo {author} {\bibfnamefont{Giuseppe}\ \bibnamefont{Bogna}}, \bibinfo {author} {\bibfnamefont{Simon}\ \bibnamefont{Heuveline}}, \bibinfo {author} {\bibfnamefont{Adam}\ \bibnamefont{Kmec}}, \bibinfo {author} {\bibfnamefont{Lionel}\ \bibnamefont{Mason}},\ and\ \bibinfo {author} {\bibfnamefont{David}\ \bibnamefont{Skinner}},\ }%
  \bibfield{title}{%
  \enquote{\bibinfo {title} {{On AdS$_{4}$ deformations of celestial symmetries}},}\ }%
  \bibfield{journal}{%
  \Doi{10.1007/JHEP07(2024)010}{\bibinfo {journal} {JHEP}}\ }%
  \textbf{\bibinfo {volume} {07}},\ \bibinfo {pages} {010} (\bibinfo {year} {2024}),\ \Eprint{http://arxiv.org/abs/2403.18011}{arXiv:2403.18011 [hep-th]}%
  \bibAnnoteFile{NoStop}{Bittleston:2024rqe}%
\bibitem{Nair:1988bq}%
  \BibitemOpen
  \bibfield{author}{%
  \bibinfo {author} {\bibfnamefont{V.~P.}\ \bibnamefont{Nair}},\ }%
  \bibfield{title}{%
  \enquote{\bibinfo {title} {{A Current Algebra for Some Gauge Theory Amplitudes}},}\ }%
  \bibfield{journal}{%
  \Doi{10.1016/0370-2693(88)91471-2}{\bibinfo {journal} {Phys. Lett. B}}\ }%
  \textbf{\bibinfo {volume} {214}},\ \bibinfo {pages} {215--218} (\bibinfo {year} {1988})%
  \bibAnnoteFile{NoStop}{Nair:1988bq}%
\bibitem{Elvang:2011fx}%
  \BibitemOpen
  \bibfield{author}{%
  \bibinfo {author} {\bibfnamefont{Henriette}\ \bibnamefont{Elvang}}, \bibinfo {author} {\bibfnamefont{Yu-tin}\ \bibnamefont{Huang}},\ and\ \bibinfo {author} {\bibfnamefont{Cheng}\ \bibnamefont{Peng}},\ }%
  \bibfield{title}{%
  \enquote{\bibinfo {title} {{On-shell superamplitudes in N\ensuremath{<}4 SYM}},}\ }%
  \bibfield{journal}{%
  \Doi{10.1007/JHEP09(2011)031}{\bibinfo {journal} {JHEP}}\ }%
  \textbf{\bibinfo {volume} {09}},\ \bibinfo {pages} {031} (\bibinfo {year} {2011}),\ \Eprint{http://arxiv.org/abs/1102.4843}{arXiv:1102.4843 [hep-th]}%
  \bibAnnoteFile{NoStop}{Elvang:2011fx}%
\bibitem{Elvang:2013cua}%
  \BibitemOpen
  \bibfield{author}{%
  \bibinfo {author} {\bibfnamefont{Henriette}\ \bibnamefont{Elvang}}\ and\ \bibinfo {author} {\bibfnamefont{Yu-tin}\ \bibnamefont{Huang}},\ }%
  \bibfield{title}{%
  \enquote{\bibinfo {title} {{Scattering Amplitudes}},}\ }%
   (\bibinfo {month} {8}\ \bibinfo {year} {2013}),\ \Eprint{http://arxiv.org/abs/1308.1697}{arXiv:1308.1697 [hep-th]}%
  \bibAnnoteFile{NoStop}{Elvang:2013cua}%
\bibitem{mol2024ads3dualsupersymmetricmhv}%
  \BibitemOpen
  \bibfield{author}{%
  \bibinfo {author} {\bibfnamefont{Igor}\ \bibnamefont{Mol}},\ }%
  \bibfield{title}{%
  \enquote{\bibinfo {title} {An $ads_{3}$ dual for supersymmetric mhv celestial amplitudes},}\ }%
   (\bibinfo {year} {2024}),\ \Eprint{http://arxiv.org/abs/2411.14311}{arXiv:2411.14311 [hep-th]},\ \url{https://arxiv.org/abs/2411.14311}%
  \bibAnnoteFile{NoStop}{mol2024ads3dualsupersymmetricmhv}%
\bibitem{Jiang:2021xzy}%
  \BibitemOpen
  \bibfield{author}{%
  \bibinfo {author} {\bibfnamefont{Hongliang}\ \bibnamefont{Jiang}},\ }%
  \bibfield{title}{%
  \enquote{\bibinfo {title} {{Celestial superamplitude in $ \mathcal{N} $ = 4 SYM theory}},}\ }%
  \bibfield{journal}{%
  \Doi{10.1007/JHEP08(2021)031}{\bibinfo {journal} {JHEP}}\ }%
  \textbf{\bibinfo {volume} {08}},\ \bibinfo {pages} {031} (\bibinfo {year} {2021}),\ \Eprint{http://arxiv.org/abs/2105.10269}{arXiv:2105.10269 [hep-th]}%
  \bibAnnoteFile{NoStop}{Jiang:2021xzy}%
\bibitem{Brandhuber:2021nez}%
  \BibitemOpen
  \bibfield{author}{%
  \bibinfo {author} {\bibfnamefont{Andreas}\ \bibnamefont{Brandhuber}}, \bibinfo {author} {\bibfnamefont{Graham~R.}\ \bibnamefont{Brown}}, \bibinfo {author} {\bibfnamefont{Joshua}\ \bibnamefont{Gowdy}}, \bibinfo {author} {\bibfnamefont{Bill}\ \bibnamefont{Spence}},\ and\ \bibinfo {author} {\bibfnamefont{Gabriele}\ \bibnamefont{Travaglini}},\ }%
  \bibfield{title}{%
  \enquote{\bibinfo {title} {{Celestial superamplitudes}},}\ }%
  \bibfield{journal}{%
  \Doi{10.1103/PhysRevD.104.045016}{\bibinfo {journal} {Phys. Rev. D}}\ }%
  \textbf{\bibinfo {volume} {104}},\ \bibinfo {pages} {045016} (\bibinfo {year} {2021}),\ \Eprint{http://arxiv.org/abs/2105.10263}{arXiv:2105.10263 [hep-th]}%
  \bibAnnoteFile{NoStop}{Brandhuber:2021nez}%
\bibitem{Bondi:1962px}%
  \BibitemOpen
  \bibfield{author}{%
  \bibinfo {author} {\bibfnamefont{H.}~\bibnamefont{Bondi}}, \bibinfo {author} {\bibfnamefont{M.~G.~J.}\ \bibnamefont{van~der Burg}},\ and\ \bibinfo {author} {\bibfnamefont{A.~W.~K.}\ \bibnamefont{Metzner}},\ }%
  \bibfield{title}{%
  \enquote{\bibinfo {title} {{Gravitational waves in general relativity. 7. Waves from axisymmetric isolated systems}},}\ }%
  \bibfield{journal}{%
  \Doi{10.1098/rspa.1962.0161}{\bibinfo {journal} {Proc. Roy. Soc. Lond. A}}\ }%
  \textbf{\bibinfo {volume} {269}},\ \bibinfo {pages} {21--52} (\bibinfo {year} {1962})%
  \bibAnnoteFile{NoStop}{Bondi:1962px}%
\bibitem{Sachs:1962wk}%
  \BibitemOpen
  \bibfield{author}{%
  \bibinfo {author} {\bibfnamefont{R.~K.}\ \bibnamefont{Sachs}},\ }%
  \bibfield{title}{%
  \enquote{\bibinfo {title} {{Gravitational waves in general relativity. 8. Waves in asymptotically flat space-times}},}\ }%
  \bibfield{journal}{%
  \Doi{10.1098/rspa.1962.0206}{\bibinfo {journal} {Proc. Roy. Soc. Lond. A}}\ }%
  \textbf{\bibinfo {volume} {270}},\ \bibinfo {pages} {103--126} (\bibinfo {year} {1962})%
  \bibAnnoteFile{NoStop}{Sachs:1962wk}%
\bibitem{Barnich:2009se}%
  \BibitemOpen
  \bibfield{author}{%
  \bibinfo {author} {\bibfnamefont{Glenn}\ \bibnamefont{Barnich}}\ and\ \bibinfo {author} {\bibfnamefont{Cedric}\ \bibnamefont{Troessaert}},\ }%
  \bibfield{title}{%
  \enquote{\bibinfo {title} {{Symmetries of asymptotically flat 4 dimensional spacetimes at null infinity revisited}},}\ }%
  \bibfield{journal}{%
  \Doi{10.1103/PhysRevLett.105.111103}{\bibinfo {journal} {Phys. Rev. Lett.}}\ }%
  \textbf{\bibinfo {volume} {105}},\ \bibinfo {pages} {111103} (\bibinfo {year} {2010}),\ \Eprint{http://arxiv.org/abs/0909.2617}{arXiv:0909.2617 [gr-qc]}%
  \bibAnnoteFile{NoStop}{Barnich:2009se}%
\bibitem{Awada:1985by}%
  \BibitemOpen
  \bibfield{author}{%
  \bibinfo {author} {\bibfnamefont{M.~A.}\ \bibnamefont{Awada}}, \bibinfo {author} {\bibfnamefont{G.~W.}\ \bibnamefont{Gibbons}},\ and\ \bibinfo {author} {\bibfnamefont{W.~T.}\ \bibnamefont{Shaw}},\ }%
  \bibfield{title}{%
  \enquote{\bibinfo {title} {{CONFORMAL SUPERGRAVITY, TWISTORS AND THE SUPER BMS GROUP}},}\ }%
  \bibfield{journal}{%
  \Doi{10.1016/S0003-4916(86)80023-9}{\bibinfo {journal} {Annals Phys.}}\ }%
  \textbf{\bibinfo {volume} {171}},\ \bibinfo {pages} {52} (\bibinfo {year} {1986})%
  \bibAnnoteFile{NoStop}{Awada:1985by}%
\bibitem{Fotopoulos:2020bqj}%
  \BibitemOpen
  \bibfield{author}{%
  \bibinfo {author} {\bibfnamefont{Angelos}\ \bibnamefont{Fotopoulos}}, \bibinfo {author} {\bibfnamefont{Stephan}\ \bibnamefont{Stieberger}}, \bibinfo {author} {\bibfnamefont{Tomasz~R.}\ \bibnamefont{Taylor}},\ and\ \bibinfo {author} {\bibfnamefont{Bin}\ \bibnamefont{Zhu}},\ }%
  \bibfield{title}{%
  \enquote{\bibinfo {title} {{Extended Super BMS Algebra of Celestial CFT}},}\ }%
  \bibfield{journal}{%
  \Doi{10.1007/JHEP09(2020)198}{\bibinfo {journal} {JHEP}}\ }%
  \textbf{\bibinfo {volume} {09}},\ \bibinfo {pages} {198} (\bibinfo {year} {2020}),\ \Eprint{http://arxiv.org/abs/2007.03785}{arXiv:2007.03785 [hep-th]}%
  \bibAnnoteFile{NoStop}{Fotopoulos:2020bqj}%
\bibitem{Banerjee:2022lnz}%
  \BibitemOpen
  \bibfield{author}{%
  \bibinfo {author} {\bibfnamefont{Nabamita}\ \bibnamefont{Banerjee}}, \bibinfo {author} {\bibfnamefont{Tabasum}\ \bibnamefont{Rahnuma}},\ and\ \bibinfo {author} {\bibfnamefont{Ranveer~Kumar}\ \bibnamefont{Singh}},\ }%
  \bibfield{title}{%
  \enquote{\bibinfo {title} {{Asymptotic symmetry algebra of N=8 supergravity}},}\ }%
  \bibfield{journal}{%
  \Doi{10.1103/PhysRevD.109.046010}{\bibinfo {journal} {Phys. Rev. D}}\ }%
  \textbf{\bibinfo {volume} {109}},\ \bibinfo {pages} {046010} (\bibinfo {year} {2024}),\ \Eprint{http://arxiv.org/abs/2212.12133}{arXiv:2212.12133 [hep-th]}%
  \bibAnnoteFile{NoStop}{Banerjee:2022lnz}%
\bibitem{Henneaux:2020ekh}%
  \BibitemOpen
  \bibfield{author}{%
  \bibinfo {author} {\bibfnamefont{Marc}\ \bibnamefont{Henneaux}}, \bibinfo {author} {\bibfnamefont{Javier}\ \bibnamefont{Matulich}},\ and\ \bibinfo {author} {\bibfnamefont{Turmoli}\ \bibnamefont{Neogi}},\ }%
  \bibfield{title}{%
  \enquote{\bibinfo {title} {{Asymptotic realization of the super-BMS algebra at spatial infinity}},}\ }%
  \bibfield{journal}{%
  \Doi{10.1103/PhysRevD.101.126016}{\bibinfo {journal} {Phys. Rev. D}}\ }%
  \textbf{\bibinfo {volume} {101}},\ \bibinfo {pages} {126016} (\bibinfo {year} {2020}),\ \Eprint{http://arxiv.org/abs/2004.07299}{arXiv:2004.07299 [hep-th]}%
  \bibAnnoteFile{NoStop}{Henneaux:2020ekh}%
\bibitem{Fuentealba:2020aax}%
  \BibitemOpen
  \bibfield{author}{%
  \bibinfo {author} {\bibfnamefont{Oscar}\ \bibnamefont{Fuentealba}}, \bibinfo {author} {\bibfnamefont{Marc}\ \bibnamefont{Henneaux}}, \bibinfo {author} {\bibfnamefont{Sucheta}\ \bibnamefont{Majumdar}}, \bibinfo {author} {\bibfnamefont{Javier}\ \bibnamefont{Matulich}},\ and\ \bibinfo {author} {\bibfnamefont{Turmoli}\ \bibnamefont{Neogi}},\ }%
  \bibfield{title}{%
  \enquote{\bibinfo {title} {{Asymptotic structure of the Rarita-Schwinger theory in four spacetime dimensions at spatial infinity}},}\ }%
  \bibfield{journal}{%
  \Doi{10.1007/JHEP02(2021)031}{\bibinfo {journal} {JHEP}}\ }%
  \textbf{\bibinfo {volume} {02}},\ \bibinfo {pages} {031} (\bibinfo {year} {2021}),\ \Eprint{http://arxiv.org/abs/2011.04669}{arXiv:2011.04669 [hep-th]}%
  \bibAnnoteFile{NoStop}{Fuentealba:2020aax}%
\bibitem{Fuentealba:2021xhn}%
  \BibitemOpen
  \bibfield{author}{%
  \bibinfo {author} {\bibfnamefont{Oscar}\ \bibnamefont{Fuentealba}}, \bibinfo {author} {\bibfnamefont{Marc}\ \bibnamefont{Henneaux}}, \bibinfo {author} {\bibfnamefont{Sucheta}\ \bibnamefont{Majumdar}}, \bibinfo {author} {\bibfnamefont{Javier}\ \bibnamefont{Matulich}},\ and\ \bibinfo {author} {\bibfnamefont{Turmoli}\ \bibnamefont{Neogi}},\ }%
  \bibfield{title}{%
  \enquote{\bibinfo {title} {{Local supersymmetry and the square roots of Bondi-Metzner-Sachs supertranslations}},}\ }%
  \bibfield{journal}{%
  \Doi{10.1103/PhysRevD.104.L121702}{\bibinfo {journal} {Phys. Rev. D}}\ }%
  \textbf{\bibinfo {volume} {104}},\ \bibinfo {pages} {L121702} (\bibinfo {year} {2021}),\ \Eprint{http://arxiv.org/abs/2108.07825}{arXiv:2108.07825 [hep-th]}%
  \bibAnnoteFile{NoStop}{Fuentealba:2021xhn}%
\bibitem{Boulanger:2023gpw}%
  \BibitemOpen
  \bibfield{author}{%
  \bibinfo {author} {\bibfnamefont{Nicolas}\ \bibnamefont{Boulanger}}, \bibinfo {author} {\bibfnamefont{Yannick}\ \bibnamefont{Herfray}},\ and\ \bibinfo {author} {\bibfnamefont{No\'emie}\ \bibnamefont{Parrini}},\ }%
  \bibfield{title}{%
  \enquote{\bibinfo {title} {{Conformal boundaries of Minkowski superspace and their super cuts}},}\ }%
  \bibfield{journal}{%
  \Doi{10.1007/JHEP02(2024)177}{\bibinfo {journal} {JHEP}}\ }%
  \textbf{\bibinfo {volume} {02}},\ \bibinfo {pages} {177} (\bibinfo {year} {2024}),\ \Eprint{http://arxiv.org/abs/2312.11222}{arXiv:2312.11222 [hep-th]}%
  \bibAnnoteFile{NoStop}{Boulanger:2023gpw}%
\bibitem{Prabhu:2021bod}%
  \BibitemOpen
  \bibfield{author}{%
  \bibinfo {author} {\bibfnamefont{Kartik}\ \bibnamefont{Prabhu}},\ }%
  \bibfield{title}{%
  \enquote{\bibinfo {title} {{Novel supersymmetric extension of BMS symmetries at null infinity}},}\ }%
  \bibfield{journal}{%
  \Doi{10.1103/PhysRevD.105.064054}{\bibinfo {journal} {Phys. Rev. D}}\ }%
  \textbf{\bibinfo {volume} {105}},\ \bibinfo {pages} {064054} (\bibinfo {year} {2022}),\ \Eprint{http://arxiv.org/abs/2112.07186}{arXiv:2112.07186 [gr-qc]}%
  \bibAnnoteFile{NoStop}{Prabhu:2021bod}%
\bibitem{Banks:2014iha}%
  \BibitemOpen
  \bibfield{author}{%
  \bibinfo {author} {\bibfnamefont{T.}~\bibnamefont{Banks}},\ }%
  \bibfield{title}{%
  \enquote{\bibinfo {title} {{The Super BMS Algebra, Scattering and Holography}},}\ }%
   (\bibinfo {month} {3}\ \bibinfo {year} {2014}),\ \Eprint{http://arxiv.org/abs/1403.3420}{arXiv:1403.3420 [hep-th]}%
  \bibAnnoteFile{NoStop}{Banks:2014iha}%
\bibitem{Banerjee:2022abf}%
  \BibitemOpen
  \bibfield{author}{%
  \bibinfo {author} {\bibfnamefont{Nabamita}\ \bibnamefont{Banerjee}}, \bibinfo {author} {\bibfnamefont{Arpita}\ \bibnamefont{Mitra}}, \bibinfo {author} {\bibfnamefont{Debangshu}\ \bibnamefont{Mukherjee}},\ and\ \bibinfo {author} {\bibfnamefont{H.~R.}\ \bibnamefont{Safari}},\ }%
  \bibfield{title}{%
  \enquote{\bibinfo {title} {{Supersymmetrization of deformed BMS algebras}},}\ }%
  \bibfield{journal}{%
  \Doi{10.1140/epjc/s10052-022-11036-y}{\bibinfo {journal} {Eur. Phys. J. C}}\ }%
  \textbf{\bibinfo {volume} {83}},\ \bibinfo {pages} {3} (\bibinfo {year} {2023}),\ \Eprint{http://arxiv.org/abs/2201.09853}{arXiv:2201.09853 [hep-th]}%
  \bibAnnoteFile{NoStop}{Banerjee:2022abf}%
\bibitem{Bagchi:2022owq}%
  \BibitemOpen
  \bibfield{author}{%
  \bibinfo {author} {\bibfnamefont{Arjun}\ \bibnamefont{Bagchi}}, \bibinfo {author} {\bibfnamefont{Daniel}\ \bibnamefont{Grumiller}},\ and\ \bibinfo {author} {\bibfnamefont{Poulami}\ \bibnamefont{Nandi}},\ }%
  \bibfield{title}{%
  \enquote{\bibinfo {title} {{Carrollian superconformal theories and super BMS}},}\ }%
  \bibfield{journal}{%
  \Doi{10.1007/JHEP05(2022)044}{\bibinfo {journal} {JHEP}}\ }%
  \textbf{\bibinfo {volume} {05}},\ \bibinfo {pages} {044} (\bibinfo {year} {2022}),\ \Eprint{http://arxiv.org/abs/2202.01172}{arXiv:2202.01172 [hep-th]}%
  \bibAnnoteFile{NoStop}{Bagchi:2022owq}%
\bibitem{Laddha:2017vfh}%
  \BibitemOpen
  \bibfield{author}{%
  \bibinfo {author} {\bibfnamefont{Alok}\ \bibnamefont{Laddha}}\ and\ \bibinfo {author} {\bibfnamefont{Prahar}\ \bibnamefont{Mitra}},\ }%
  \bibfield{title}{%
  \enquote{\bibinfo {title} {{Asymptotic Symmetries and Subleading Soft Photon Theorem in Effective Field Theories}},}\ }%
  \bibfield{journal}{%
  \Doi{10.1007/JHEP05(2018)132}{\bibinfo {journal} {JHEP}}\ }%
  \textbf{\bibinfo {volume} {05}},\ \bibinfo {pages} {132} (\bibinfo {year} {2018}),\ \Eprint{http://arxiv.org/abs/1709.03850}{arXiv:1709.03850 [hep-th]}%
  \bibAnnoteFile{NoStop}{Laddha:2017vfh}%
\bibitem{Fotopoulos:2019vac}%
  \BibitemOpen
  \bibfield{author}{%
  \bibinfo {author} {\bibfnamefont{Angelos}\ \bibnamefont{Fotopoulos}}, \bibinfo {author} {\bibfnamefont{Stephan}\ \bibnamefont{Stieberger}}, \bibinfo {author} {\bibfnamefont{Tomasz~R.}\ \bibnamefont{Taylor}},\ and\ \bibinfo {author} {\bibfnamefont{Bin}\ \bibnamefont{Zhu}},\ }%
  \bibfield{title}{%
  \enquote{\bibinfo {title} {{Extended BMS Algebra of Celestial CFT}},}\ }%
  \bibfield{journal}{%
  \Doi{10.1007/JHEP03(2020)130}{\bibinfo {journal} {JHEP}}\ }%
  \textbf{\bibinfo {volume} {03}},\ \bibinfo {pages} {130} (\bibinfo {year} {2020}),\ \Eprint{http://arxiv.org/abs/1912.10973}{arXiv:1912.10973 [hep-th]}%
  \bibAnnoteFile{NoStop}{Fotopoulos:2019vac}%
\bibitem{Ferber:1977qx}%
  \BibitemOpen
  \bibfield{author}{%
  \bibinfo {author} {\bibfnamefont{Alan}\ \bibnamefont{Ferber}},\ }%
  \bibfield{title}{%
  \enquote{\bibinfo {title} {{Supertwistors and Conformal Supersymmetry}},}\ }%
  \bibfield{journal}{%
  \Doi{10.1016/0550-3213(78)90257-2}{\bibinfo {journal} {Nucl. Phys. B}}\ }%
  \textbf{\bibinfo {volume} {132}},\ \bibinfo {pages} {55--64} (\bibinfo {year} {1978})%
  \bibAnnoteFile{NoStop}{Ferber:1977qx}%
\bibitem{He:2014cra}%
  \BibitemOpen
  \bibfield{author}{%
  \bibinfo {author} {\bibfnamefont{Temple}\ \bibnamefont{He}}, \bibinfo {author} {\bibfnamefont{Prahar}\ \bibnamefont{Mitra}}, \bibinfo {author} {\bibfnamefont{Achilleas~P.}\ \bibnamefont{Porfyriadis}},\ and\ \bibinfo {author} {\bibfnamefont{Andrew}\ \bibnamefont{Strominger}},\ }%
  \bibfield{title}{%
  \enquote{\bibinfo {title} {{New Symmetries of Massless QED}},}\ }%
  \bibfield{journal}{%
  \Doi{10.1007/JHEP10(2014)112}{\bibinfo {journal} {JHEP}}\ }%
  \textbf{\bibinfo {volume} {10}},\ \bibinfo {pages} {112} (\bibinfo {year} {2014}),\ \Eprint{http://arxiv.org/abs/1407.3789}{arXiv:1407.3789 [hep-th]}%
  \bibAnnoteFile{NoStop}{He:2014cra}%
\bibitem{Polchinski:1998rr}%
  \BibitemOpen
  \bibfield{author}{%
  \bibinfo {author} {\bibfnamefont{J.}~\bibnamefont{Polchinski}},\ }%
  \Doi{10.1017/CBO9780511618123}{\emph{\bibinfo {title} {{String theory. Vol. 2: Superstring theory and beyond}}}},\ Cambridge Monographs on Mathematical Physics\ (\bibinfo {publisher} {Cambridge University Press},\ \bibinfo {year} {2007})\ ISBN \bibinfo {isbn} {978-0-511-25228-0, 978-0-521-63304-8, 978-0-521-67228-3}%
  \bibAnnoteFile{NoStop}{Polchinski:1998rr}%
\bibitem{Marotta:2019cip}%
  \BibitemOpen
  \bibfield{author}{%
  \bibinfo {author} {\bibfnamefont{Raffaele}\ \bibnamefont{Marotta}}\ and\ \bibinfo {author} {\bibfnamefont{Mritunjay}\ \bibnamefont{Verma}},\ }%
  \bibfield{title}{%
  \enquote{\bibinfo {title} {{Soft Theorems from Compactification}},}\ }%
  \bibfield{journal}{%
  \Doi{10.1007/JHEP02(2020)008}{\bibinfo {journal} {JHEP}}\ }%
  \textbf{\bibinfo {volume} {02}},\ \bibinfo {pages} {008} (\bibinfo {year} {2020}),\ \Eprint{http://arxiv.org/abs/1911.05099}{arXiv:1911.05099 [hep-th]}%
  \bibAnnoteFile{NoStop}{Marotta:2019cip}%
\bibitem{Marotta:2021oiw}%
  \BibitemOpen
  \bibfield{author}{%
  \bibinfo {author} {\bibfnamefont{Raffaele}\ \bibnamefont{Marotta}}, \bibinfo {author} {\bibfnamefont{Massimo}\ \bibnamefont{Taronna}},\ and\ \bibinfo {author} {\bibfnamefont{Mritunjay}\ \bibnamefont{Verma}},\ }%
  \bibfield{title}{%
  \enquote{\bibinfo {title} {{Revisiting higher-spin gyromagnetic couplings}},}\ }%
  \bibfield{journal}{%
  \Doi{10.1007/JHEP06(2021)168}{\bibinfo {journal} {JHEP}}\ }%
  \textbf{\bibinfo {volume} {06}},\ \bibinfo {pages} {168} (\bibinfo {year} {2021}),\ \Eprint{http://arxiv.org/abs/2102.13180}{arXiv:2102.13180 [hep-th]}%
  \bibAnnoteFile{NoStop}{Marotta:2021oiw}%
\bibitem{Miller:2022fvc}%
  \BibitemOpen
  \bibfield{author}{%
  \bibinfo {author} {\bibfnamefont{Noah}\ \bibnamefont{Miller}}, \bibinfo {author} {\bibfnamefont{Andrew}\ \bibnamefont{Strominger}}, \bibinfo {author} {\bibfnamefont{Adam}\ \bibnamefont{Tropper}},\ and\ \bibinfo {author} {\bibfnamefont{Tianli}\ \bibnamefont{Wang}},\ }%
  \bibfield{title}{%
  \enquote{\bibinfo {title} {{Soft gravitons in the BFSS matrix model}},}\ }%
  \bibfield{journal}{%
  \Doi{10.1007/JHEP11(2023)174}{\bibinfo {journal} {JHEP}}\ }%
  \textbf{\bibinfo {volume} {11}},\ \bibinfo {pages} {174} (\bibinfo {year} {2023}),\ \Eprint{http://arxiv.org/abs/2208.14547}{arXiv:2208.14547 [hep-th]}%
  \bibAnnoteFile{NoStop}{Miller:2022fvc}%
\bibitem{Tropper:2023fjr}%
  \BibitemOpen
  \bibfield{author}{%
  \bibinfo {author} {\bibfnamefont{Adam}\ \bibnamefont{Tropper}}\ and\ \bibinfo {author} {\bibfnamefont{Tianli}\ \bibnamefont{Wang}},\ }%
  \bibfield{title}{%
  \enquote{\bibinfo {title} {{Lorentz symmetry and IR structure of the BFSS matrix model}},}\ }%
  \bibfield{journal}{%
  \Doi{10.1007/JHEP07(2023)150}{\bibinfo {journal} {JHEP}}\ }%
  \textbf{\bibinfo {volume} {07}},\ \bibinfo {pages} {150} (\bibinfo {year} {2023}),\ \Eprint{http://arxiv.org/abs/2303.14200}{arXiv:2303.14200 [hep-th]}%
  \bibAnnoteFile{NoStop}{Tropper:2023fjr}%
\bibitem{Banerjee:2021cly}%
  \BibitemOpen
  \bibfield{author}{%
  \bibinfo {author} {\bibfnamefont{Shamik}\ \bibnamefont{Banerjee}}, \bibinfo {author} {\bibfnamefont{Sudip}\ \bibnamefont{Ghosh}},\ and\ \bibinfo {author} {\bibfnamefont{Sai~Satyam}\ \bibnamefont{Samal}},\ }%
  \bibfield{title}{%
  \enquote{\bibinfo {title} {{Subsubleading soft graviton symmetry and MHV graviton scattering amplitudes}},}\ }%
  \bibfield{journal}{%
  \Doi{10.1007/JHEP08(2021)067}{\bibinfo {journal} {JHEP}}\ }%
  \textbf{\bibinfo {volume} {08}},\ \bibinfo {pages} {067} (\bibinfo {year} {2021}),\ \Eprint{http://arxiv.org/abs/2104.02546}{arXiv:2104.02546 [hep-th]}%
  \bibAnnoteFile{NoStop}{Banerjee:2021cly}%
\bibitem{Himwich:2023njb}%
  \BibitemOpen
  \bibfield{author}{%
  \bibinfo {author} {\bibfnamefont{Elizabeth}\ \bibnamefont{Himwich}}\ and\ \bibinfo {author} {\bibfnamefont{Monica}\ \bibnamefont{Pate}},\ }%
  \bibfield{title}{%
  \enquote{\bibinfo {title} {{w$_{1+\infty}$ in 4D gravitational scattering}},}\ }%
  \bibfield{journal}{%
  \Doi{10.1007/JHEP07(2024)180}{\bibinfo {journal} {JHEP}}\ }%
  \textbf{\bibinfo {volume} {07}},\ \bibinfo {pages} {180} (\bibinfo {year} {2024}),\ \Eprint{http://arxiv.org/abs/2312.08597}{arXiv:2312.08597 [hep-th]}%
  \bibAnnoteFile{NoStop}{Himwich:2023njb}%
\bibitem{Ball:2024oqa}%
  \BibitemOpen
  \bibfield{author}{%
  \bibinfo {author} {\bibfnamefont{Adam}\ \bibnamefont{Ball}},\ }%
  \bibfield{title}{%
  \enquote{\bibinfo {title} {{Currents in Celestial CFT}},}\ }%
   (\bibinfo {month} {7}\ \bibinfo {year} {2024}),\ \Eprint{http://arxiv.org/abs/2407.13558}{arXiv:2407.13558 [hep-th]}%
  \bibAnnoteFile{NoStop}{Ball:2024oqa}%
\bibitem{Guevara:2024ixn}%
  \BibitemOpen
  \bibfield{author}{%
  \bibinfo {author} {\bibfnamefont{Alfredo}\ \bibnamefont{Guevara}}, \bibinfo {author} {\bibfnamefont{Yangrui}\ \bibnamefont{Hu}},\ and\ \bibinfo {author} {\bibfnamefont{Sabrina}\ \bibnamefont{Pasterski}},\ }%
  \bibfield{title}{%
  \enquote{\bibinfo {title} {{Multiparticle Contributions to the Celestial OPE}},}\ }%
   (\bibinfo {month} {2}\ \bibinfo {year} {2024}),\ \Eprint{http://arxiv.org/abs/2402.18798}{arXiv:2402.18798 [hep-th]}%
  \bibAnnoteFile{NoStop}{Guevara:2024ixn}%
\bibitem{Ball:2022bgg}%
  \BibitemOpen
  \bibfield{author}{%
  \bibinfo {author} {\bibfnamefont{Adam}\ \bibnamefont{Ball}},\ }%
  \bibfield{title}{%
  \enquote{\bibinfo {title} {{Celestial locality and the Jacobi identity}},}\ }%
  \bibfield{journal}{%
  \Doi{10.1007/JHEP01(2023)146}{\bibinfo {journal} {JHEP}}\ }%
  \textbf{\bibinfo {volume} {01}},\ \bibinfo {pages} {146} (\bibinfo {year} {2023}),\ \Eprint{http://arxiv.org/abs/2211.09151}{arXiv:2211.09151 [hep-th]}%
  \bibAnnoteFile{NoStop}{Ball:2022bgg}%
\bibitem{Himwich:2021dau}%
  \BibitemOpen
  \bibfield{author}{%
  \bibinfo {author} {\bibfnamefont{Elizabeth}\ \bibnamefont{Himwich}}, \bibinfo {author} {\bibfnamefont{Monica}\ \bibnamefont{Pate}},\ and\ \bibinfo {author} {\bibfnamefont{Kyle}\ \bibnamefont{Singh}},\ }%
  \bibfield{title}{%
  \enquote{\bibinfo {title} {{Celestial operator product expansions and w$_{1+\infty}$ symmetry for all spins}},}\ }%
  \bibfield{journal}{%
  \Doi{10.1007/JHEP01(2022)080}{\bibinfo {journal} {JHEP}}\ }%
  \textbf{\bibinfo {volume} {01}},\ \bibinfo {pages} {080} (\bibinfo {year} {2022}),\ \Eprint{http://arxiv.org/abs/2108.07763}{arXiv:2108.07763 [hep-th]}%
  \bibAnnoteFile{NoStop}{Himwich:2021dau}%
\bibitem{jiang2022holographic}%
  \BibitemOpen
  \bibfield{author}{%
  \bibinfo {author} {\bibfnamefont{Hongliang}\ \bibnamefont{Jiang}},\ }%
  \bibfield{title}{%
  \enquote{\bibinfo {title} {Holographic chiral algebra: supersymmetry, infinite ward identities, and efts},}\ }%
  \bibfield{journal}{%
  \bibinfo {journal} {Journal of High Energy Physics}\ }%
  \textbf{\bibinfo {volume} {2022}},\ \bibinfo {pages} {1--40} (\bibinfo {year} {2022})%
  \bibAnnoteFile{NoStop}{jiang2022holographic}%
\bibitem{Bu:2021avc}%
  \BibitemOpen
  \bibfield{author}{%
  \bibinfo {author} {\bibfnamefont{Wei}\ \bibnamefont{Bu}},\ }%
  \bibfield{title}{%
  \enquote{\bibinfo {title} {{Supersymmetric celestial OPEs and soft algebras from the ambitwistor string worldsheet}},}\ }%
  \bibfield{journal}{%
  \Doi{10.1103/PhysRevD.105.126029}{\bibinfo {journal} {Phys. Rev. D}}\ }%
  \textbf{\bibinfo {volume} {105}},\ \bibinfo {pages} {126029} (\bibinfo {year} {2022}),\ \Eprint{http://arxiv.org/abs/2111.15584}{arXiv:2111.15584 [hep-th]}%
  \bibAnnoteFile{NoStop}{Bu:2021avc}%
\bibitem{Guevara:2021abz}%
  \BibitemOpen
  \bibfield{author}{%
  \bibinfo {author} {\bibfnamefont{Alfredo}\ \bibnamefont{Guevara}}, \bibinfo {author} {\bibfnamefont{Elizabeth}\ \bibnamefont{Himwich}}, \bibinfo {author} {\bibfnamefont{Monica}\ \bibnamefont{Pate}},\ and\ \bibinfo {author} {\bibfnamefont{Andrew}\ \bibnamefont{Strominger}},\ }%
  \bibfield{title}{%
  \enquote{\bibinfo {title} {{Holographic symmetry algebras for gauge theory and gravity}},}\ }%
  \bibfield{journal}{%
  \Doi{10.1007/JHEP11(2021)152}{\bibinfo {journal} {JHEP}}\ }%
  \textbf{\bibinfo {volume} {11}},\ \bibinfo {pages} {152} (\bibinfo {year} {2021}),\ \Eprint{http://arxiv.org/abs/2103.03961}{arXiv:2103.03961 [hep-th]}%
  \bibAnnoteFile{NoStop}{Guevara:2021abz}%
\bibitem{Mago:2021wje}%
  \BibitemOpen
  \bibfield{author}{%
  \bibinfo {author} {\bibfnamefont{Jorge}\ \bibnamefont{Mago}}, \bibinfo {author} {\bibfnamefont{Lecheng}\ \bibnamefont{Ren}}, \bibinfo {author} {\bibfnamefont{Akshay~Yelleshpur}\ \bibnamefont{Srikant}},\ and\ \bibinfo {author} {\bibfnamefont{Anastasia}\ \bibnamefont{Volovich}},\ }%
  \bibfield{title}{%
  \enquote{\bibinfo {title} {{Deformed $w_{1+\infty}$ Algebras in the Celestial CFT}},}\ }%
  \bibfield{journal}{%
  \Doi{10.3842/SIGMA.2023.044}{\bibinfo {journal} {SIGMA}}\ }%
  \textbf{\bibinfo {volume} {19}},\ \bibinfo {pages} {044} (\bibinfo {year} {2023}),\ \Eprint{http://arxiv.org/abs/2111.11356}{arXiv:2111.11356 [hep-th]}%
  \bibAnnoteFile{NoStop}{Mago:2021wje}%
\bibitem{Bakas:1989xu}%
  \BibitemOpen
  \bibfield{author}{%
  \bibinfo {author} {\bibfnamefont{I.}~\bibnamefont{Bakas}},\ }%
  \bibfield{title}{%
  \enquote{\bibinfo {title} {{The Large n Limit of Extended Conformal Symmetries}},}\ }%
  \bibfield{journal}{%
  \Doi{10.1016/0370-2693(89)90525-X}{\bibinfo {journal} {Phys. Lett. B}}\ }%
  \textbf{\bibinfo {volume} {228}},\ \bibinfo {pages} {57} (\bibinfo {year} {1989})%
  \bibAnnoteFile{NoStop}{Bakas:1989xu}%
\bibitem{Hoppe:1988gk}%
  \BibitemOpen
  \bibfield{author}{%
  \bibinfo {author} {\bibfnamefont{Jens}\ \bibnamefont{Hoppe}},\ }%
  \bibfield{title}{%
  \enquote{\bibinfo {title} {{Diffeomorphism Groups, Quantization and SU(infinity)}},}\ }%
  \bibfield{journal}{%
  \Doi{10.1142/S0217751X89002235}{\bibinfo {journal} {Int. J. Mod. Phys. A}}\ }%
  \textbf{\bibinfo {volume} {4}},\ \bibinfo {pages} {5235} (\bibinfo {year} {1989})%
  \bibAnnoteFile{NoStop}{Hoppe:1988gk}%
\bibitem{Penrose:1976js}%
  \BibitemOpen
  \bibfield{author}{%
  \bibinfo {author} {\bibfnamefont{R.}~\bibnamefont{Penrose}},\ }%
  \bibfield{title}{%
  \enquote{\bibinfo {title} {{Nonlinear Gravitons and Curved Twistor Theory}},}\ }%
  \bibfield{journal}{%
  \Doi{10.1007/BF00762011}{\bibinfo {journal} {Gen. Rel. Grav.}}\ }%
  \textbf{\bibinfo {volume} {7}},\ \bibinfo {pages} {31--52} (\bibinfo {year} {1976})%
  \bibAnnoteFile{NoStop}{Penrose:1976js}%
\bibitem{Park:1989fz}%
  \BibitemOpen
  \bibfield{author}{%
  \bibinfo {author} {\bibfnamefont{Q-Han}\ \bibnamefont{Park}},\ }%
  \bibfield{title}{%
  \enquote{\bibinfo {title} {{Extended Conformal Symmetries in Real Heavens}},}\ }%
  \bibfield{journal}{%
  \Doi{10.1016/0370-2693(90)90378-J}{\bibinfo {journal} {Phys. Lett. B}}\ }%
  \textbf{\bibinfo {volume} {236}},\ \bibinfo {pages} {429--432} (\bibinfo {year} {1990})%
  \bibAnnoteFile{NoStop}{Park:1989fz}%
\bibitem{Boyer:1985aj}%
  \BibitemOpen
  \bibfield{author}{%
  \bibinfo {author} {\bibfnamefont{C.~P.}\ \bibnamefont{Boyer}}\ and\ \bibinfo {author} {\bibfnamefont{J.~F.}\ \bibnamefont{Plebanski}},\ }%
  \bibfield{title}{%
  \enquote{\bibinfo {title} {{AN INFINITE HIERARCHY OF CONSERVATION LAWS AND NONLINEAR SUPERPOSITION PRINCIPLES FOR SELFDUAL EINSTEIN SPACES}},}\ }%
  \bibfield{journal}{%
  \Doi{10.1063/1.526652}{\bibinfo {journal} {J. Math. Phys.}}\ }%
  \textbf{\bibinfo {volume} {26}},\ \bibinfo {pages} {229--234} (\bibinfo {year} {1985})%
  \bibAnnoteFile{NoStop}{Boyer:1985aj}%
\bibitem{Adamo:2021lrv}%
  \BibitemOpen
  \bibfield{author}{%
  \bibinfo {author} {\bibfnamefont{Tim}\ \bibnamefont{Adamo}}, \bibinfo {author} {\bibfnamefont{Lionel}\ \bibnamefont{Mason}},\ and\ \bibinfo {author} {\bibfnamefont{Atul}\ \bibnamefont{Sharma}},\ }%
  \bibfield{title}{%
  \enquote{\bibinfo {title} {{Celestial $w_{1+\infty}$ Symmetries from Twistor Space}},}\ }%
  \bibfield{journal}{%
  \Doi{10.3842/SIGMA.2022.016}{\bibinfo {journal} {SIGMA}}\ }%
  \textbf{\bibinfo {volume} {18}},\ \bibinfo {pages} {016} (\bibinfo {year} {2022}),\ \Eprint{http://arxiv.org/abs/2110.06066}{arXiv:2110.06066 [hep-th]}%
  \bibAnnoteFile{NoStop}{Adamo:2021lrv}%
\bibitem{Mason:2007ct}%
  \BibitemOpen
  \bibfield{author}{%
  \bibinfo {author} {\bibfnamefont{L.~J.}\ \bibnamefont{Mason}}\ and\ \bibinfo {author} {\bibfnamefont{Martin}\ \bibnamefont{Wolf}},\ }%
  \bibfield{title}{%
  \enquote{\bibinfo {title} {{Twistor Actions for Self-Dual Supergravities}},}\ }%
  \bibfield{journal}{%
  \Doi{10.1007/s00220-009-0732-5}{\bibinfo {journal} {Commun. Math. Phys.}}\ }%
  \textbf{\bibinfo {volume} {288}},\ \bibinfo {pages} {97--123} (\bibinfo {year} {2009}),\ \Eprint{http://arxiv.org/abs/0706.1941}{arXiv:0706.1941 [hep-th]}%
  \bibAnnoteFile{NoStop}{Mason:2007ct}%
\bibitem{Wolf:2007tx}%
  \BibitemOpen
  \bibfield{author}{%
  \bibinfo {author} {\bibfnamefont{Martin}\ \bibnamefont{Wolf}},\ }%
  \bibfield{title}{%
  \enquote{\bibinfo {title} {{Self-Dual Supergravity and Twistor Theory}},}\ }%
  \bibfield{journal}{%
  \Doi{10.1088/0264-9381/24/24/010}{\bibinfo {journal} {Class. Quant. Grav.}}\ }%
  \textbf{\bibinfo {volume} {24}},\ \bibinfo {pages} {6287--6328} (\bibinfo {year} {2007}),\ \Eprint{http://arxiv.org/abs/0705.1422}{arXiv:0705.1422 [hep-th]}%
  \bibAnnoteFile{NoStop}{Wolf:2007tx}%
\bibitem{Inami:1988xy}%
  \BibitemOpen
  \bibfield{author}{%
  \bibinfo {author} {\bibfnamefont{T.}~\bibnamefont{Inami}}, \bibinfo {author} {\bibfnamefont{Y.}~\bibnamefont{Matsuo}},\ and\ \bibinfo {author} {\bibfnamefont{I.}~\bibnamefont{Yamanaka}},\ }%
  \bibfield{title}{%
  \enquote{\bibinfo {title} {{Extended Conformal Algebras With $N=1$ Supersymmetry}},}\ }%
  \bibfield{journal}{%
  \Doi{10.1016/0370-2693(88)90045-7}{\bibinfo {journal} {Phys. Lett. B}}\ }%
  \textbf{\bibinfo {volume} {215}},\ \bibinfo {pages} {701--705} (\bibinfo {year} {1988})%
  \bibAnnoteFile{NoStop}{Inami:1988xy}%
\bibitem{Bergshoeff:1990yd}%
  \BibitemOpen
  \bibfield{author}{%
  \bibinfo {author} {\bibfnamefont{E.}~\bibnamefont{Bergshoeff}}, \bibinfo {author} {\bibfnamefont{C.~N.}\ \bibnamefont{Pope}}, \bibinfo {author} {\bibfnamefont{L.~J.}\ \bibnamefont{Romans}}, \bibinfo {author} {\bibfnamefont{E.}~\bibnamefont{Sezgin}},\ and\ \bibinfo {author} {\bibfnamefont{X.}~\bibnamefont{Shen}},\ }%
  \bibfield{title}{%
  \enquote{\bibinfo {title} {{The Super $W$(infinity) Algebra}},}\ }%
  \bibfield{journal}{%
  \Doi{10.1016/0370-2693(90)90672-S}{\bibinfo {journal} {Phys. Lett. B}}\ }%
  \textbf{\bibinfo {volume} {245}},\ \bibinfo {pages} {447--452} (\bibinfo {year} {1990})%
  \bibAnnoteFile{NoStop}{Bergshoeff:1990yd}%
\bibitem{Buffon:1996dv}%
  \BibitemOpen
  \bibfield{author}{%
  \bibinfo {author} {\bibfnamefont{L.~O.}\ \bibnamefont{Buffon}}, \bibinfo {author} {\bibfnamefont{D.}~\bibnamefont{Dalmazi}},\ and\ \bibinfo {author} {\bibfnamefont{A.}~\bibnamefont{Zadra}},\ }%
  \bibfield{title}{%
  \enquote{\bibinfo {title} {{Classical and quantum N=1 super W infinity algebras}},}\ }%
  \bibfield{journal}{%
  \Doi{10.1142/S0217732396002332}{\bibinfo {journal} {Mod. Phys. Lett. A}}\ }%
  \textbf{\bibinfo {volume} {11}},\ \bibinfo {pages} {2339--2350} (\bibinfo {year} {1996}),\ \Eprint{http://arxiv.org/abs/hep-th/9607122}{arXiv:hep-th/9607122}%
  \bibAnnoteFile{NoStop}{Buffon:1996dv}%
\bibitem{Ahn:2021erj}%
  \BibitemOpen
  \bibfield{author}{%
  \bibinfo {author} {\bibfnamefont{Changhyun}\ \bibnamefont{Ahn}},\ }%
  \bibfield{title}{%
  \enquote{\bibinfo {title} {{Towards a supersymmetric w1+\ensuremath{\infty} symmetry in the celestial conformal field theory}},}\ }%
  \bibfield{journal}{%
  \Doi{10.1103/PhysRevD.105.086028}{\bibinfo {journal} {Phys. Rev. D}}\ }%
  \textbf{\bibinfo {volume} {105}},\ \bibinfo {pages} {086028} (\bibinfo {year} {2022}),\ \Eprint{http://arxiv.org/abs/2111.04268}{arXiv:2111.04268 [hep-th]}%
  \bibAnnoteFile{NoStop}{Ahn:2021erj}%
\bibitem{Ahn:2022oor}%
  \BibitemOpen
  \bibfield{author}{%
  \bibinfo {author} {\bibfnamefont{Changhyun}\ \bibnamefont{Ahn}},\ }%
  \bibfield{title}{%
  \enquote{\bibinfo {title} {{A deformed supersymmetric $w_{1+\infty }$ symmetry in the celestial conformal field theory}},}\ }%
  \bibfield{journal}{%
  \Doi{10.1140/epjc/s10052-022-10582-9}{\bibinfo {journal} {Eur. Phys. J. C}}\ }%
  \textbf{\bibinfo {volume} {82}},\ \bibinfo {pages} {630} (\bibinfo {year} {2022}),\ \Eprint{http://arxiv.org/abs/2202.02949}{arXiv:2202.02949 [hep-th]}%
  \bibAnnoteFile{NoStop}{Ahn:2022oor}%
\bibitem{Ahn:2024kpv}%
  \BibitemOpen
  \bibfield{author}{%
  \bibinfo {author} {\bibfnamefont{Changhyun}\ \bibnamefont{Ahn}}\ and\ \bibinfo {author} {\bibfnamefont{Man~Hea}\ \bibnamefont{Kim}},\ }%
  \bibfield{title}{%
  \enquote{\bibinfo {title} {{A supersymmetric extension of w$_{1+\infty}$ algebra in the celestial holography}},}\ }%
  \bibfield{journal}{%
  \Doi{10.1007/JHEP09(2024)081}{\bibinfo {journal} {JHEP}}\ }%
  \textbf{\bibinfo {volume} {09}},\ \bibinfo {pages} {081} (\bibinfo {year} {2024}),\ \Eprint{http://arxiv.org/abs/2407.05601}{arXiv:2407.05601 [hep-th]}%
  \bibAnnoteFile{NoStop}{Ahn:2024kpv}%
\bibitem{Adamo:2015ina}%
  \BibitemOpen
  \bibfield{author}{%
  \bibinfo {author} {\bibfnamefont{Tim}\ \bibnamefont{Adamo}},\ }%
  \bibfield{title}{%
  \enquote{\bibinfo {title} {{Gravity with a cosmological constant from rational curves}},}\ }%
  \bibfield{journal}{%
  \Doi{10.1007/JHEP11(2015)098}{\bibinfo {journal} {JHEP}}\ }%
  \textbf{\bibinfo {volume} {11}},\ \bibinfo {pages} {098} (\bibinfo {year} {2015}),\ \Eprint{http://arxiv.org/abs/1508.02554}{arXiv:1508.02554 [hep-th]}%
  \bibAnnoteFile{NoStop}{Adamo:2015ina}%
\bibitem{VanProeyen:1999ni}%
  \BibitemOpen
  \bibfield{author}{%
  \bibinfo {author} {\bibfnamefont{Antoine}\ \bibnamefont{Van~Proeyen}},\ }%
  \bibfield{title}{%
  \enquote{\bibinfo {title} {{Tools for supersymmetry}},}\ }%
  \bibfield{journal}{%
  \bibinfo {journal} {Ann. U. Craiova Phys.}\ }%
  \textbf{\bibinfo {volume} {9}},\ \bibinfo {pages} {1--48} (\bibinfo {year} {1999}),\ \Eprint{http://arxiv.org/abs/hep-th/9910030}{arXiv:hep-th/9910030}%
  \bibAnnoteFile{NoStop}{VanProeyen:1999ni}%
\bibitem{Taylor:2023bzj}%
  \BibitemOpen
  \bibfield{author}{%
  \bibinfo {author} {\bibfnamefont{Tomasz~R.}\ \bibnamefont{Taylor}}\ and\ \bibinfo {author} {\bibfnamefont{Bin}\ \bibnamefont{Zhu}},\ }%
  \bibfield{title}{%
  \enquote{\bibinfo {title} {{Celestial Supersymmetry}},}\ }%
  \bibfield{journal}{%
  \Doi{10.1007/JHEP06(2023)210}{\bibinfo {journal} {JHEP}}\ }%
  \textbf{\bibinfo {volume} {06}},\ \bibinfo {pages} {210} (\bibinfo {year} {2023}),\ \Eprint{http://arxiv.org/abs/2302.12830}{arXiv:2302.12830 [hep-th]}%
  \bibAnnoteFile{NoStop}{Taylor:2023bzj}%
\bibitem{Stieberger:2022zyk}%
  \BibitemOpen
  \bibfield{author}{%
  \bibinfo {author} {\bibfnamefont{Stephan}\ \bibnamefont{Stieberger}}, \bibinfo {author} {\bibfnamefont{Tomasz~R.}\ \bibnamefont{Taylor}},\ and\ \bibinfo {author} {\bibfnamefont{Bin}\ \bibnamefont{Zhu}},\ }%
  \bibfield{title}{%
  \enquote{\bibinfo {title} {{Celestial Liouville theory for Yang-Mills amplitudes}},}\ }%
  \bibfield{journal}{%
  \Doi{10.1016/j.physletb.2022.137588}{\bibinfo {journal} {Phys. Lett. B}}\ }%
  \textbf{\bibinfo {volume} {836}},\ \bibinfo {pages} {137588} (\bibinfo {year} {2023}),\ \Eprint{http://arxiv.org/abs/2209.02724}{arXiv:2209.02724 [hep-th]}%
  \bibAnnoteFile{NoStop}{Stieberger:2022zyk}%
\bibitem{Stieberger:2023fju}%
  \BibitemOpen
  \bibfield{author}{%
  \bibinfo {author} {\bibfnamefont{Stephan}\ \bibnamefont{Stieberger}}, \bibinfo {author} {\bibfnamefont{Tomasz~R.}\ \bibnamefont{Taylor}},\ and\ \bibinfo {author} {\bibfnamefont{Bin}\ \bibnamefont{Zhu}},\ }%
  \bibfield{title}{%
  \enquote{\bibinfo {title} {{Yang-Mills as a Liouville theory}},}\ }%
  \bibfield{journal}{%
  \Doi{10.1016/j.physletb.2023.138229}{\bibinfo {journal} {Phys. Lett. B}}\ }%
  \textbf{\bibinfo {volume} {846}},\ \bibinfo {pages} {138229} (\bibinfo {year} {2023}),\ \Eprint{http://arxiv.org/abs/2308.09741}{arXiv:2308.09741 [hep-th]}%
  \bibAnnoteFile{NoStop}{Stieberger:2023fju}%
\bibitem{Melton:2024akx}%
  \BibitemOpen
  \bibfield{author}{%
  \bibinfo {author} {\bibfnamefont{Walker}\ \bibnamefont{Melton}}, \bibinfo {author} {\bibfnamefont{Atul}\ \bibnamefont{Sharma}}, \bibinfo {author} {\bibfnamefont{Andrew}\ \bibnamefont{Strominger}},\ and\ \bibinfo {author} {\bibfnamefont{Tianli}\ \bibnamefont{Wang}},\ }%
  \bibfield{title}{%
  \enquote{\bibinfo {title} {{Celestial Dual for Maximal Helicity Violating Amplitudes}},}\ }%
  \bibfield{journal}{%
  \Doi{10.1103/PhysRevLett.133.091603}{\bibinfo {journal} {Phys. Rev. Lett.}}\ }%
  \textbf{\bibinfo {volume} {133}},\ \bibinfo {pages} {091603} (\bibinfo {year} {2024}),\ \Eprint{http://arxiv.org/abs/2403.18896}{arXiv:2403.18896 [hep-th]}%
  \bibAnnoteFile{NoStop}{Melton:2024akx}%
\bibitem{Mol:2024etg}%
  \BibitemOpen
  \bibfield{author}{%
  \bibinfo {author} {\bibfnamefont{Igor}\ \bibnamefont{Mol}},\ }%
  \bibfield{title}{%
  \enquote{\bibinfo {title} {{A Holographic Construction of MHV Graviton Amplitudes in Celestial CFT}},}\ }%
   (\bibinfo {month} {8}\ \bibinfo {year} {2024}),\ \Eprint{http://arxiv.org/abs/2408.10944}{arXiv:2408.10944 [hep-th]}%
  \bibAnnoteFile{NoStop}{Mol:2024etg}%
\bibitem{Mol:2024onu}%
  \BibitemOpen
  \bibfield{author}{%
  \bibinfo {author} {\bibfnamefont{Igor}\ \bibnamefont{Mol}},\ }%
  \bibfield{title}{%
  \enquote{\bibinfo {title} {{Comments on Celestial CFT and $AdS_{3}$ String Theory}},}\ }%
   (\bibinfo {month} {10}\ \bibinfo {year} {2024}),\ \Eprint{http://arxiv.org/abs/2410.02620}{arXiv:2410.02620 [hep-th]}%
  \bibAnnoteFile{NoStop}{Mol:2024onu}%
\bibitem{Mol:2024qct}%
  \BibitemOpen
  \bibfield{author}{%
  \bibinfo {author} {\bibfnamefont{Igor}\ \bibnamefont{Mol}},\ }%
  \bibfield{title}{%
  \enquote{\bibinfo {title} {{Partial Differential Equations for MHV Celestial Amplitudes in Liouville Theory}},}\ }%
   (\bibinfo {month} {9}\ \bibinfo {year} {2024}),\ \Eprint{http://arxiv.org/abs/2409.05936}{arXiv:2409.05936 [hep-th]}%
  \bibAnnoteFile{NoStop}{Mol:2024qct}%
\bibitem{Srednicki:2007qs}%
  \BibitemOpen
  \bibfield{author}{%
  \bibinfo {author} {\bibfnamefont{M.}~\bibnamefont{Srednicki}},\ }%
  \emph{\bibinfo {title} {{Quantum field theory}}}\ (\bibinfo {publisher} {Cambridge University Press},\ \bibinfo {year} {2007})\ ISBN \bibinfo {isbn} {978-0-521-86449-7, 978-0-511-26720-8}%
  \bibAnnoteFile{NoStop}{Srednicki:2007qs}%
\bibitem{Mizera:2023tfe}%
  \BibitemOpen
  \bibfield{author}{%
  \bibinfo {author} {\bibfnamefont{Sebastian}\ \bibnamefont{Mizera}},\ }%
  \bibfield{title}{%
  \enquote{\bibinfo {title} {{Physics of the analytic S-matrix}},}\ }%
  \bibfield{journal}{%
  \Doi{10.1016/j.physrep.2023.10.006}{\bibinfo {journal} {Phys. Rept.}}\ }%
  \textbf{\bibinfo {volume} {1047}},\ \bibinfo {pages} {1--92} (\bibinfo {year} {2024}),\ \Eprint{http://arxiv.org/abs/2306.05395}{arXiv:2306.05395 [hep-th]}%
  \bibAnnoteFile{NoStop}{Mizera:2023tfe}%
\end{thebibliography}%
\bibliographystyle{apsrev4-1long}

\end{document}